\documentclass[11pt, a4paper]{article}
\pdfoutput=1
\usepackage{jcappub}
\usepackage{amsmath}
\usepackage{amssymb}
\usepackage{mathrsfs}
\usepackage[toc,page]{appendix}
\usepackage{calligra}

\DeclareMathAlphabet{\mathpzc}{OT1}{pzc}{m}{it} 
\DeclareMathAlphabet{\mathcalligra}{T1}{calligra}{m}{n}
\DeclareFontShape{T1}{calligra}{m}{n}{<->s*[2.2]callig15}{}

\newcommand{\rr}{\ensuremath{\mathcalligra{r}}}

\title{The virialization density of peaks with general density profiles under spherical collapse}

\author[a]{Douglas Rubin}
\author[b]{and Abraham Loeb}

\affiliation[a]{Department of Physics, Harvard University, \\
Cambridge, MA 02138, USA}
\affiliation[b]{Department of Astronomy, Harvard University, \\
Cambridge, MA 02138, USA}

\emailAdd{dsrubin@physics.harvard.edu}
\emailAdd{aloeb@cfa.harvard.edu}

\abstract{We calculate the non-linear virialization density, $\Delta_c$, of halos under spherical collapse from peaks with an arbitrary initial and final density profile.  This is in contrast to the standard calculation of $\Delta_c$ which assumes top-hat profiles.  Given our formalism, the non-linear halo density can be calculated once the shape of the initial peak's density profile and the shape of the virialized halo's profile are provided.  We solve for $\Delta_c$ for halos in an Einstein de-Sitter and $\Lambda$CDM universe.  As examples, we consider power-law initial profiles as well as spherically averaged peak profiles calculated from the statistics of a Gaussian random field.  

We find that, depending on the profiles used, $\Delta_c$ is smaller by a factor of a few to as much as a factor of 10 as compared to the density given by the standard calculation ($\approx 200$).  Using our results, we show that, for halo finding algorithms that identify halos through an over-density threshold, the halo mass function measured from cosmological simulations can be enhanced at all halo masses by a factor of a few.  This difference could be important when using numerical simulations to assess the validity of analytic models of the halo mass function.}


\arxivnumber{arxiv number goes here}

\begin{document}
\maketitle
\flushbottom

\section{Introduction}

%The processes underlying the formation and abundance of cosmological dark matter halos has received considerable attention for many years.  Analytic models , this has been explored models such as...    both analytically and numerically 

%The physics underlying the abundance of cosmological dark matter halos has received considerable attention for many years.  The problem has been studied both with analytic models such as the excursion set formalism and with high resolution numerical simulations.  Both of these approaches rely on a threshold in the density field above which a halo is defined.  In the excursion set formalism, the abundance of halos can be predicted by setting a linearized density barrier \cite{bond, sheth:2001, sheth:2002}, used to calculate the fraction of particular ``trajectories" in a diffusion process.  In numerical simulations, certain halo finding algorithms directly measure the mass fuction of dark matter halos (e.g., \cite{sheth:1999, jenkins, springel, warren, reed, tinker, crocce, angulo, watson}) by searching the density field for a predetermined non-linear density threshold (known as the spherical over-density method of \cite{lacey}).

The physics underlying the abundance of cosmological dark matter halos has received considerable attention for several decades.  The problem has been studied both with analytic models such as the excursion set formalism and with high resolution numerical simulations.  In the excursion set formalism, the abundance of halos can be predicted by setting a linearized density barrier, used to calculate the fraction of particular ``trajectories" in a diffusion process \cite{bond, sheth:2001, sheth:2002}.  In numerical simulations, the halo mass function is directly measured by searching the cosmological density field for halos identified by halo finding algorithms (e.g., \cite{sheth:1999, jenkins, springel, warren, reed, tinker, crocce, angulo}).  The success of an analytic model is often based on its agreement with the halo mass function as measured from simulations.  The measured mass function, however, is dependent on the method used to identify halos.  A commonly used method is the friends-of-friends algorithm \cite{davis} which assigns particles to a particular halo when they are separated by less than a linking length.  Another method is the spherical over-density algorithm \cite{lacey:1994} which identifies halos through a predetermined non-linear density threshold.  The halo mass function measured with this method of course depends on the threshold used \cite{watson}.  In order to compare these mass functions to the predictions of analytic models, it is therefore necessary to provide an appropriate value for the expected non-linear density of a halo.

The non-linear density of a halo is also useful in estimating physical properties of dark matter halos such as the virial radius, virial temperature and circular velocity (see for example \cite{loeb:2012}).  It is also frequently used to normalize formulas for the density profiles of halos.  For example, the normalization constant for a Navarro, Frenk \& White (NFW) profile, $\rho = \rho_o/ [ c w (1+cw)^2]$ (where $w$ is the radial position scaled by halo radius), \cite{navarro} is given by
\begin{equation}
\label{eq:norm_nfw}
\rho_o = \frac{\rho_c(z_{vir})\Delta_c(z_{vir})}{3} \frac{c^3}{\ln(1+c)-\frac{c}{1+c}},
\end{equation}
where $\rho_c(z_{vir})$ is the critical density of the universe at the time of halo virialization, $c$ is the concentration parameter, and $\Delta_c(z_{vir})$ is the volume-averaged, non-linear density of the halo at virialization in units of the critical density of the universe.

The value of $\Delta_c$ is typically taken to be $\approx 200$, derived by considering the dynamics of spherical collapse \cite{gunn}.  It is calculated by assuming that at an early time, the initial density profile of the nascent halo is uniform out to its edge (also known as a ``top-hat").  According to the spherical collapse model, for this profile, all shells within the density perturbation will have self-similar trajectories, and will therefore turn-around at the same time.  At turn-around, therefore, the kinetic energy of the system is zero.  One then assumes that the energy of the system is conserved between the turn-around time and the time it takes the system to reach virial equilibrium.  To calculate the halo's potential energy at virialization it is customary to assume a top-hat density profile.  By employing the virial thereom and setting the energies at these times equal to each other, it is possible to solve for the ratio of the halo's virial radius to its turn-around radius.  This value cubed gives the collapsing sphere's fractional change in volume. To calculate $\Delta_c$, it is left to multiply by the ratio of the turn-around mean density to the critical density of the universe at virialization.  This ratio is found with the spherical collapse solution, assuming that virialization occurs at twice the turn-around time.

For an Einstein de-Sitter universe (E-dS), $R_{vir}/R_{ta} = 1/2$ and $\Delta_c=18 \pi^2 \approx178$ \cite{gunn}.  For cosmologies including a cosmological constant, the calculation is slightly more complex since the cosmological constant contributes gravitational energy and must be included in the virial thereom \cite{lahav}.  Further, the equations for spherical collapse are more complicated and must be solved numerically.  This calculation has been done for various cosmologies by \cite{eke, lacey, bryan} and for general cosmologies by \cite{rubin:2012, lokas}.  They find that, depending on the cosmology, the value of $\Delta_c$ can be larger or smaller than the E-dS value by about a factor of 2.

The assumptions of top-hat density profiles at the initial time and virialization lead to several simplifications which make the calculation relatively easy.  Since the assumption of an initial top-hat results in zero kinetic energy at turn-around, the total kinetic energy of the system at this time need not be calculated.  Moreover, for this initial density profile, the sphere maintains its top-hat shape during collapse, so that the calculation of its potential energy at turn-around (due to gravity and the cosmological constant for cosmologies with $\Lambda \neq 0$) is simply that of a homogenous sphere.  Finally, for an initial top-hat profile, since the shells are on self-similiar trajectories, they do not cross before turn-around.  This simplifies the spherical collapse problem since the mass within each shell remains constant in time.  By assuming
a top-hat density profile at virialization, the calculation of the halo's potential energy due to gravity and $\Lambda$ is also simplified to that of a homogenous sphere.

To calculate more realistic values of the non-linear over-density of a halo at virialization, we repeat this calculation, generalized for any density profiles.  We re-derive the formulas for $R_{vir}/R_{ta}$ and $\Delta_c$ to allow for non-zero kinetic energy at turn-around as well potential energies for an arbitrary density profile sphere.  Given an initial, realistic density profile, we use the spherical collapse equations to analytically calculate the velocity and density profiles at turn-around (and thus the total kinetic energy and potential energies due to gravity and $\Lambda$).  For initial density profiles which result in shell crossings before turn-around, we employ a one dimensional code to numerically solve the equations of spherical collapse.  By specifying realistic density profiles for the virialized halo, we are able to calculate realistic potential energies due to gravity and $\Lambda$ at virialization.  We non-dimensionalize our equations in such a way that, as with the standard calculation, $R_{vir}/R_{ta}$ and $\Delta_c$ are independent of halo mass and only depend on the redshift of virialization. 

Our calculation still relies on an idealized spherical geometry in the Newtonian limit.  However, the formalism we derive allows us to utilize realistic density profiles and thus calculate more representative estimates of the non-linear over-density of a virialized halo.  This is favorable when using the halo mass function measured from numerical simulations with the spherical over-density algorithm to compare to analytic models.  It also provides us with better estimates of the physical parameters of virialized halos as mentioned above.  Moreover, our calculation allows us to normalize halo density profiles in a self consistent manner.  For example, to normalize an NFW profile with Eqn.~\ref{eq:norm_nfw}, we can calculate $\Delta_c(z_{vir})$ using an actual NFW profile at virialization, rather than a top-hat as in the standard calculation.  We can do this because, in deriving our formalism, we non-dimensionalize all equations such that we do not require the normalized density, but only its shape. 

In \S~\ref{sec:spherical_collapse}, we present relevant equations from the spherical collapse model which are used for derivations in the rest of this paper.  In \S~\ref{sec:eds_cosmology}, we consider an E-dS cosmology and derive the formula for the non-linear halo over-density for any initial and final density profiles (\S~\ref{sec:eds_formulas}).  Using spherical collapse dynamics in an E-dS universe, we solve for the velocity and density profiles at turn-around and use these to calculate the total kinetic and potential energies in \S~\ref{sec:eds_conditions}.  We present our E-dS results for several density profiles in \S~\ref{sec:eds_results}. In \S~\ref{sec:gen_cosmology} we perform the same derivations as in the E-dS case, but keep our equations general to allow for a cosmological constant and curvature.  In \S~\ref{sec:results_gen}, we show results for a $\Lambda$CDM cosmology with highly realistic initial density profiles as calculated from peak statistics in a Gaussian random field.  

\section{Relevant Results from the Spherical Collapse Model}
\label{sec:spherical_collapse}

In this section, we summarize a few key results from the spherical collapse model that will be of use later in this paper.  For a more detailed treatment of spherical collapse, we refer the reader to \cite{gunn, peebles:1980, rubin:2012, loeb:2006, peebles, bertschinger_1, bertschinger_2, lokas, sheth:2004}.  According to the spherical collapse model, the evolution of a spherical perturbation in the cosmic density field is understood as a series of thin, concentric shells of mass whose positions vary with time.  Solving for the evolution of the perturbation is then simply reduced to a problem of kinematics.  In the Newtonian limit, the evolution of a shell is governed by the following partial differential equation, found by integrating the shell's equation of motion and assuming an initial velocity given by linear theory \cite{loeb:2006}:
\begin{equation}
\label{eq:eq_of_motion}
\frac{1}{H_o^2} \left(\frac{\partial \mathpzc{x}}{\partial t}\right)^2 = \frac{\Omega_m}{\mathpzc{x}}+\Omega_{\Lambda}\mathpzc{x}^2+\Omega_k-\frac{5}{3}\frac{\bar \delta_i(r_i)}{a_i}\Omega_m,
\end{equation}
where the radius of the shell, $r(t, r_i)$, is non-dimensionalized with
\begin{equation}
\label{eq:def_x}
\mathpzc{x} \equiv \frac{r a_i}{r_i}. 
\end{equation}
Here, $r_i$ is the initial position of the shell when the scale factor is $a_i$, $H_o$ is the present-day Hubble parameter, and $\Omega_m$, $\Omega_{\Lambda}$ and $\Omega_k$ ($=1-\Omega_m-\Omega_{\Lambda}$) are the present-day matter, vacuum and curvature energy densities respectively.  The initial density profile of the perturbation is parameterized by $\delta_i(r_i)$, defined as $\rho_i(r_i)/\bar \rho_m(a_i)-1$, where $\bar \rho_m(a_i)$ is the mean matter density of the universe at $a_i$.  The ``bar" over the delta denotes a volume average: 
\begin{equation}
\label{eq:vol_avg}
\bar \delta(r) = \frac{3}{r^3} \int_0^r \delta(r^\prime)r^{\prime 2}dr^\prime.  
\end{equation}
Notice that for a top-hat profile, $\bar \delta_i(r_i)=const$ and Eqn.~\ref{eq:eq_of_motion} is independent of initial position so that the trajectories of each shell are self-similar.  Equation~\ref{eq:eq_of_motion} is strictly valid only if the mass within $r$ is constant in time (i.e. there are no shell crossings for the shell in question).  In a perturbation consisting of dark matter, shell crossing is a legitimate concern since the matter is collisionless, and shells can therefore slide past each other unencumbered.  It should also be noted that the equation was derived assuming that both $|\bar \delta_i|$ and $a_i$ are $\ll 1$.  Indeed, we make this assumption in our derivations throughout the rest of this paper.

Eqn.~\ref{eq:eq_of_motion} can be further integrated to find the time, $t H_o$, at which a shell has reached a position $\mathpzc{x}(t)$ \cite{rubin:2012}, 
\begin{equation}
\label{eq:int_eq_of_mo}
tH_o=\left \{
 \begin{array}{lcl}
\mathcal I \left[0, \mathpzc{x}(t), \frac{\bar \delta_i(r_i)}{a_i} \right] &\mbox{for } ~~~t H_o \le t_{\scriptscriptstyle{\mathscr{TA}}}H_o \\
\mathcal I \left[0, \mathpzc{x}_{\scriptscriptstyle{\mathscr{TA}}}, \frac{\bar \delta_i(r_i)}{a_i} \right] +\mathcal I \left[\mathpzc{x}(t), \mathpzc{x}_{\scriptscriptstyle{\mathscr{TA}}}, \frac{\bar \delta_i(r_i)}{a_i} \right] &\mbox{for } ~~~tH_o > t_{\scriptscriptstyle{\mathscr{TA}}} H_o
\end{array} \right.,
\end{equation}
with 
\begin{equation}
\label{eq:def_I}
\mathcal{I}(l, u, d)\equiv \int_l^{u} d \lambda \left [1+\Omega_m \left ( \frac{1}{\lambda}-1-\frac{5}{3}d\right ) +\Omega_{\Lambda}\left(\lambda^2-1\right)\right ]^{-1/2}.
\end{equation}
The first line of Eqn.~\ref{eq:int_eq_of_mo} applies to shells which have yet to turn-around (denoted by ${\scriptscriptstyle{\mathscr{TA}}}$), and the second applies to shells which have already turned around.  The first integral in the second line represents the amount of time that it takes for a shell to reach turn-around, and the second integral represents the amount of time between turn-around and $t$.  Of course if the energy of the shell is greater than zero, it will be on an unbound orbit and will never collapse\footnote{This is neglecting the effect of shells crossing from its exterior to its interior.}.  In this paper, however, we only concern ourselves with shells with bound trajectories. 

For an E-dS universe the trajectory can be written in parametric form.  This is obtained by first integrating Eqn.~\ref{eq:eq_of_motion} with $\Omega_k=\Omega_{\Lambda}=0$ and $\Omega_m=1$ to find
\begin{equation}
\label{eq:param_int}
 t = \frac{1}{H_o}\int_0^ \mathpzc{x} \frac{\sqrt{ \mathpzc{x^\prime}} d\mathpzc{x^\prime}}{\sqrt{1\pm \frac{5}{3}\frac{|\bar \delta_i|}{a_i}\mathpzc{x^\prime}}},
 \end{equation}
where the plus and minus sign correspond to open $(\bar \delta_i<0)$ and closed $(\bar \delta_i>0)$ trajectories respectively.  Since in this paper we are considing collapsing halos, we derive the parametric trajectory for only the closed (although the derivation for the open case is almost exactly same, using hyperbolic geometry).  By defining, $\sin^2(\Theta/2) \equiv (5/3) (\bar \delta_i/a_i)\mathpzc{x}$, we can take advantage of the Pythagorean trigonometric identity in the denominator, so that the integral may be computed analytically\footnote{We may use this definition since $\mathpzc{x} \in [0, (3/5)(a_i/\bar \delta_i)]$, where the maximal value of $\mathpzc{x}$ can quickly be verified by solving for the extremum of Eqn.~\ref{eq:eq_of_motion}.}.  Switching the variable of integration to $\Theta$ results in
\begin{align}
\label{eq:eq_of_motion_param2}
t &= \frac{1}{H_o}\left[\frac{5}{3}\frac{\bar \delta_i(r_i)}{a_i} \right]^{-3/2} \int_0^{\Theta} \sin^2 \left (\frac{\Theta^\prime}{2} \right)d \Theta^{\prime} \nonumber \\
&=\frac{1}{2H_o} \left[\frac{5}{3}\frac{\bar \delta_i(r_i)}{a_i} \right]^{-3/2}\left [\Theta(r_i)-\sin \Theta(r_i)\right].
\end{align}
The solution for $\mathpzc{x}$ as a function of $\Theta$ may also be simplified with the double-angle formula,
\begin{align}
\label{eq:eq_of_motion_param1}
\mathpzc{x} &= \left[\frac{5}{3}\frac{\bar \delta_i(r_i)}{a_i} \right]^{-1} \sin^2 \frac{\Theta}{2} \nonumber \\
&=\frac{1}{2} \left[\frac{5}{3}\frac{\bar \delta_i(r_i)}{a_i} \right]^{-1}\left [1-\cos \Theta(r_i)\right ].
\end{align}
We have written the so-called ``development angle", $\Theta$, as a function of $r_i$ to make explicit that each shell in the perturbation moves independently of one another, and so has its own development angle parameterizing its motion.  The values of $\Theta=0$, $\pi$ and $2\pi$ correspond to the the initial, turn-around and collapse times for a shell respectively. This parametric solution is valid until shell crossing, which, for a top-hat initial perturbation, occurs at $\Theta= 2 \pi$, when the shells have collapse to a singularity, cross each other, and then re-expand (of course the Newtonian approximation will break down at this point).  Evaluating Eqn.~\ref{eq:eq_of_motion_param2} at $\Theta=\pi$, we find that $t_{ta} H_o = \pi /2 [(5/3) (\bar \delta_i/a_i)]^{-3/2}$, and thus, the turn-around time decreases monotonically with $\bar \delta_i$.  In fact, one can show that $t_{ta}$ decreases monotonically with $\bar \delta_i$ for a general cosmology by evaluating $\mathcal I [0, \mathpzc{x}_{\scriptscriptstyle{\mathscr{TA}}}(\bar \delta_i/a_i), \bar \delta_i/a_i]$  ($=t_{ta} H_o$), with $\mathpzc{x}_{\scriptscriptstyle{\mathscr{TA}}}(\bar \delta_i/a_i)$ found by solving Eqn.~\ref{eq:cubic_ta}, and plotting $t_{ta}H_o$ vs. $\bar \delta_i/a_i$.   Thus, for a perturbation with a monotonically decreasing $\bar \delta_i$ profile, collapse proceeds from the inside out, and we do not have to worry about shell crossing until the innermost shell reaches the center of the sphere and crosses itself.

%For an Einstein de Sitter universe (E-dS), the solution can be written in parametric form:
%\begin{equation}
%\label{eq:eq_of_motion_param1}
%\mathpzc{x}=\frac{1}{2} \left[\frac{5}{3}\frac{\bar \delta_i(r_i)}{a_i} \right]^{-1}\left [1-\cos \Theta(r_i)\right ],
%\end{equation}
%and
%\begin{equation}
%\label{eq:eq_of_motion_param2}
%t=\frac{1}{2H_o} \left[\frac{5}{3}\frac{\bar \delta_i(r_i)}{a_i} \right]^{-3/2}\left [\Theta(r_i)-\sin \Theta(r_i)\right],
%\end{equation}
%where the development angle $\Theta(r_i) \in [0, 2\pi]$.  The values of $\Theta=0$, $\pi$ and $2\pi$ %correspond to the the initial, turn-around and collapse times for a shell respectively.  We have written $\Theta$ as a function of $r_i$ to make explicit that (in the absence of shell crossing) each moves independently of each other, and so has its own development angle.  

The value of $\mathpzc{x}$ for a shell at turn-around, $\mathpzc{x}_{\scriptscriptstyle{\mathscr{TA}}}$, can be found by setting the velocity in Eqn.~\ref{eq:eq_of_motion} to $0$, resulting in the following cubic:
\begin{equation}
\label{eq:cubic_ta}
\Omega_{\Lambda}\mathpzc{x}_{\scriptscriptstyle{\mathscr{TA}}}^3+\left [\Omega_k-\frac{5}{3}\frac{\bar \delta_i(r_i)}{a_i}\Omega_m\right]\mathpzc{x}_{\scriptscriptstyle{\mathscr{TA}}}+\Omega_m=0.
\end{equation} 
For a general cosmology, $\mathpzc{x}_{\scriptscriptstyle{\mathscr{TA}}}$ must either be solved numerically by taking the smallest, positive, pure real root (if one exists)\footnote{The proper solution is the \textit{smallest} positive, pure real root since for the case of two pure real, positive roots, an expanding sphere first reaches the smaller root, turns around and collapses to zero.  Any turn-around solution after this time is spurious since Eqn.~\ref{eq:eq_of_motion} is no longer valid because shell crossing at the origin has occurred.} or by using the closed form solution of $\mathpzc{x}_{\scriptscriptstyle{\mathscr{TA}}}$ presented in \cite{rubin:2012} (their Eqns. 2.13-2.15).  For an E-dS universe, Eqn.~\ref{eq:cubic_ta} simplifies to 
\begin{equation}
\label{eq:xta_eds}
\mathpzc{x}_{\scriptscriptstyle{\mathscr{TA}}} = \left [\frac{5}{3}\frac{\bar \delta_i(r_i)}{a_i}\right]^{-1},
\end{equation}
which is the same turn-around solution obtained by setting $\Theta=\pi$ in Eqn.~\ref{eq:eq_of_motion_param1}.  Throughout this paper, we will refer to two turn-around times: the turn-around time of a particular shell (which we denote with a script font, ``$\scriptstyle{\mathscr{TA}}$", subscript as in the previous two equations), and the turn-around time of the outermost shell of a spherical perturbation (which we denote with a, ``$ta$", subscript).  For example, for a particular shell with radius $\mathpzc{x}$ the value, $\mathpzc{x}_{\scriptscriptstyle{\mathscr{TA}}}$, refers to the radius at the time that this shell turns around, and the value $\mathpzc{x}_{ta}$ refers to the radius at the time that the outermost shell turns around.  Unless we explicitly state otherwise, from this point on, whenever we refer to ``turn-around" in the text, we are referring to the time at which the outermost shell turns around.

\section{Einstein-de Sitter Universe}
\label{sec:eds_cosmology}
In this section we re-derive the equations for calculating the over-density of a halo at collapse for an E-dS universe with $\Omega=1$, leaving the initial and virialized density profiles completely general.  We then use this formalism to calculate $\Delta_c$ for several examples of reasonable initial and virialized density profiles.

\subsection{$R_{vir}/R_{ta}$ and $\Delta_c$}
\label{sec:eds_formulas}

The gravitational potential energy of a spherically symmetric object of mass $M$, radius $R$, and density profile $\rho(r)$ is
\begin{equation}
\label{eq:pot_uniform}
U = -\int_0^R G M(r) 4 \pi r  \rho(r) dr,
\end{equation}
where $M(r) = 4 \pi \int_0^{R} r^2 \rho(r) dr$.  Note that throughout this paper, we reserve upper case ``R"s to denote the edge of the sphere, and lower case ``r"s to denote the position variable.  For a uniform sphere, Eqn.~\ref{eq:pot_uniform} can be integrated to show that its potential energy is $U=-(3/5) G M^2/R$.  For a sphere with an arbitrary density profile, we re-write Eqn.~\ref{eq:pot_uniform}  as
\begin{equation}
\label{eq:pot_uniform_non_dim}
U =- \frac{3}{5} \frac{G M^2}{R} \mathscr{U}.
\end{equation}
The factor, $\mathscr{U}$, is a geometric correction factor accounting for the deviation of the sphere from complete homogeneity (and can also be viewed as the non-dimensionalized binding energy of the sphere), and is given by 
\begin{equation}
\label{eq:zeta}
\mathscr{U}=5\int_0^1 \mathscr{M}(\mathpzc{r}) \varrho(\mathpzc{r}) \mathpzc{r} d\mathpzc{r},
\end{equation}
with $\mathpzc{r} \equiv r/R$,
\begin{equation}
\label{eq:def_m}
\mathscr{M}=\frac{M(\mathpzc{r})}{M},
\end{equation}
and
\begin{equation}
\label{eq:def_rho}
\varrho \equiv \frac{\rho(\mathpzc{r})}{M/(\frac{4}{3}\pi R^3)}.
\end{equation}
%The value of $ \mathscr{U}$ runs from $0$ for a completely centrally concentrated sphere (a point mass of mass $M$ at the origin), to $1$ for a uniform sphere to $5/3$ for a completely outwardly concentrated sphere (a shell of mass $M$ at $R$.  In this work, we consider density profiles that decrease monotonically with $r$ so that our values of $\ \mathscr{U}$ typically run from 0 to 1.

For a dark matter sphere, the energy at turn-around can be related to the potential energy at virialization by employing the virial theorem and assuming energy conservation: $KE_{ta}+U_{ta} = E_{vir}=U_{vir}/2$ (for a universe with no cosmological constant).  Replacing the potential energies with Eqn.~\ref{eq:pot_uniform_non_dim} results in
\begin{equation}
KE_{ta}-\frac{3}{5}\frac{GM^2}{R_{ta}} \mathscr{U}_{ta} = -\frac{3}{10}\frac{GM^2}{R_{vir}} \mathscr{U}_{vir},
\end{equation}
which when solved for $R_{vir}/R_{ta}$ yields
\begin{equation}
\label{eq:rvir_rta}
\frac{R_{vir}}{R_{ta}} = \frac{1}{2} \left(\frac{ \mathscr{U}_{vir}}{ \mathscr{U}_{ta}-\mathscr{K}_{ta}} \right).
\end{equation}
Here,
\begin{equation}
\label{eq:kappa_def}
\mathscr{K}_{ta} \equiv \frac{KE_{ta}}{3 GM^2/(5R_{ta})},
\end{equation}
which represents the non-dimensionalized kinetic energy at turn-around.

As mentioned in the introduction, the density of a halo at virialization is typically parameterized by $\Delta_c$, the volume averaged density of the halo at virialization in units of the critical density of the universe at virialization: $\Delta_c = \bar \rho_{vir}/\rho_c(z_{vir})$.  It is customary to assume that a halo virializes at its collapse time, defined as twice the turn-around time of the edge of the halo.  In an E-dS universe, we can use the parametric solution in conjunction with the formula $a(t) = (3 H_o t/2)^{2/3}$ (valid for an E-dS cosmology) to find that
\begin{align}
\label{eq:delta_c}
\Delta_c &= \left (\frac{3 \pi}{2}\right)^2 \left( \frac{R_{ta}}{R_{vir}} \right)^3 \nonumber \\
& =18 \pi^2\left(\frac{ \mathscr{U}_{ta}-\mathscr{K}_{ta}}{ \mathscr{U}_{vir}}\right)^3.
\end{align}
Notice that in the limit of a homogenous sphere, $\mathscr{U}_{vir}$ and $\mathscr{U}_{ta}\rightarrow 1$, $\mathscr{K}_{ta} \rightarrow 0$, and Eqns.~\ref{eq:rvir_rta} and \ref{eq:delta_c} reduce to the familiar results that $R_{vir}/R_{ta} = 1/2$ and $\Delta_c=18 \pi^2$ in an E-dS universe.  Since $\Delta_c$ depends on the cube of $(\mathscr{U}_{ta}-\mathscr{K}_{ta})/ \mathscr{U}_{vir}$, it is possible that the even slight deviations from homogeneity cause significant deviation from the standard value of $18 \pi^2$.

\subsection{Conditions at Turn-around}
\label{sec:eds_conditions}

According to the spherical collapse model, once the initial density and velocity profile of a perturbation is specified, the complete kinematics of each shell within the perturbation is known at any time up until shell crossing.  In this section we express the density and velocity profiles (and hence potential and kinetic energies) at turn-around as a simple mapping of position from the initial time, to the turn-around time.  We then use the shell kinematics of the spherical collapse model to solve for the mapping, so that, given an initial density profile, the physical conditions of the sphere at turn-around are completely specified.

\subsubsection{Density and Potential Energy}

Assuming that the mass within each shell is conserved (i.e., no shell crossings) from the initial time to the turn-around time,
\begin{equation}
4 \pi \int_0^{r_{ta}} r_{ta}^{\prime 2} \rho_{ta}(r_{ta}^{\prime}) d r_{ta}^{\prime}=4 \pi \int_0^{r_{i}} r_{i}^{\prime 2} \rho_{i}(r_{i}^{\prime}) d r_{i}^{\prime},
\end{equation}
from which $\rho_{ta}(r_{ta})$ may be solved by taking a derivative with respect to $r_{ta}$:
\begin{equation}
\rho_{ta}(r_{ta}) = \rho_i(r_i) \frac{r_i^2}{r_{ta}^2}\frac{dr_i}{dr_{ta}}.
\end{equation}
We re-write this as 
\begin{equation}
\label{eq:varrho_ta}
\varrho_{ta}(y) = \frac{x^2}{y^2}\frac{dx}{dy},
\end{equation}
where $\varrho$ is defined by Eqn.~\ref{eq:def_rho}, and where
\begin{equation}
\label{eq:y}
y \equiv \frac{r_{ta}}{R_{ta}},
\end{equation}
and
\begin{equation}
\label{eq:z}
x \equiv \frac{r_i}{R_i},
\end{equation}
with both $y$ and $x$ $\in [0, 1]$.  In Eqn.~\ref{eq:varrho_ta}, we have set $\varrho_i$ to unity since
\begin{align}
\varrho_i &= \rho_i \frac{4/3 \pi R_i^3}{M} \nonumber \\
& = \frac{\bar \rho_m (a_i)[1+\delta_i(r_i) ]4/3 \pi R_i^3}{4/3 \pi \bar \rho_m (a_i) R_i^3 [1+\bar \delta_i(R_i)]} \nonumber \\
& \cong 1.
\end{align}
Using these expressions, $\mathscr{U}_{ta}$ (Eqn.~\ref{eq:zeta}) can be re-written as
\begin{equation}
\label{eq:u_int}
\mathscr{U}_{ta} = 5 \int_0^1 \mathscr{M}(y)\frac{x^2(y)}{y}\frac{dx}{dy}(y)dy
\end{equation}
with
\begin{equation}
\label{eq:m_int}
 \mathscr{M}_{ta}(y) = 3\int_0^{y} x^2(y^{\prime})\frac{dx}{dy^{\prime}}(y^{\prime})dy^{\prime}.
\end{equation}
Notice that we have just expressed the density and potential energy at turn-around as a simple mapping from the initial shell positions, $x$, to the positions at turn-around, $y$.

\subsubsection{Velocity and Kinetic Energy}
We now solve for the non-dimensionalized kinetic energy at turn-around, $\mathscr{K}_{ta}$.   Since,
\begin{align}
\label{eq:mass}
M&=\frac{4}{3}\pi R_i^3 \bar \rho_m(a_i) \left [1+\bar \delta_i(R_i) \right] \nonumber \\
&\cong \frac{R_i^3 H_o^2 \Omega_m}{2 a_i^3 G},
\end{align}
we get rid of one power of $M$ in Eqn.~\ref{eq:kappa_def}, resulting in
\begin{equation}
\mathscr{K}_{ta} = \frac{10}{3} \frac{\mathpzc{X_{ta}}^3 \frac{1}{2}\int_0^{R_{ta}} 4 \pi r_{ta}^2 \rho_{ta}(r_{ta})v_{ta}^2(r_{ta})dr_{ta}}{M H_o^2 R_{ta}^2 \Omega_m},
\end{equation}
where
\begin{equation}
\label{eq:x_ta_def}
\mathpzc{X_{ta}}\equiv \frac{R_{ta}a_i}{R_i},
\end{equation}
(the non-dimensionalized turn-around radius of the outermost shell) and where we have computed the total kinetic energy at turn-around by integrating throughout the sphere.  Using Eqns.~\ref{eq:def_rho},~\ref{eq:y} and~\ref{eq:varrho_ta}, this can be rewritten as
\begin{align}
\label{eq:k_non_dim}
\mathscr{K}_{ta}  &= 5\int_0^1 y^2 \varrho_{ta}(y) \mathpzc{v_{ta}}^2(y) dy \nonumber \\
& = 5 \int_0^1 x^2(y) \frac{dx}{dy}(y) \mathpzc{v_{ta}}^2(y)dy,
\end{align}
where the non-dimensionalized velocity (squared) profile is:
\begin{equation}
\label{eq:v_def}
\mathpzc{v}_{ta}^2(y) \equiv \frac{v_{ta}^2(y) \mathpzc{X}_{ta}^3}{R_{ta}^2 H_o^2 \Omega_m}.
\end{equation}
In an E-dS universe, this velocity profile can be found from the spherical collapse model by setting $\Omega_m=1$ and $\Omega_{\Lambda}=0$ in Eqn.~\ref{eq:eq_of_motion}.  Using the definitions given by Eqns.~\ref{eq:y}, \ref{eq:z}, \ref{eq:x_ta_def} and \ref{eq:v_def}, and after a bit of algebra, this equation becomes:
\begin{equation}
\label{eq:norm_velocity}
\mathpzc{v}_{ta}^2(y) = \frac{x^3}{y}-\frac{5}{3}\frac{\bar \delta_i(x)}{a_i}x^2 \mathpzc{X_{ta}}.
\end{equation}
Since in an E-dS universe
\begin{equation}
\label{eq:X_ta_eds}
 \mathpzc{X_{ta}} = \left[\frac{5}{3} \frac{\bar \delta_i(R_i)}{a_i} \right]^{-1},
\end{equation}
(see Eqn.~\ref{eq:xta_eds}) the non-dimensionalized kinetic energy can finally be written as:
\begin{equation}
\label{eq:k_int}
\mathscr{K}_{ta} = 5 \int_0^1 x^2(y) \frac{dx}{dy}(y)\left [\frac{x^3(y)}{y}-x^2(y)\frac{\bar \delta_i(x)}{\bar \delta_i(R_i)}\right ]dy.
\end{equation}
We have just expressed the velocity profile and kinetic energy at turn-around in terms of the shape of the initial density profile ($\bar \delta_i(x)/\bar \delta_i(R_i)$), and in terms of the mapping from $x$ to $y$.

\subsubsection{Solving the Mapping Using Spherical Collapse}
\label{sec:mapping_eds}
In this section, we use the spherical collapse model to solve for the mapping from the initial to turn-around positions: $x$ to $y$.  This then allows us to solve for the density profile, potential energy, velocity profile and kinetic energy at turn-around.  The relationship between $x$ and $y$ is given by
\begin{equation}
\label{eq:y_z_mapping_expr}
y=x\frac{\mathpzc{x}_{ta}(x)}{\mathpzc{X}_{ta}},
\end{equation}
where this equality can be shown by writing out $\mathpzc{x}_{ta}(x)$ and $\mathpzc{X}_{ta}$ explicitly with Eqns~\ref{eq:def_x} and \ref{eq:x_ta_def}.  Using Eqns.~\ref{eq:eq_of_motion_param1} and \ref{eq:X_ta_eds}, this equation becomes:
\begin{equation}
\label{eq:y_function}
y=x\frac{1}{2}\left [\frac{\bar \delta_i(x)}{\bar \delta_i(R_i)} \right ]^{-1}\left [1-\cos \Theta_{ta}(x) \right],
\end{equation}
where $\Theta_{ta}(x)$ is the development angle for a shell initially at $x$ at the time when the outermost shell turns around.

We solve for $\Theta_{ta}(x)$ by matching the parametric solution for the turn-around time of the outermost shell (Eqn.~\ref{eq:eq_of_motion_param2} with $\Theta(R_i)=\pi$) with the parametric solution for the time of a shell starting at position $x$ ($\Theta=\Theta_{ta}(x) $):
\begin{equation}
\frac{1}{2 H_o}\left [\frac{5}{3}\frac{\bar \delta_i(R_i)}{a_i} \right ]^{-3/2} \pi = \frac{1}{2 H_o}\left [\frac{5}{3}\frac{\bar \delta_i(x)}{a_i} \right ]^{-3/2}\left [\Theta_{ta}(x) - \sin \Theta_{ta}(x) \right ].
\end{equation} 
Simplifying this expression leads to a transcendental equation:
\begin{equation}
\label{eq:trans}
\Theta_{ta}(x)-\sin \Theta_{ta}(x) =  \pi \left [\frac{\bar \delta_i(x)}{\bar \delta_i(R_i)} \right ]^{3/2},
\end{equation}
which, when $\bar \delta_i(x) / \bar \delta_i(R_i)$ is specified, must be solved numerically.  The mapping, $y(x)$, is then found by plugging $\Theta_{ta}(x)$ into Eqn.~\ref{eq:y_function}.

Equation~\ref{eq:trans} places a constraint on the steepness of the initial density profile that we can use before our model fails.  For the sake of realism, and to avoid shell crossing before turn-around within the bulk of the sphere, we only consider monotonically decreasing initial density profiles.  Therefore, the first shell to undergo crossing will be the innermost shell when it crosses itself at the origin and re-expands to cross incoming shells.  Since our goal is to calculate the total kinetic and potential energies at turn-around with the formalism we have just presented, and since this formalism fails at shell crossing, the maximum that $\Theta_{ta}(x=0)$ can be is $2 \pi$.  Evaluating Eqn.~\ref{eq:trans} at $x=0$, the maximum that the left hand side of this equation can be for $0 \le \Theta_{ta}(0) \le 2 \pi$ is $2\pi$, and therefore $[\bar \delta_i(0)/ \bar \delta_i(R_i)]_{max}= 2^{2/3} \cong1.587$.  The amplitude of the initial density profiles that we utilize at the origin is therefore constrained by $1 \le \bar \delta_i(0)/ \bar \delta_i(R_i) \le 1.587$.

\subsection{Procedure}
Once a $\bar \delta_i(x)/ \bar \delta_i(R_i)$ profile is specified, $R_{vir}/R_{ta}$ and $\Delta_c$ can be found in the following manner.  The function $\Theta_{ta}(x)$ can be built up by solving Eqn.~\ref{eq:trans} for values of $x$ from 0 to 1.  The function $y(x)$ can then be found by plugging $\Theta_{ta}(x)$ into Eqn.~\ref{eq:y_function}, and $dy/dx(x)$ can then by found by taking the derivative numerically.  Eqns.~\ref{eq:u_int}, \ref{eq:m_int} and \ref{eq:k_int} can then be numerically integrated to solve for $\mathscr{U}_{ta}$, $\mathscr{M}_{ta}$ and $\mathscr{K}_{ta}$.  Once a density profile at virialization is specified, Eqn.~\ref{eq:zeta} can be integrated to solve for $\mathscr{U}_{vir}$.  The quantities $R_{vir}/R_{ta}$ and $\Delta_c$ can be found by plugging $\mathscr{U}_{ta}$, $\mathscr{K}_{ta}$ and $\mathscr{U}_{vir}$ into Eqns.~\ref{eq:rvir_rta} and \ref{eq:delta_c} respectively.  Note that the only dependence of $R_{vir}/R_{ta}$ and $\Delta_c$ is on the initial, normalized density profile, $\bar \delta_i(x)/ \bar \delta_i(R_i)$, and on the final, virialized density profile.

\subsection{E-dS Results}
\label{sec:eds_results}
\begin{figure}
\centering
\includegraphics[clip, width=4in]{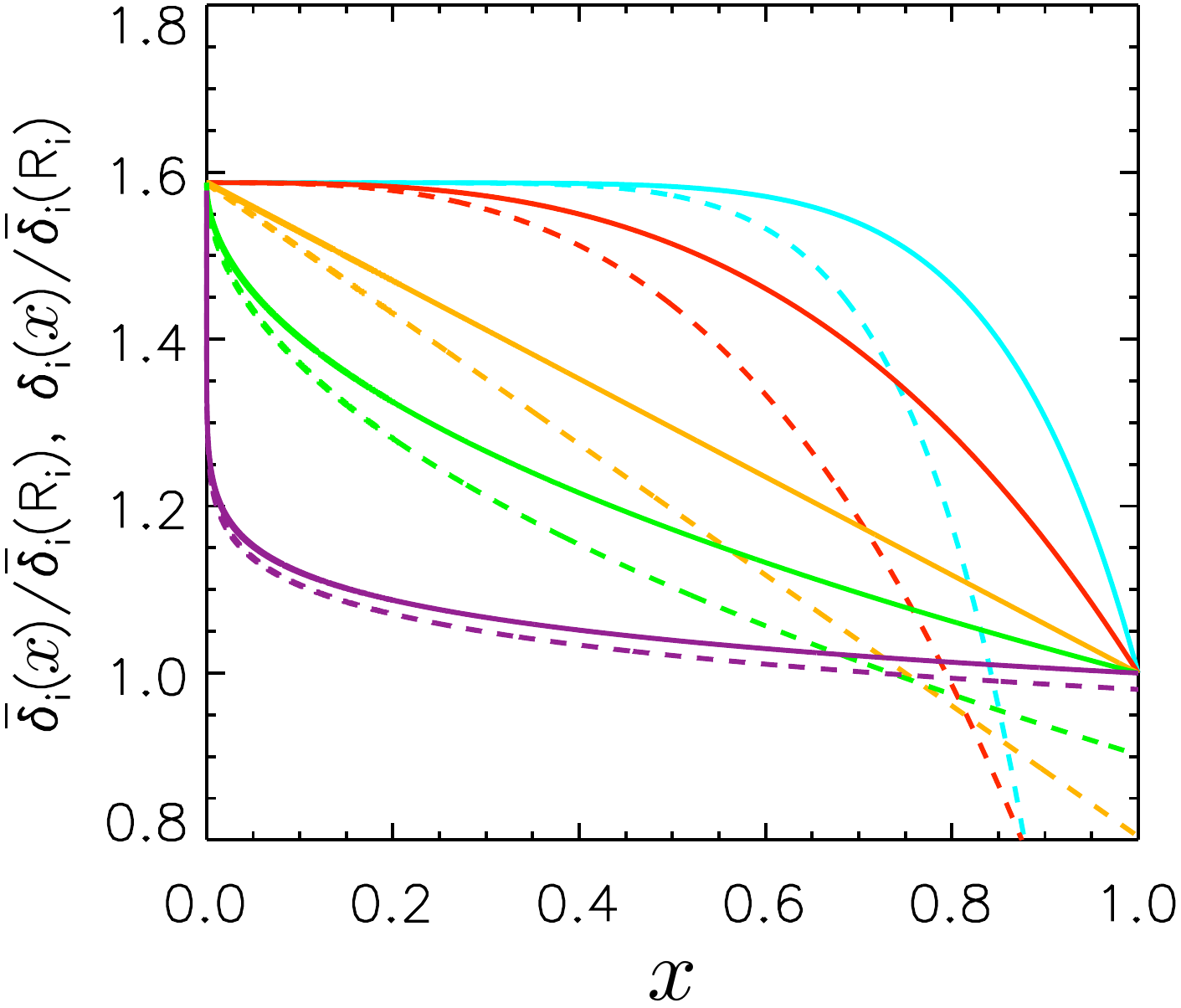}
\caption{ \label{fig:delta_v_z} Examples of the volume averaged, $\bar \delta_i(x)/\bar \delta_i(R_i)$, (solid lines) and local, $\delta_i(x)/\bar \delta_i(R_i)$, initial density profiles that we use in our calculations.  The turquoise, red, orange, light green and purple lines (top line to bottom line) correspond to $\beta=7$, 3, 1, 0.5 and 0.1 respectively, with $\beta$ defined by Eqn.~\ref{eq:initial_density_profile}.} 
\end{figure}

\begin{figure}
\centering
\includegraphics[clip, width=4in]{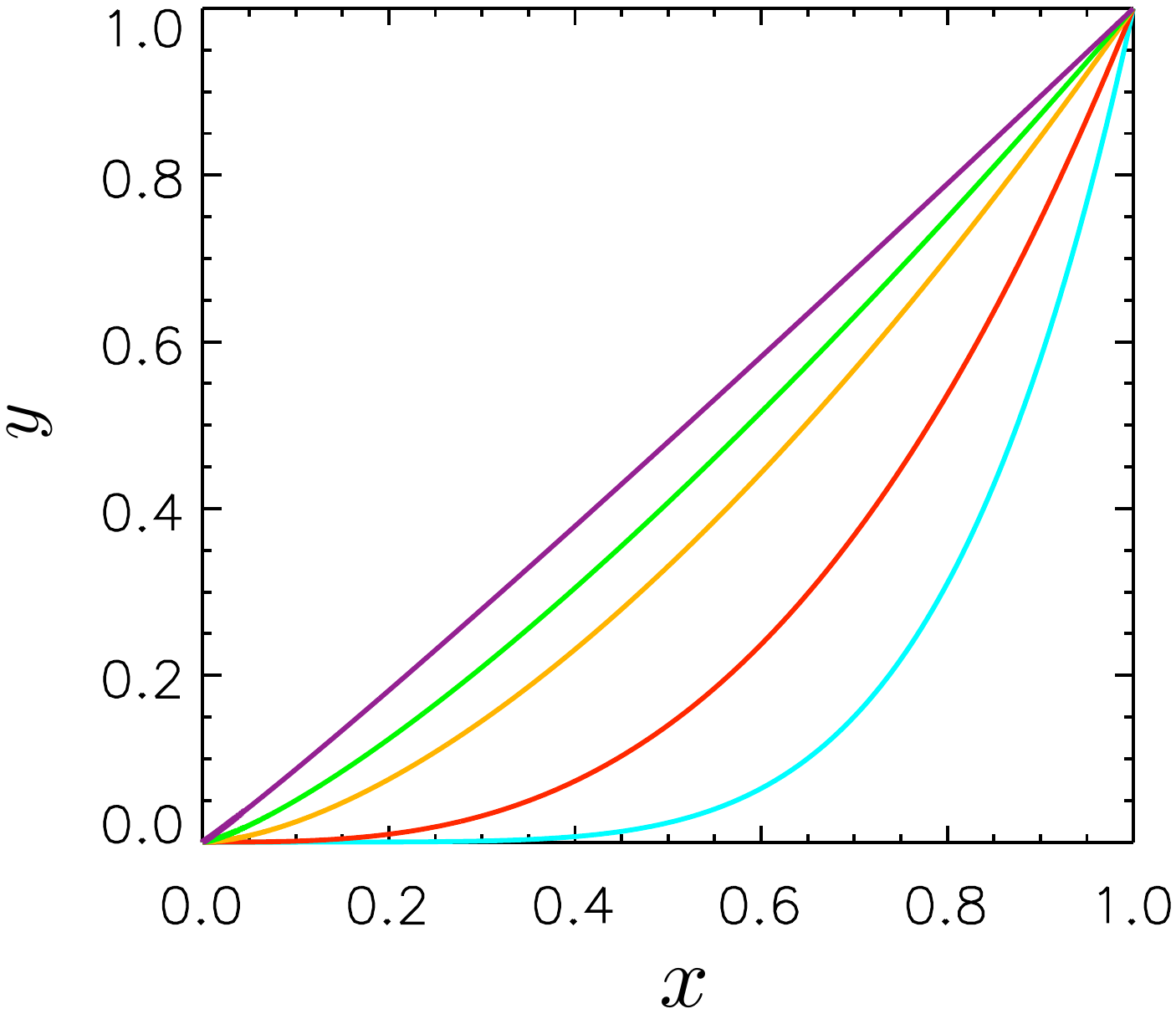}
\caption{ \label{fig:y_v_z}The normalized position at turn-around, $y$ as a function of the initial normalized position, $x$ for the same density profiles as in Fig.~\ref{fig:delta_v_z} (from bottom line to top line: $\beta=$ 7, 3, 1, 0.5 and 0.1).} 
\end{figure}

We now present our results in an E-dS universe by adopting reasonable density profiles at the initial time and at virialization.  We use a power-law initial density profile of the form, 
\begin{equation}
\label{eq:initial_density_profile}
\frac{\bar \delta_i(x)}{\bar \delta_i(R_i)}= \frac{\delta_i(0)}{\bar \delta_i(R_i)}\left(1-A x^\beta \right),
\end{equation}
since it both decreases monotonically and is quite pliable depending on the value of $\beta$.  We set $A$ such that $\bar \delta_i(x=1)/\bar \delta_i(R_i)=1$ ($A=1-[\delta_i(0)/\bar \delta_i(R_i)]^{-1}$).  The local (not volume averaged) density profile, $\delta_i(x)/\bar \delta_i(R_i)$, is given by the same function with $A \rightarrow A(\beta+3)/3$.  This can be found with the formula, $\delta(r) = (r/3) d\bar \delta/dr+\bar \delta$, derived by taking the derivative with respect to $r$ of Eqn.~\ref{eq:vol_avg}.  To maximize the effects that a non-uniform initial density profile has in our calculations, we use $\bar \delta_i(0)/\bar \delta_i(R_i)=2^{2/3}$.  We show several examples of the volume averaged initial density profile in Fig.~\ref{fig:delta_v_z}, along with the corresponding local density profiles for comparison.
In Fig.~\ref{fig:y_v_z} we show the $x$ to $y$ mapping for the same initial density profiles following the procedures outlined in the previous section.  As a practical matter, the mapping becomes very difficult to solve numerically for small $x$ at large values of $\beta$ ($\gtrsim 5$), as explained in Appendix~\ref{app:approx_form}.  In this regime, we use a highly accurate analytic approximation formula to calculate $y$ as a function of $x$, which we derive in the same appendix.  We show the physical conditions within the dark matter sphere at turn-around by plotting the non-dimensionalized velocity, density and interior mass profiles as given by Eqns.~\ref{eq:norm_velocity}, \ref{eq:varrho_ta} and \ref{eq:m_int}.  The panels show that, in contrast to the case of an initially uniform sphere, the velocity profile within the sphere at turn-around is non-zero, and the density is profile can be far from uniform.
\begin{figure}
\centerline{\mbox{  \includegraphics[clip, width=2.00in]{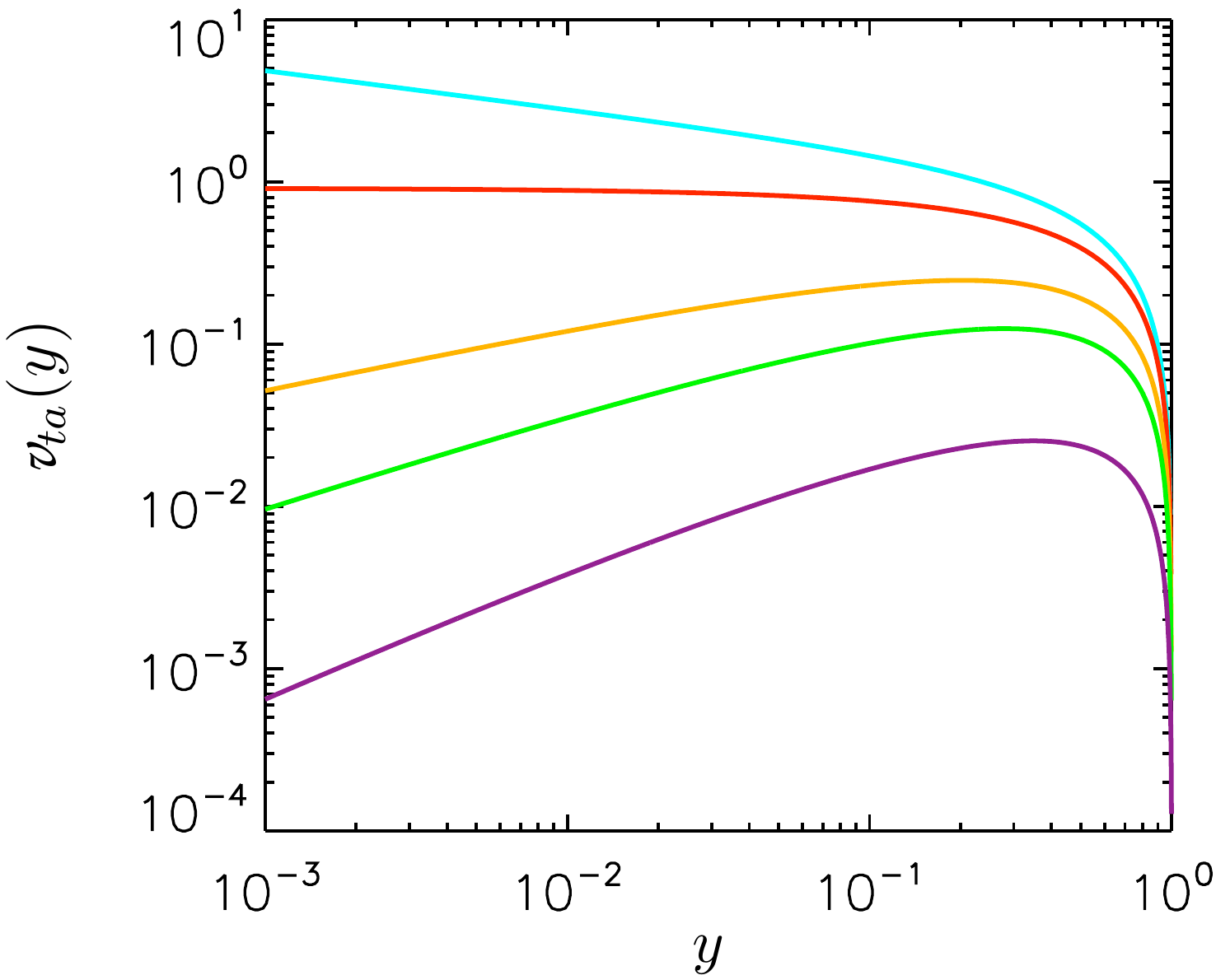}}
\mbox{  \includegraphics[clip, width=1.93in]{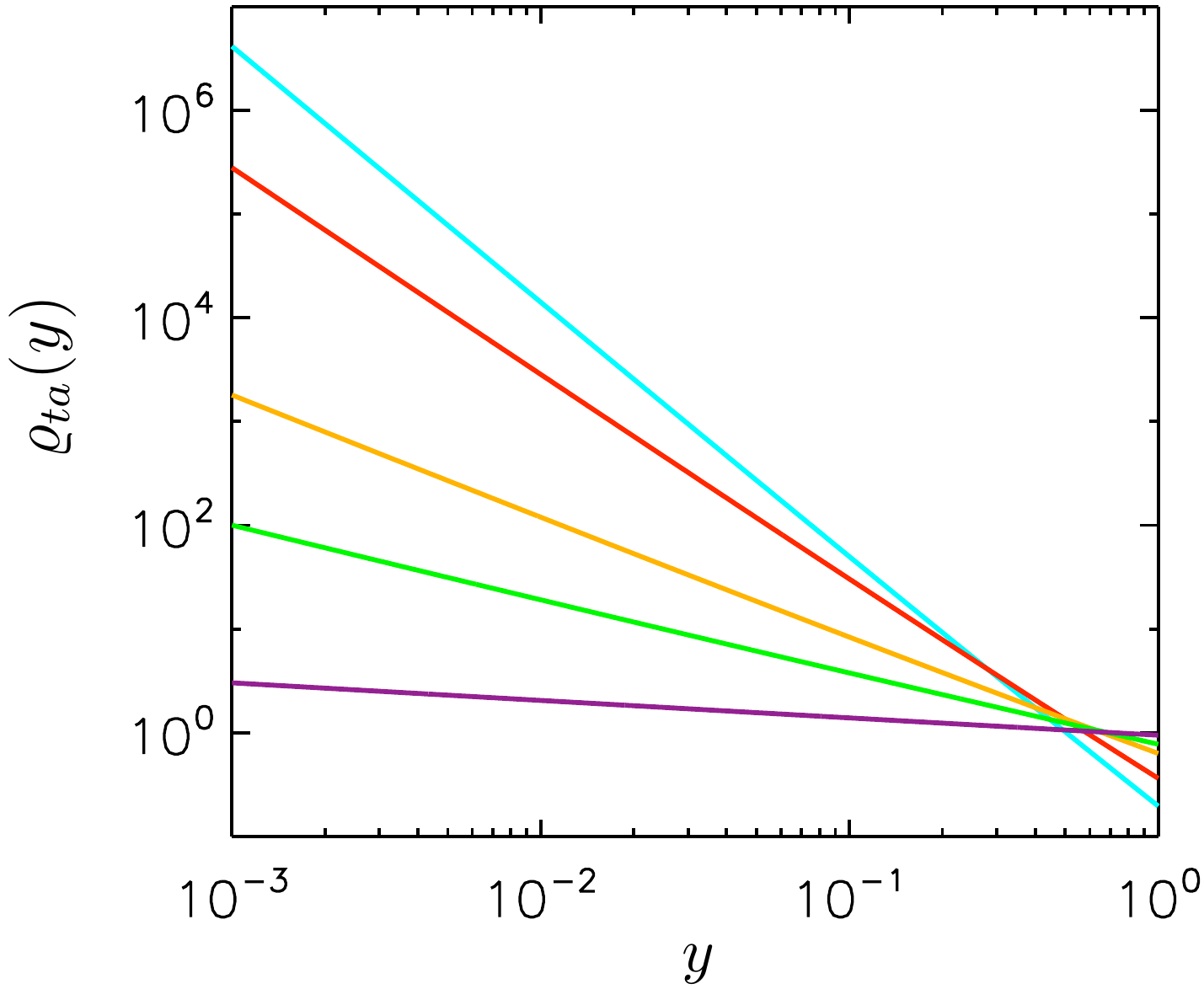}}
\mbox{  \includegraphics[clip, width=1.95in]{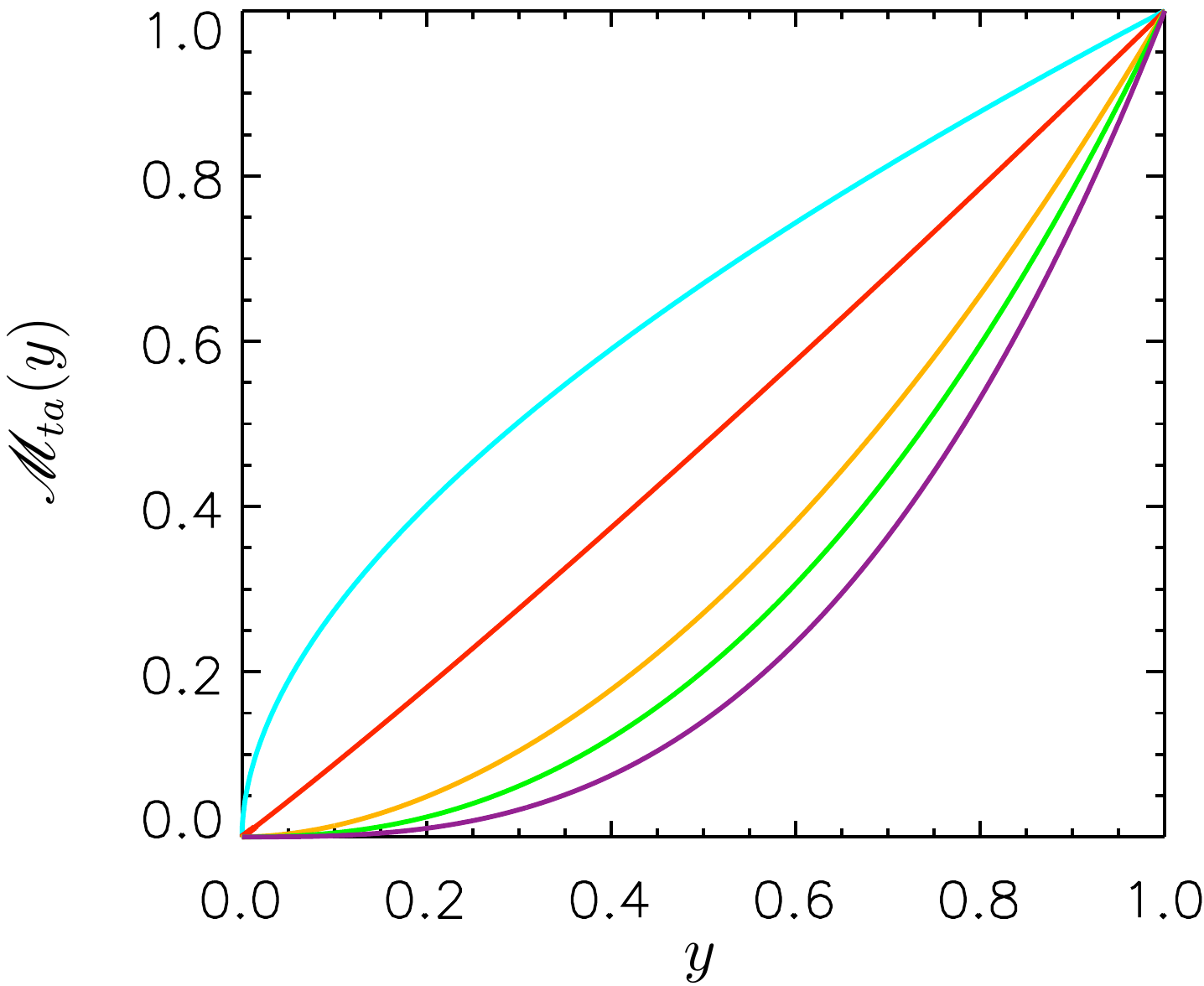}}}
\caption{ \label{fig:physical_cond_at_ta}The normalized velocity, density and interior mass profiles (defined by Eqns.~\ref{eq:v_def}, \ref{eq:def_rho}, \ref{eq:def_m} respectively) within a dark matter sphere at turn-around for the same density profiles as in Fig.~\ref{fig:delta_v_z} (for all panels, from top line to bottom line: $\beta=$ 7, 3, 1, 0.5 and 0.1).} 
\end{figure}

To calculate the non-dimensionalized binding energy of the dark matter halo at virialization, we use an NFW density profile \cite{navarro}.  For an NFW profile, 
\begin{equation}
\label{eq:varrho_NFW}
\varrho(w) = \frac{1}{3}\frac{c^2}{w(1+cw)^2} \frac{1}{\ln(1+c)- \frac{c}{1+c}}
\end{equation}
and
\begin{equation}
\label{eq:M_NFW}
\mathscr{M}(w) = \frac{\ln(1+c w) - \frac{cw}{1+cw}}{\ln(1+c)- \frac{c}{1+c}},
\end{equation}
where $w \equiv r_{vir}/R_{vir}$, and where $c$, the concentration parameter, depends on the recent merger history of the halo.  Using these expressions we calculate the non-dimsionalized binding energy at virialization with Eqn.~\ref{eq:zeta}, where the integral can be evaluated analytically:
\begin{align}
\label{eq:u_vir}
\mathscr{U}_{vir} &= \frac{5}{3}\frac{c^2}{\left [\ln(1+c) - \frac{c}{1+c}\right]^2} \int_0^1 \frac{\ln(1+cw)-\frac{cw}{1+cw}}{(1+cw)^2}dw \nonumber \\ 
&= \frac{5}{6} \frac{c[c(2+c)-2(1+c) \ln(1+c)]}{(1+c)^2\left [\ln(1+c)-\frac{c}{1+c}\right ]^2}.
\end{align}
\begin{figure}
\centering
\includegraphics[clip, width=3.6in]{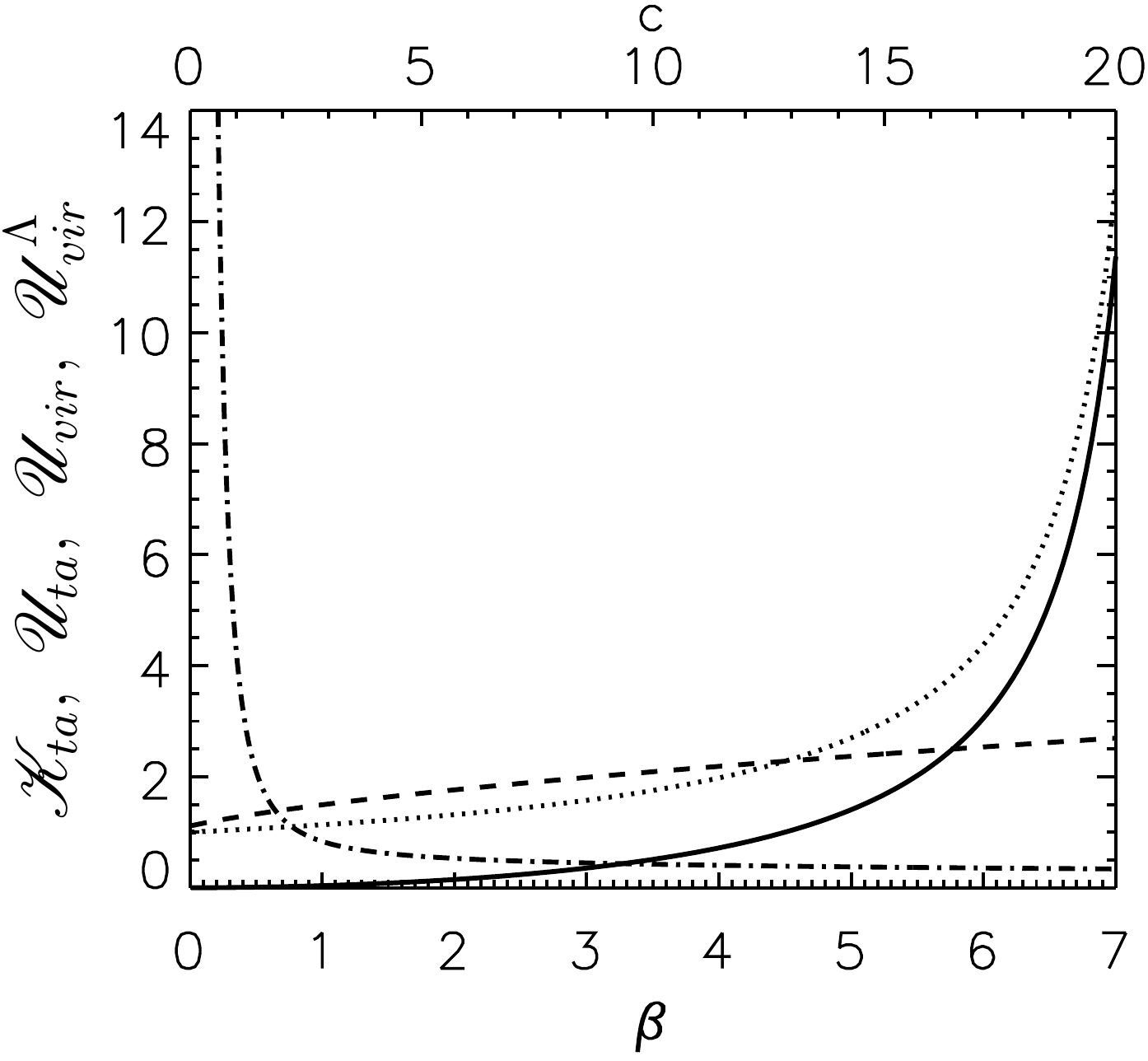}
\caption{ \label{fig:ke_u_v_beta}The non-dimensionalized kinetic (solid line) and potential energy (dotted line) at turn-around as a function of $\beta$ for an E-dS cosmology (bottom x-axis), as well as the non-dimensionalized potential energy at virialization due to gravity (dashed line) and dark energy (dot-dashed line) for an NFW density profile as a function of concentration parameter, c (top x-axis).  The former two quantities are independent of cosmology since they are computed directly from the NFW profile.  The non-dimensionalized potential energy at virialization due dark energy, $\mathscr{U}_{vir}^{\Lambda}$, is defined in \S~\ref{sec:results_gen}.} 
\end{figure}
In Fig.~\ref{fig:ke_u_v_beta} we show $\mathscr{U}_{vir} $ as a function of the concentration parameter, as well as $\mathscr{K}_{ta}$ and $\mathscr{U}_{ta}$ as a function of $\beta$, calculated with Eqns.~\ref{eq:k_int} and~\ref{eq:u_int} respectively.  For comparison, we note that for an initially uniform sphere, $\mathscr{K}_{ta}=0$ and $\mathscr{U}_{ta}=1$ and for a uniform sphere at virialization $\mathscr{U}_{vir}=1$.

\begin{figure}
\centerline{\mbox{  \includegraphics[clip, width=3.2in]{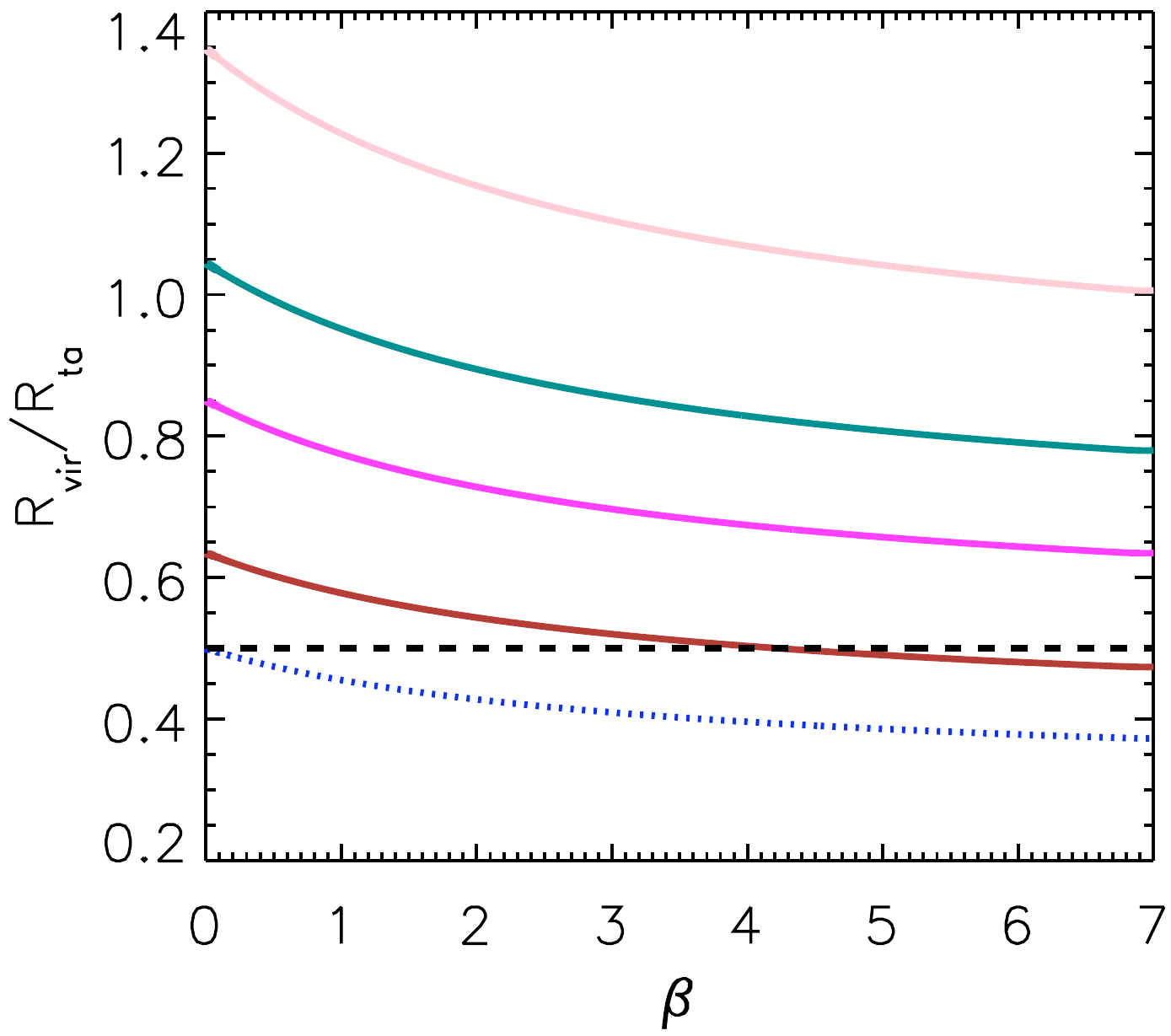}}
\mbox{  \includegraphics[clip, width=3.22in]{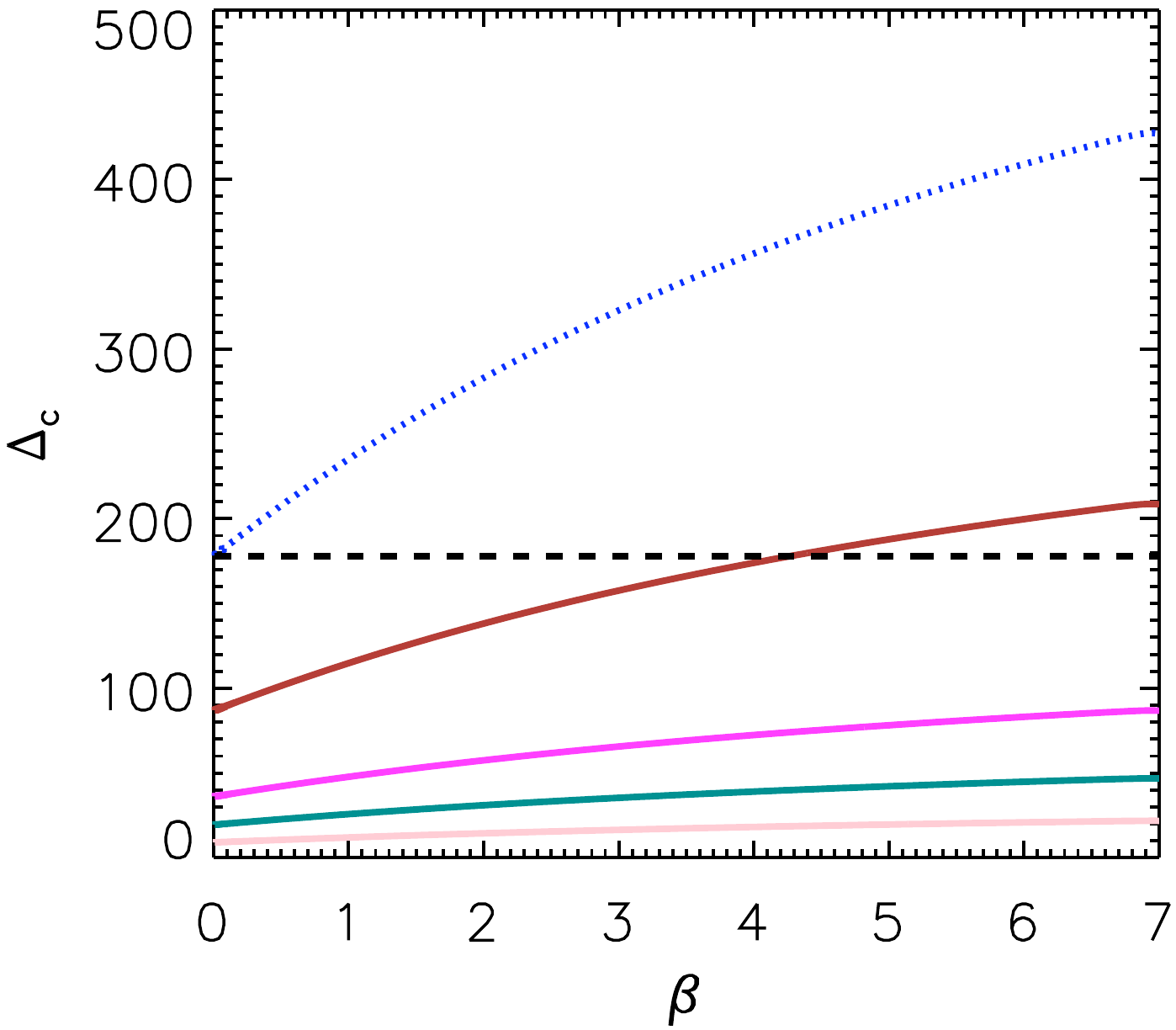}}}
\caption{ \label{fig:delta_c_v_beta} $R_{vir}/R_{ta}$ and $\Delta_c$ as a function of $\beta$ for different values of $c $.  The brown, magenta, teal and pink lines (bottom line to top line in the left panel and top line to bottom line in the right panel) correspond to $c=1$, 5, 10 and 20 respectively.  The blue dotted line has $\mathscr{U}_{vir}$ set to unity (corresponding to a uniform sphere).  The black dashed lines show that standard results for an initially uniform density profile.} 
\end{figure}

Having calculated $\mathscr{K}_{ta}$, $\mathscr{U}_{ta}$ and $\mathscr{U}_{vir}$, we may now calculate $R_{vir}/R_{ta}$ and $\Delta_c$ with Eqns.~\ref{eq:rvir_rta} and \ref{eq:delta_c}.  We show $R_{vir}/R_{ta}$ and $\Delta_c$ as a function of $\beta$ for several values of $c$ in Fig.~\ref{fig:delta_c_v_beta}.  For comparison, we also show the results for the standard uniform sphere calculation in an E-dS cosmology ($R_{vir}/R_{ta}=1/2$ and $\Delta_c=18 \pi^2$) with the black dashed line.  Since, for an NFW profile, $\mathscr{U}_{vir}$ never equals unity (regardless of the value of $c$ used) we also plot a curve with $\mathscr{U}_{vir}$ set to unity (blue dotted line) to show that in the limit that $\beta$ goes to zero, our results reduce to the uniform sphere calculation.  The figure shows that $R_{vir}/R_{ta}$ and $\Delta_c$ can deviate significantly from the standard values, with a slightly stronger dependence on the density profile at virialization than on the initial density profile (i.e., a stronger dependence on $c$ than on $\beta$).  By taking into account non-uniform density profiles, the non-linear density at virialization is typically smaller by a factor of a few to more than a factor of 10 for halos with the highest concentration parameter.  We note, however, that for the highest concentration parameters shown, these halos can be quite rare.  In fact, when considering Fig.~\ref{fig:delta_c_v_beta} (as well as Figs.~\ref{fig:results_gen} and \ref{fig:results_realistic}) one should note that the concentration parameter for recently formed halos is typically $c\approx 4$ \cite{zhao:2009}.  It is interesting note that for the highest values of $c$ (the steepest NFW profiles), the virial radius can be bigger than the turn-around radius.  

\section{Cosmologies with a Cosmological Constant and Curvature}
The universe in which we live has a non-zero dark energy component which significantly affects the formation of cosmological structure at low redshifts.  To accurately describe the non-linear collapse of halos at low redshift, we repeat our calculation of $\Delta_c$, but include the dynamical effects of a cosmological constant.  Since the derivations are not any more difficult when curvature is included, we leave our equations general to allow for any cosmology with matter, curvature and vacuum energy components. % This allows us to consider the collapse of halos in regions of our universe with large-scale over/under-densities (such as voids or superclusters) which are effectively described by a change in the matter and curvature components of the background cosmology.

\label{sec:gen_cosmology}

\subsection{$R_{vir}/R_{ta}$ and $\Delta_c$}
\label{sec:gen_cos_formalism}
To solve for $R_{vir}/R_{ta}$ and $\Delta_c$ in a cosmology with a cosmological constant, one must include the effective gravitational potential energy due to dark energy.  The gravitational density associated with dark energy is  $\rho_{\Lambda}+3P/c^2 = -2 \rho_{\Lambda} = -3H_o^2 \Omega_\Lambda/(4 \pi G)$, from which it can be found that the contribution to the potential energy of a sphere from dark energy is:
\begin{equation}
U_{\Lambda} = -\frac{1}{2} \Omega_{\Lambda} H_o^2 \int_0^R 4 \pi r^4 \rho(r)dr.
\end{equation}
For a uniform sphere, $U_{\Lambda} = (-3/10)\Omega_{\Lambda}H_o^2 M R^2$, motivating our re-expression of the previous equation,
\begin{equation}
\label{eq:pot_uniform_non_dim_L}
U_{\Lambda} =-\frac{3}{10} \Omega_{\Lambda} H_o^2 M R^2 \mathscr{U}^{\Lambda},
\end{equation}
where,
\begin{equation}
\label{eq:U_lam_non_dim}
\mathscr{U}^{\Lambda} = 5\int_0^1 \varrho(\mathpzc{r})\mathpzc{r}^4 d\mathpzc{r}.
\end{equation}
Analogous to the case of $\mathscr{U}$ introduced in Eqn.~\ref{eq:pot_uniform_non_dim}, $\mathscr{U}^{\Lambda}$ can either be viewed as a geometric correction factor accounting for the deviation of a sphere from complete homogeneity, or as the non-dimensionalized binding energy of a sphere due to dark energy.

According to the virial theorem, for potential energies of the form $U \propto R^n$, $KE=(n/2)U$, where the energies are time averaged.  Since $U_{\Lambda}\propto R^2$ and $U \propto R^{-1}$ (see Eqns.~\ref{eq:pot_uniform_non_dim} and \ref{eq:pot_uniform_non_dim_L}), for a dark matter halo at virialization, $KE_{vir} = -U_{vir}/2+U_{vir}^{\Lambda}$, and the total energy is therefore $E_{vir}=KE_{vir} +U_{vir}+U_{vir}^{\Lambda}=U_{vir}/2+2U_{vir}^{\Lambda}$.  If energy is conserved between turn-around virialization then $KE_{ta}+U_{ta}+U_{ta}^{\Lambda} =U_{vir}/2+2U_{vir}^{\Lambda}$.  Replacing the potential energies in this equation with Eqns.~\ref{eq:pot_uniform_non_dim} and \ref{eq:pot_uniform_non_dim_L} results in: 
\begin{equation}
KE_{ta}-\frac{3}{5}\frac{GM^2}{R_{ta}}\mathscr{U}_{ta}-\frac{3}{10}\Omega_{\Lambda}H_o^2MR_{ta}^2\mathscr{U}_{ta}^\Lambda=-\frac{3}{10}\frac{GM^2}{R_{vir}}\mathscr{U}_{vir}-\frac{3}{5}\Omega_{\Lambda}H_o^2MR_{vir}^2\mathscr{U}_{vir}^\Lambda.
\end{equation}
Using Eqns.~\ref{eq:mass} and \ref{eq:x_ta_def}, and after a bit of algebra, one finds the following cubic in $R_{vir}/R_{ta}$:
\begin{equation}
\label{eq:cubic_gen}
4 \zeta \mathscr{U}_{vir}^{\Lambda}\left(\frac{R_{vir}}{R_{ta}} \right)^3-2 \left[\mathscr{U}_{ta}-\mathscr{K}_{ta} +\zeta \mathscr{U}_{ta}^{\Lambda}\right] \frac{R_{vir}}{R_{ta}}+\mathscr{U}_{vir}=0,
\end{equation}
with 
\begin{equation}
\label{eq:zeta_gen}
\zeta[\bar \delta_i(R_i)/a_i] \equiv \frac{\Omega_\Lambda}{\Omega_m} \mathpzc{X}_{ta}^3[\bar \delta_i(R_i)/a_i],
\end{equation}
and where $\mathscr{K}_{ta}$ is still defined by Eqn.~\ref{eq:kappa_def}.  The proper solution of $R_{vir}/R_{ta}$ is the smallest, positive, pure real root of the cubic (if a positive, pure real root exists).  A physical solution will not exist if the initial seed of the perturbation, $\bar \delta_i/a_i$, is less than a critical value, given by \cite{rubin:2012}:
%\footnote{In the case that two positive, pure real solutions exist, the \textit{smaller} of the two is the proper solution, since, energetically, this is the most stable state of the system.  Since, at virialization, $U_{vir} \sim -(1/R_{vir}+1/R_{vir}^2)$, for a given $R_{ta}$, the global minimum in the potential energy of the system is given when $R_{vir}$ is minimized.}. 
\begin{equation}
\label{eq:di_ai_cr}
\left (\frac{\bar \delta_i}{a_i}\right)_{cr}=\frac{3}{10 \Omega_m}\left[2 \Omega_k+3\left (2 \Omega_\Lambda \Omega_m^{2} \right)^{1/3} \right].
\end{equation}
The equation implies that the presence of curvature and/or a cosmological constant can prevent a perturbation from ever turning around to eventually form a virialized halo even if $\bar \delta_i/a_i>0$.

Having solved for $R_{vir}/R_{ta}$, we may now solve for $\Delta_c$.  This parameter can be written as 
\begin{equation}
\label{eq:delta_c_gen}
\Delta_c(z_c) = \left(\frac{R_{vir}}{R_{ta}}\right)^{-3} \frac{a_c^3 \Omega_m(z_c)}{ \mathpzc{X}_{ta}^3},
\end{equation}
where we have used Eqns.~\ref{eq:mass}, \ref{eq:x_ta_def} and
\begin{equation}
\rho_c(z_c) = \frac{3 H_o^2 \Omega_m}{8 \pi G \Omega_m(z_c)a_c^3},
\end{equation}
and where
\begin{equation}
\Omega_m(z) = \frac{\Omega_m(1+z)^3}{\Omega_m(1+z)^3+\Omega_\Lambda+\Omega_k(1+z)^2}.
\end{equation}
In the limit of a homogenous sphere, $\mathscr{U}_{vir}$ and $\mathscr{U}_{ta}\rightarrow 1$ and $\mathscr{K}_{ta} \rightarrow 0$, Eqns.~\ref{eq:cubic_gen} and \ref{eq:delta_c_gen} reduce to the equations for $R_{vir}/R_{ta}$ and $\Delta_c$ derived assuming an initially uniform sphere (for example, see Eqns. 2.18 and 2.23 of \cite{rubin:2012}).

Equations.~\ref{eq:cubic_gen} and \ref{eq:delta_c_gen} show that one of the main differences between the calculation of $R_{vir}/R_{ta}$ and $\Delta_c$ in an E-dS cosmology and in a general cosmology is that in the former case, these quantities are constant, while in the latter case, they are functions of the collapse redshift of the halo.  This is because these parameters depend on $\mathpzc{X}_{ta}$ (through Eqn.~\ref{eq:zeta_gen}), which is a function of the initial seed, $\bar \delta_i(R_i)/a_i$, which uniquely determines the collapse redshift of a halo.  Once a value of $\bar \delta_i(R_i)/a_i$ is specified, the halo collapse time (defined as twice the turn-around time) can be found by integrating Eqn.~\ref{eq:eq_of_motion}:
\begin{equation}
\label{eq:collapse_time}
H_o t_{c}= 2 \, \mathcal{I}\left[0, \mathpzc{X}_{ta}, \frac{\bar \delta_i(R_i)}{a_i} \right],
\end{equation}
where $\mathpzc{X}_{ta}$ is also solely a function of $\bar \delta_i(R_i)/a_i$ (see Eqn.~\ref{eq:cubic_ta}), and where $\mathcal{I}$ is defined by Eqn.~\ref{eq:def_I}.  The collapse time can then be converted to a collapse redshift by integrating the Friedmann equation:
\begin{equation}
\label{eq:friedmann}
H_o t_c =\int_0^{a_c} da^{\prime} \left[1+\Omega_m\left (\frac{1}{a^{\prime}}-1\right) +\Omega_{\Lambda}(a^{\prime 2} -1)\right]^{-1/2},
\end{equation}
where $a_c=(1+z_c)^{-1}$, and matching the times.  Thus, by setting the previous two equations equal to each other, one may numerically build up $\bar \delta_i(R_i)/a_i$ as a function of $z_c$.

A very accurate fitting formula for $\bar \delta_i(R_i)/a_i$ given a collapse redshift, $z_c$, is given by \cite{rubin:2012}:
\begin{equation}
\label{eq:fitting_di_ai}
\frac{\bar \delta_i (R_i)}{a_i} = \frac{0.674588 (1+z_c)} {\Omega_m^{0.9945}(z_c)}\left \{ \Omega_m^{4/7}(z_c)+ \Omega_\Lambda(z_c)+\left [1+ \frac{\Omega_m(z_c)}{2}\right ] \left [1+ \frac{\Omega_\Lambda(z_c)}{70}\right ] \right \},
\end{equation}
where this expression is strictly valid for a flat cosmology.  This formula was found by inserting a fitting formula for the linear theory growth factor \cite{carroll} and a fitting formula for the linear theory over-density at collapse \cite{mo} into the linear theory relation, $\bar \delta_i/a_i=\bar \delta_c(z_c)/D(z_c)$ (where it is assumed that $a_i<<1$ so that $D(a_i)\rightarrow a_i$).

%can be found by inserting two fitting formulas (Mo 2011 and Carol 1992) into the linear theory relation, $\bar \delta_i=\bar \delta_c(z_c) D(a_i)/D(z_c)$, where $\bar \delta_c$ is the linear theory over-density at the collapse time and $D$ is the linear theory ``growth factor", as done in Rubin and Loeb 2012:
%\begin{equation}
%\label{eq:fitting_di_ai}
%\frac{\bar \delta_i (R_i)}{a_i} = \frac{0.674588 (1+z_c)} {\Omega_m^{0.9945}(z_c)}\left \{ \Omega_m^{4/7}(z_c)+ \Omega_\Lambda(z_c)+\left [1+ \frac{\Omega_m(z_c)}{2}\right ] \left [1+ \frac{\Omega_\Lambda(z_c)}{70}\right ] \right \}.
%\end{equation}
%This  and we have also assumed that $a_i<<1$ ($D(a_i)\rightarrow a_i$).  We check this expression against the exact numerical results from Eqns.~\ref{eq:collapse_time} and~\ref{eq:friedmann} for a cosmology with $\Omega_m = 0.27$ and $\Omega_\Lambda=0.73$ and find that it is accurate to within $\sim 0.25\%$ at $z_c \approx 0$, and increases in accuracy by orders of magnitude for higher $z_c$.  We have provided this simple formula for the convenience of the reader, as throughout the rest of this paper, whenever we need to solve for $\bar \delta_i(R_i)/a_i$ given $z_c$ we use the exact numerical results from Eqns.~\ref{eq:collapse_time} and~\ref{eq:friedmann}.

\subsection{Conditions at Turn-around}
\label{sec:conditions_general}
As we have done for the E-dS calculation, we now express the conditions at turn-around in terms of the mapping, $x$ to $y$.  We then solve for the mapping using the shell kinematics as given by the spherical collapse model.

\subsubsection{Density and Potential Energy}
In the case of a general cosmology, the non-dimesionalized density, binding energy and interior mass profile is still given by Eqns.~\ref{eq:varrho_ta}, \ref{eq:u_int} and \ref{eq:m_int} respectively, with the caveat that the function $x(y)$ must be re-solved to include curvature and a cosmological constant (we cover this in \S~\ref{sec:mapping_gen}).  The non-dimensionalized binding energy at turn-around associated with dark energy is found simply by inserting Eqn.~\ref{eq:varrho_ta} into Eqn.~\ref{eq:U_lam_non_dim}:

\begin{equation}
\label{eq:u_lam}
\mathscr{U}_{ta}^{\Lambda} = 5 \int_0^1 y^2 \frac{dx}{dy}(y)x^2(y)dy.
\end{equation}

\subsubsection{Velocity and Kinetic Energy}
The non-dimensionalized velocity profile at turn-around (defined in Eqn.~\ref{eq:v_def}) for a dark matter sphere in a general cosmology can be found by rearranging Eqn.~\ref{eq:eq_of_motion},
\begin{equation}
\label{eq:norm_velocity_gen}
\mathpzc{v}_{ta}^2(y) = \frac{x^3}{y}+\zeta y^2 +x^2 \mathpzc{X}_{ta} \left [ \frac{\Omega_k}{\Omega_m} - \frac{5}{3} \left (\frac{\bar \delta_i(x)}{\bar \delta_i(R_i)}\right) \left( \frac{\bar \delta_i(R_i)}{a_i} \right) \right ].
\end{equation}
Inserting this expression into Eqn.~\ref{eq:k_non_dim}, we find the non-dimensionalized kinetic energy of the sphere at turn-around:
\begin{equation}
\label{eq:k_int_gen}
\mathscr{K}_{ta} = 5 \int_0^1 x^2(y) \frac{dx}{dy}(y)\left \{ \frac{x^3(y)}{y}+\zeta y^2 +x^2(y) \mathpzc{X}_{ta} \left [ \frac{\Omega_k}{\Omega_m} - \frac{5}{3} \left (\frac{\bar \delta_i(x)}{\bar \delta_i(R_i)}\right) \left ( \frac{\bar \delta_i(R_i)}{a_i} \right) \right ] \right \} dy,
\end{equation}
which reduces to Eqn.~\ref{eq:k_int} in the limit that $\Omega_k, \Omega_\Lambda \rightarrow 0$ and $\Omega_m \rightarrow 1$. 

\subsubsection{Solving the Mapping Using Spherical Collapse}
\label{sec:mapping_gen}

To solve for the $x$ to $y$ mapping in a general cosmology, we must first specify the halo collapse redshift of interest, $z_c$.  The corresponding $\bar \delta_i(R_i)/a_i$ value can then be calculated as explained in \S~\ref{sec:gen_cos_formalism}.  The non-dimensionalized turn-around radius of the outermost shell, $\mathpzc{X}_{ta}$, is then found by solving Eqn.~\ref{eq:cubic_ta}, and the turn-around time is found by evaluating the following integral:
\begin{equation}
\label{eq:t_ta_1}
t_{ta} H_o = \mathcal{I}\left [0, \mathpzc{X}_{ta}\left(\frac{\bar \delta_i(R_i)}{a_i} \right), \frac{\bar \delta_i(R_i)}{a_i} \right ].
\end{equation}
To solve for the position of a shell at time $t_{ta} H_o$ which starts at $x$, we re-write Eqn.~\ref{eq:eq_of_motion} as 
\begin{equation}
\frac{d \mathpzc{x}}{d (t H_o)} = \pm \sqrt{\frac{\Omega_m}{\mathpzc{x}}+\Omega_{\Lambda}\mathpzc{x}^2+\Omega_k-\frac{5}{3}\left (\frac{\bar \delta_i(x)}{\bar \delta_i(R_i)}\right) \left(\frac{\bar \delta_i(R_i)}{a_i}\right)\Omega_m},
\end{equation}
where the velocity is taken to be positive if $t H_o<t_{\scriptscriptstyle{\mathscr{TA}}} H_o$, and negative if $t H_o>t_{\scriptscriptstyle{\mathscr{TA}}} H_o$ (keeping in mind that $t_{\scriptscriptstyle{\mathscr{TA}}}$ corresponds to the turn-around time of a shell starting at $x$).  Thus, if the initial normalized density profile, $\bar \delta_i(x)/\bar \delta_i(R_i)$, is known, a value of  $x$ is specified and the differential equation is integrated numerically with $ \mathpzc{x}(t=0)=0$ until $t H_o =t_{ta} H_o$ to find $\mathpzc{x}_{ta}(x)$.  The corresponding $y$ value is then found with Eqn.~\ref{eq:y_z_mapping_expr}.  By repeating this procedure for values of $x \in [0, 1]$, we build up the function $x(y)$.

\begin{figure}
\centering
\includegraphics[clip, width=4in]{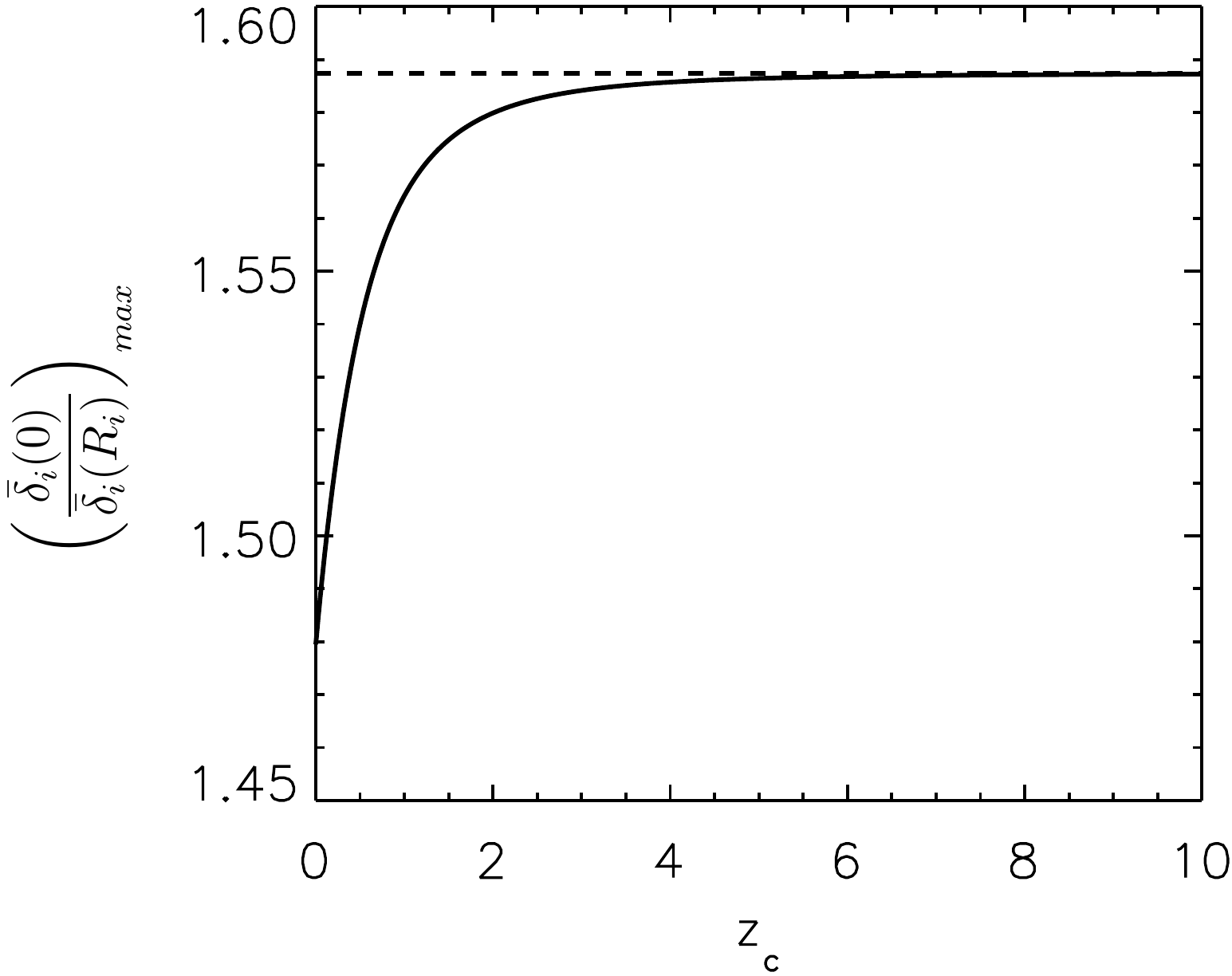}
\caption{ \label{fig:d_max_v_zc} The maximum value of $\bar \delta_i(0)/\bar \delta_i(R_i)$ allowed for a monotonically decreasing density profile before our model breaks down due to shell crossing.  The solid line is for a cosmology with $\Omega_m=0.27$ and $\Omega_\Lambda=0.73$, and the dashed line is for an E-dS cosmology $(\bar \delta_i(0)/\bar \delta_i(R_i))_{max} = 2^{2/3}$.} 
\end{figure}

As noted in \S~\ref{sec:mapping_eds}, there is a maximum value of $\bar \delta_i (0)/\bar \delta_i(R_i)$ for which our analysis is valid due to the innermost shell undergoing shell crossing at the origin before the outermost shell turns around.  We can solve for this value numerically by matching the time it takes for the innermost shell to reach the origin with with the time it takes for the outermost shell to turn-around (Eqn.~\ref{eq:t_ta_1}):
\begin{equation}
\label{eq:t_ta_max}
2 \, \mathcal{I}\left [0, \mathpzc{x}_{\scriptscriptstyle{\mathscr{TA}}}\left (\frac{\bar \delta_i(0)}{a_i}\right), \left (\frac{\bar \delta_i(0)}{\bar \delta_i(R_i)}\right)_{max} \left( \frac{\bar \delta_i(R_i)}{a_i} \right) \right] =\mathcal{I}\left [0, \mathpzc{X}_{ta}\left(\frac{\bar \delta_i(R_i)}{a_i} \right), \frac{\bar \delta_i(R_i)}{a_i} \right ].
\end{equation}
The parameter $\mathpzc{x}_{\scriptscriptstyle{\mathscr{TA}}} (\bar \delta_i(0) /a_i )$ is found from Eqn.~\ref{eq:cubic_ta} with the substitution $\bar \delta_i(0)/a_i \rightarrow [\bar \delta_i(0)/\bar \delta_i(R_i)] [\bar \delta_i(R_i)/a_i]$.  We solve for  $(\bar \delta_i(0)/\bar \delta_i(R_i))_{max}$ as a function of $z_c$ for a cosmology with $\Omega_m=0.27$ and $\Omega_\Lambda=0.73$, shown in Fig.~\ref{fig:d_max_v_zc}.  At high redshift, when the cosmology approaches an E-dS cosmology, $(\bar \delta_i(0)/\bar \delta_i(R_i))_{max}$ approaches the E-dS value of $2^{2/3}$.

\subsection{Procedure}
The values of $R_{vir}/R_{ta}$ and $\Delta_c$ as functions of the collapse redshift for a specified cosmology are found in the following manner.  First, we specify a collapse redshift, $z_c$, and normalized initial density profile, $\bar \delta_i(x)/\bar \delta_i(R_i)$.  The value of $\bar \delta_i(R_i)/a_i$ corresponding to $z_c$ is then found either by equating Eqns.~\ref{eq:collapse_time} and \ref{eq:friedmann} and solving for $\bar \delta_i(R_i)/a_i$ numerically, or by simply using Eqn.~\ref{eq:fitting_di_ai} (for a flat cosmology).  We then solve for the $x$ to $y$ mapping as explained in \S~\ref{sec:mapping_gen}.  Once the $x(y)$ function is found, the non-dimensionalized kinetic energy, interior mass profile and potential energies associated with gravity and dark energy at turn-around are found by integrating Eqns.~\ref{eq:k_int_gen}, \ref{eq:m_int}, \ref{eq:u_int} and \ref{eq:u_lam}, respectively.  By specifying a density profile at virialization, the non-dimensionalized potential energies associated with gravity and dark energy at virialization can be found by integrating Eqns.~\ref{eq:zeta} and \ref{eq:U_lam_non_dim}, respectively.  The ratio $R_{vir}/R_{ta}$ is then found by numerically solving the cubic in Eqn.~\ref{eq:cubic_gen}, and $\Delta_c$ is found with Eqn.~\ref{eq:delta_c_gen}.  This entire procedure is then repeated for different values of $z_c$ so that we may build up the functions, $R_{vir}/R_{ta} (z_c)$ and $\Delta_c(z_c)$.

\subsection{$\Lambda$CDM Results}
\label{sec:results_gen}
In this section we present results for a $\Lambda$CDM cosmology with ($\Omega_m, \Omega_{\Lambda})=(0.27, 0.73)$, consistent with the WMAP7+BAO+$H_o$ cosmological parameters of \cite{komatsu}.  We note that although these parameters may differ slightly from those as measured by the Planck satellite, slight differences in $(\Omega_m, \Omega_{\Lambda})$ do not produce any appreciable differences in our results.  We consider two different prescriptions to define the initial density profile.  As a simple example, we use the same initial density profile as in \S~\ref{sec:eds_results} (Eqn.~\ref{eq:initial_density_profile}), with the normalization at $\mathpzc z =0$ chosen to avoid shell crossing before $z_c=0$ in a $\Lambda$CDM cosmology.  We also calculate results for more realistic initial density profiles, derived from the statistics of peaks in an initial density field characterized by the linear theory matter power spectrum.  In either case, we assume an NFW profile as the final halo density profile, so that $\mathscr{U}_{vir}$ is still given by Eqn.~\ref{eq:u_vir}.  The non-dimensionalized binding energy associated with dark energy, $\mathscr{U}_{vir}^{\Lambda}$, is found by integrating Eqn.~\ref{eq:U_lam_non_dim}, with $\varrho(\mathpzc{r})$ given by Eqn.~\ref{eq:varrho_NFW}.  The integral can be computed analytically, and is
\begin{align}
\label{eq:u_vir_nfw}
\mathscr{U}_{vir}^{\Lambda} &= \frac{5}{3}\frac{c^2}{\ln(1+c)-\frac{c}{1+c}}\int_0^1\frac{w^3}{(1+cw)^2}dw \nonumber \\
& = \frac{5}{6}\frac{c[c(c-3)-6]+6(1+c)\ln(1+c)}{c^2(1+c) \left[ \ln (1+c) -\frac{c}{1+c} \right]},
\end{align}
which is plotted as a function of $c$ in Fig.~\ref{fig:ke_u_v_beta}.

\subsubsection{Power-law Initial Density Profile}
\label{sec:power_law_gen}
For an initial density profile given by Eqn.~\ref{eq:initial_density_profile}, as with the E-dS case, we choose the value of $\bar \delta_i(0)/\bar \delta_i(R_i)$ to avoid shell crossing before the outermost shell turns around.  The maximum that $\bar \delta_i(0)/\bar \delta_i(R_i)$ can be in order to avoid shell crossing before $t_{ta}$ for a $\Lambda$CDM cosmology, given a halo collapse redshift, is shown in Fig.~\ref{fig:d_max_v_zc}.  The value of $(\bar \delta_i(0)/\bar \delta_i(R_i))_{max}$ at $z_c=0$ is about 1.479450.  Since $(\bar \delta_i(0)/\bar \delta_i(R_i))_{max}$ increases monotonically with $z_c$, we can avoid shell crossing at all values of $z_c$ by choosing $\bar \delta_i(0)/\bar \delta_i(R_i)$ just below this value.  We therefore use a constant value of $\bar \delta_i(0)/\bar \delta_i(R_i) = 1.479$.  By using a constant value of $\bar \delta_i(0)/\bar \delta_i(R_i)$, rather than using the $(\bar \delta_i(0)/\bar \delta_i(R_i))_{max}$ value for the $z_c$ under consideration, we keep interpretation of our results simple.  That is, we keep the initial density profile constant so that we may be able to clearly see how our results vary with only $z_c$.

%  To choose an appropriate value of the normalized initial density at the center of the sphere, $\bar \delta_i(0)/\bar \delta_i(R_i)$, we consider Fig.~\ref{fig:d_max_v_zc}, the maximum allowed initial central density before our model fails.  Since $(\bar \delta_i(0)/\bar \delta_i(R_i))_{max}$ increases monotonically with $z_c$, the minimum value of $(\bar \delta_i(0)/\bar \delta_i(R_i))_{max}$ occurs at $z_c=0$, and is about 1.479450.  We therefore use a constant value just below this number, 1.479, for $\bar \delta_i(0)/\bar \delta_i(R_i)$, which prevents shell crossing at all $z_c$ that we consider.  It is true that we could use increasing values of $\bar \delta_i(0)/\bar \delta_i(R_i)$ at higher collapse redshift, and since the initial density profile would therefore be even less like a top-hat, our results would differ from the standard top-hat results more dramatically.  To keep interpretation of our results simple, however, we choose to use a constant value of $1.479$ so that we may be able to clearly see how our results vary with only $z_c$ and $\beta$. 
  
\begin{figure}
\centering
\includegraphics[clip, width=4in]{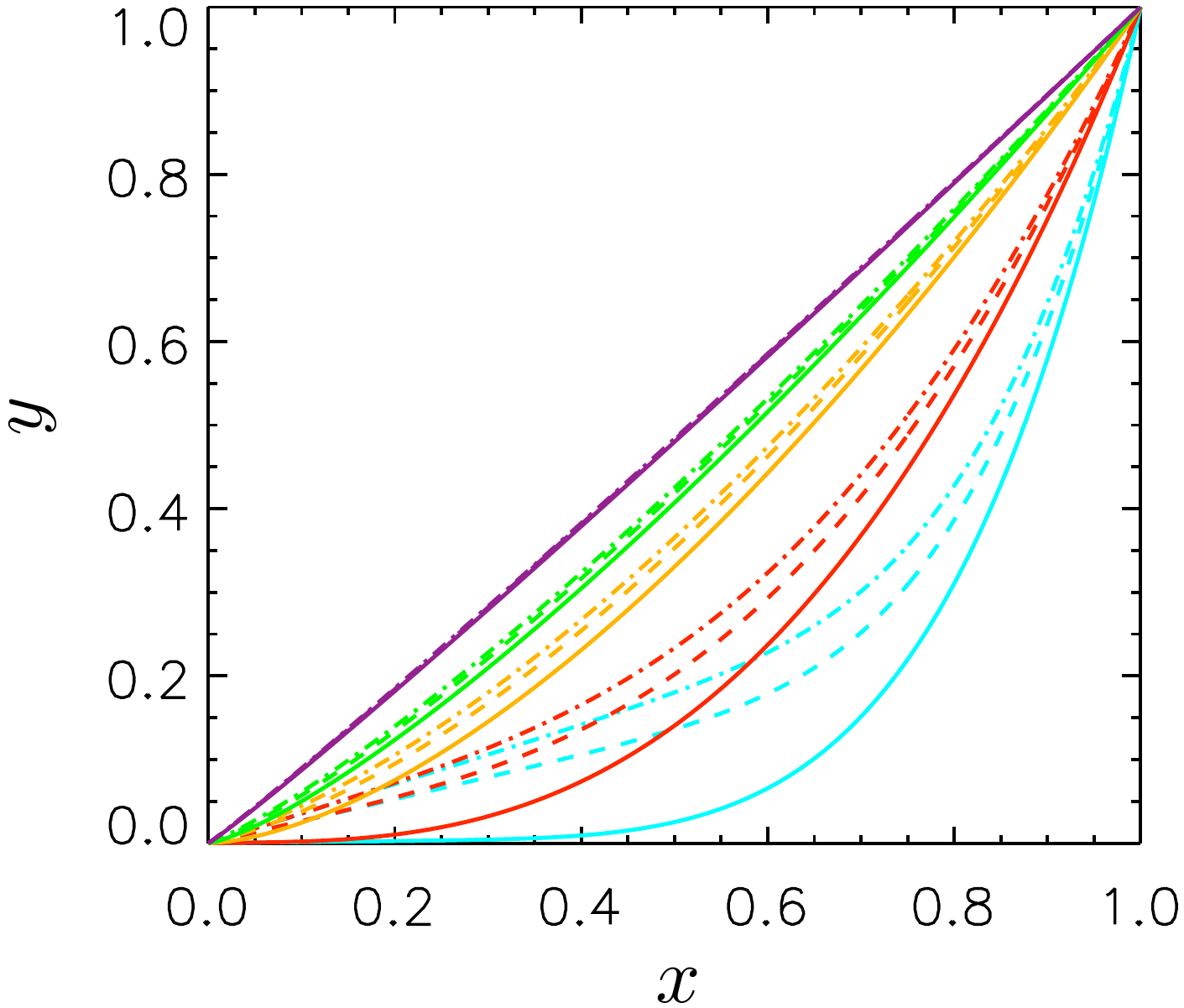}
\caption{ \label{fig:y_z_mapping_gen} The $x$ - $y$ mapping for a $\Lambda$CDM cosmology with an initial density profile given by Eqn.~\ref{eq:initial_density_profile} with $\bar \delta_i(0)/ \bar \delta_i(R_i) =1.479$.  The different colors correspond to the same $\beta$ values as in Fig.~\ref{fig:delta_v_z} (from bottom lines to top lines: $\beta=$ 7, 3, 1, 0.5 and 0.1).  The solid, dashed, and dot-dashed lines correspond to collapse redshifts of 0, 0.5 and 3 respectively.} 
\end{figure}

\begin{figure*}
\centerline{\mbox{  \includegraphics[clip, width=2.2in]{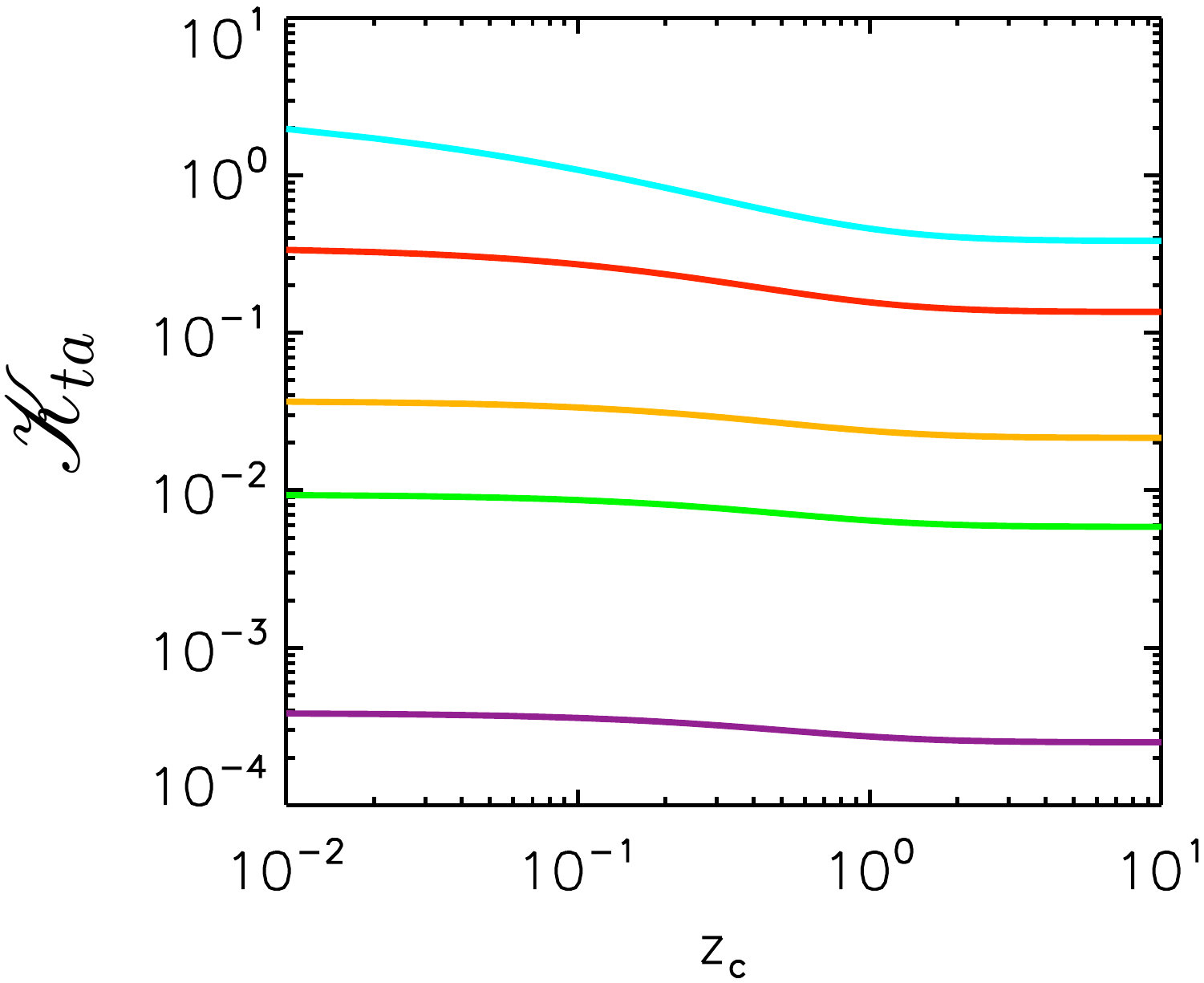}}
\mbox{  \includegraphics[clip, width=2.13in]{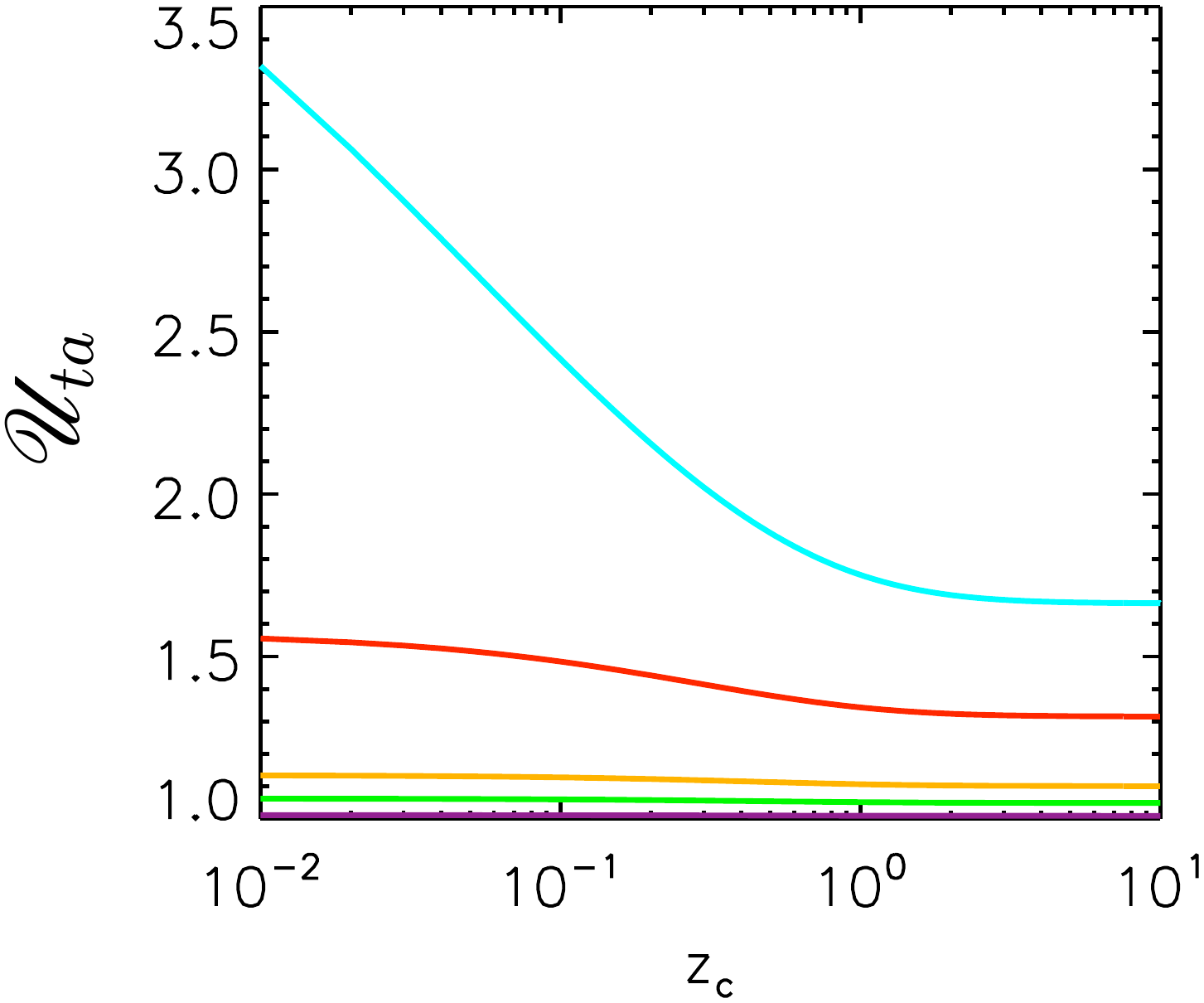}}
\mbox{  \includegraphics[clip, width=2.2in]{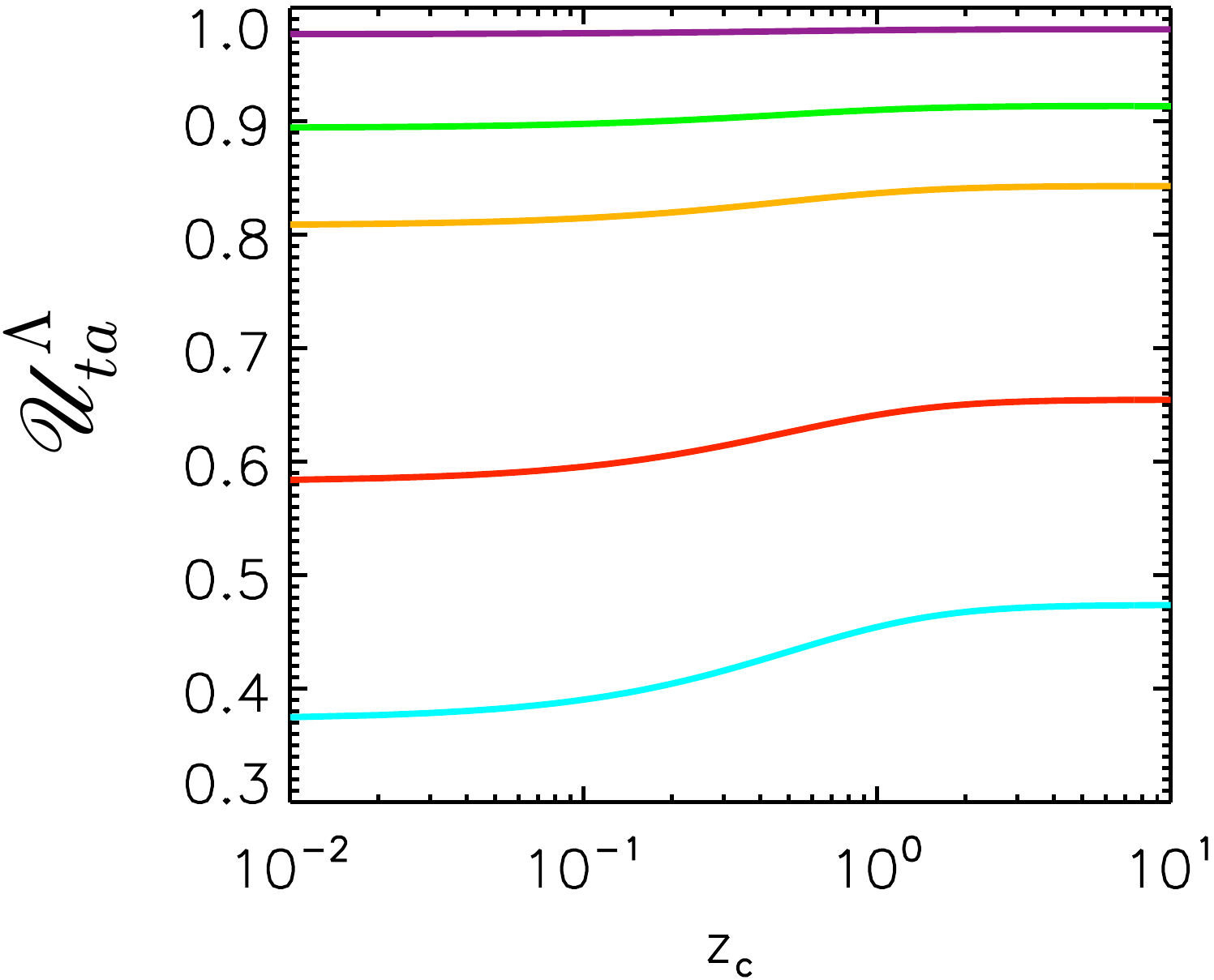}}}
\caption{ \label{fig:non_dim_energies}The non-dimensionalized kinetic and potential energies associated with gravity and dark energy at turn-around as a function of collapse time for a $\Lambda$CDM cosmology.  In each panel, the different color lines correspond to the same $\beta$ values as in Fig.~\ref{fig:delta_v_z} ($\beta=$ 7, 3, 1, 0.5 and 0.1 from top line to the bottom line for the left two panels and from bottom line to top line for the rightmost panel).} 
\end{figure*}

In Fig.~\ref{fig:y_z_mapping_gen}, we show the $x$ - $y$ mapping for several values of $\beta$ (where $\beta$ is defined in Eqn.~\ref{eq:initial_density_profile}) for halos collapsing at different redshifts.  The redshift of collapse clearly affects the mapping, especially at high values of $\beta$.  In Fig.~\ref{fig:non_dim_energies} we show the non-dimensionalized kinetic and binding energies at turn-around as a function of collapse redshift for the same values of $\beta$.  For comparison, for a top-hat, the non-dimensionalized kinetic and binding energies at turn-around should be 0 and unity respectively.  We note that we do not run into the same numerical issues for high values of $\beta$ as with the E-dS case (see App.~\ref{app:approx_form}), so that we do not need to resort to an approximation formula.  This is because our value of $\bar \delta_i(0)/\bar \delta_i(R_i)=1.479$ is sufficiently below $(\bar \delta_i(0)/\bar \delta_i(R_i))_{max}$ at $z_c=0$ (about 1.479450) that we avoid having to resolve prohibitively small differences in our calculations.

\begin{figure*}
\centerline{\mbox{  \includegraphics[clip, width=3.2in]{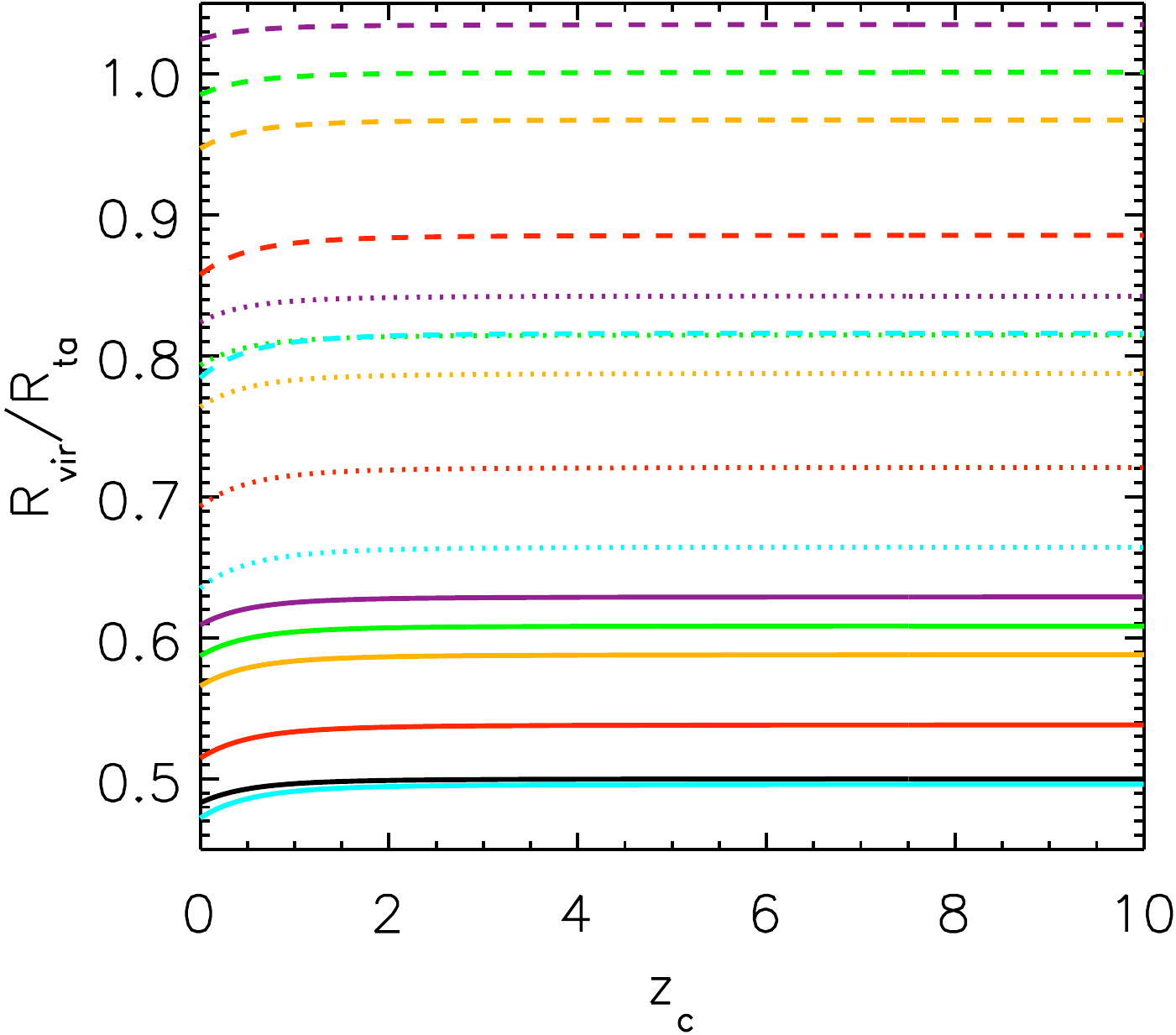}}
\mbox{  \includegraphics[clip, width=3.23in]{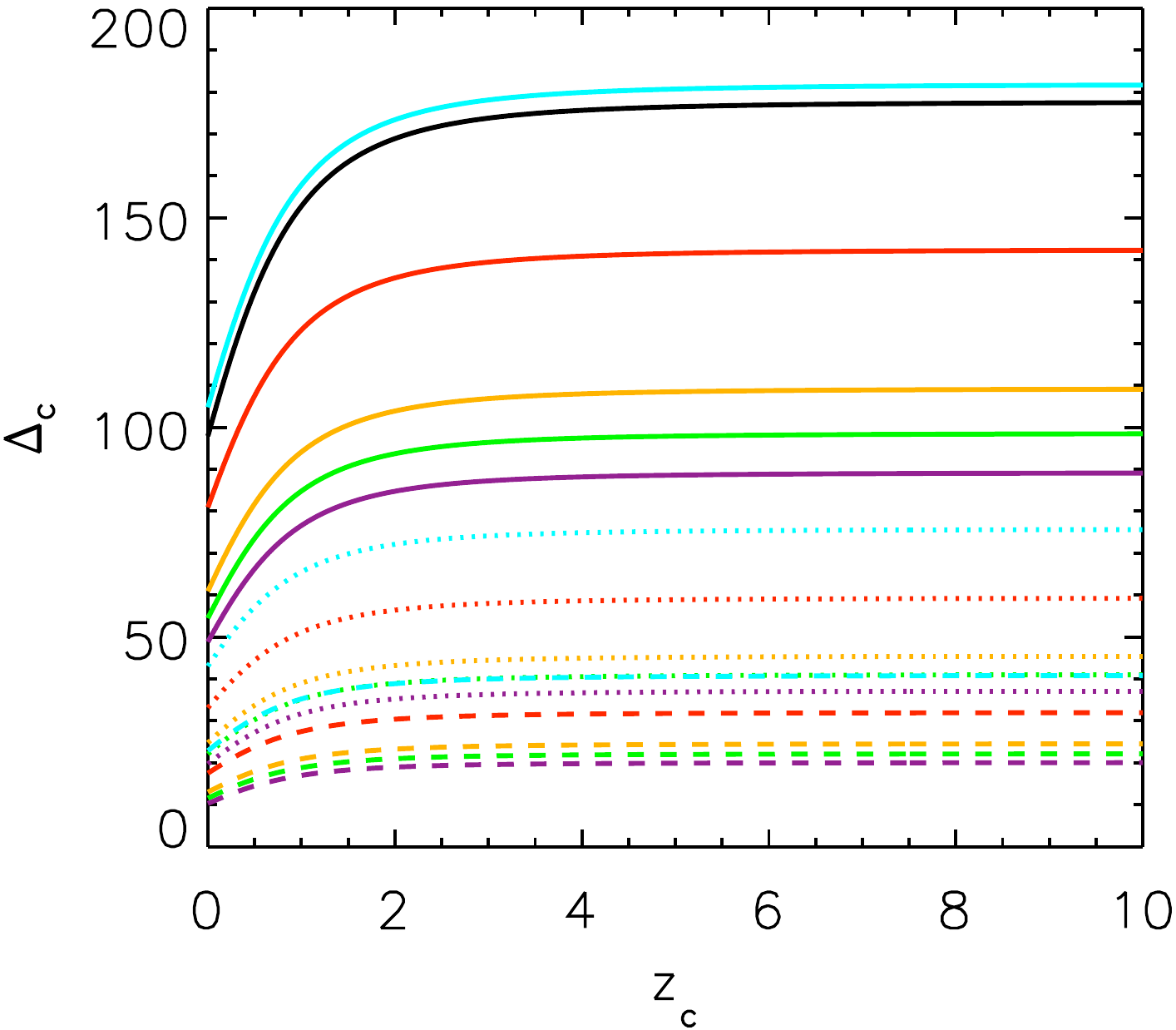}}}
\caption{ \label{fig:results_gen}$R_{vir}/R_{ta}$ and $\Delta_c$ as functions of the collapse redshift for different values of $\beta$ and concentration parameter.  In each panel, the different color lines correspond to the same $\beta$ values as in Fig.~\ref{fig:delta_v_z}.  The solid, dotted, and dashed lines correspond to $c=1$, 5 and 10 respectively.  The solid black line corresponds to the standard initial top-hat result.} 
\end{figure*}

We show our results for $R_{vir}/R_{ta}$ and $\Delta_c$ as functions of $z_c$ for the same values of $\beta$, and for $c=1$, 5, and 10 in Fig.~\ref{fig:results_gen}.  For comparison, we show the standard initial top-hat results for a $\Lambda$CDM cosmology (solid black lines in both panels).  We see that $\Delta_c$ is typically lower than the top-hat case by a factor of a few to as much a factor of about 10.  As with the E-dS results, non-uniformity can allow the virial radius to be bigger than the turn-around radius.

\subsubsection{Peak Statistics Initial Density Profiles}
\begin{figure*}
\centering
\includegraphics[clip, width=6.2in]{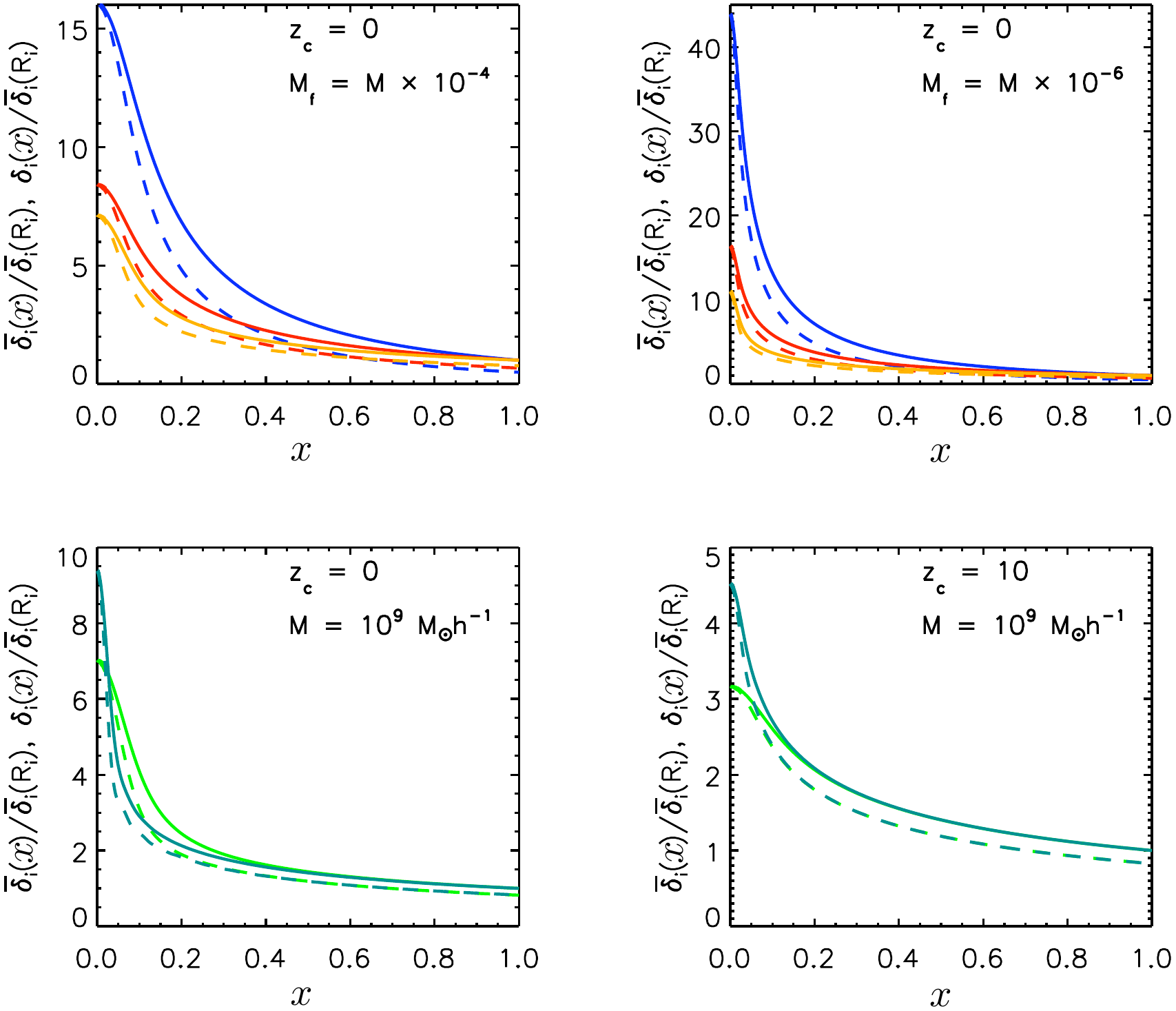}
\caption{ \label{fig:density_profiles_gen} Examples of the volume averaged $\bar \delta_i(x)/\bar \delta_i(R_i)$ (solid lines) and local $\delta_i(x)/\bar \delta_i(R_i)$ (dashed lines) initial density profiles calculated from the peak statistics formalism presented in App.~\ref{app:density_profiles}.  The top two panels show profiles with halo masses of $M=10^{15}$ (blue), $10^{13}$ (red) and $10^{11}$M$_\odot h^{-1}$ (orange) (top set of lines to bottom set of lines) for $z_c=0$ and $M_f = 10^{-4}$ and $10^{-6}$ times the halo mass.  The bottom two panels show profiles for a $10^9$M$_\odot h^{-1}$ halo with $M_f$=$10^5$ (teal) and $M_f  = 10^3$M$_\odot h^{-1}$ (light green) (top set of lines to bottom set of lines) collapsing at $z_c=0$ and $z_c=10$.} 
\end{figure*}

In this section we consider the collapse of halos initially seeded by highly realistic density profiles calculated from the statistics of a Gaussian random field.  With a given linear theory matter power spectrum, one may calculate average halo density profiles while still in the linear regime \cite{bardeen}.  In App.~\ref{app:density_profiles} we summarize this formalism, and cover how we use it to calculate the initial, normalized, volume averaged density profiles, $\bar \delta_i(x)/\bar \delta_i(R_i)$, needed for our calculations.  The profiles are parameterized by halo mass, $M$, collapse redshift, $z_c$, and a co-moving smoothing scale, $M_f$.  In Fig.~\ref{fig:density_profiles_gen} we show examples of these profiles along with the local (non-volume averaged) profiles for several combinations of halo mass, collapse redshift and smoothing scale.  

From Fig.~\ref{fig:density_profiles_gen} it is clearly evident that for these density profiles, shell crossing within the halo will occur before the outermost shell turns around since the density of the innermost shell can far exceed $\sim 1.48$ (as previously discussed).  We therefore may not use the shell kinematic formalism derived in \S~\ref{sec:conditions_general} to calculate the physical conditions at turn-around (i.e., $\mathscr{K}_{ta} $, $\mathscr{U}_{ta}$ and $\mathscr{U}_{ta}^{\Lambda}$).  To calculate these quantities, we employ a one dimensional Lagrangian simulation (described in detail in App.~\ref{sec:simulation}) up until the outermost shell turns around.  These quantities are then found by summing across all shells at the end of the simulation using Eqns.~\ref{eq:u_ta_sim}-\ref{eq:u_ta_lam_sim} (where the symbols in these equations are defined throughout App.~\ref{sec:simulation}).

\begin{figure*}
\centerline{\mbox{\includegraphics[clip, width=3.2in]{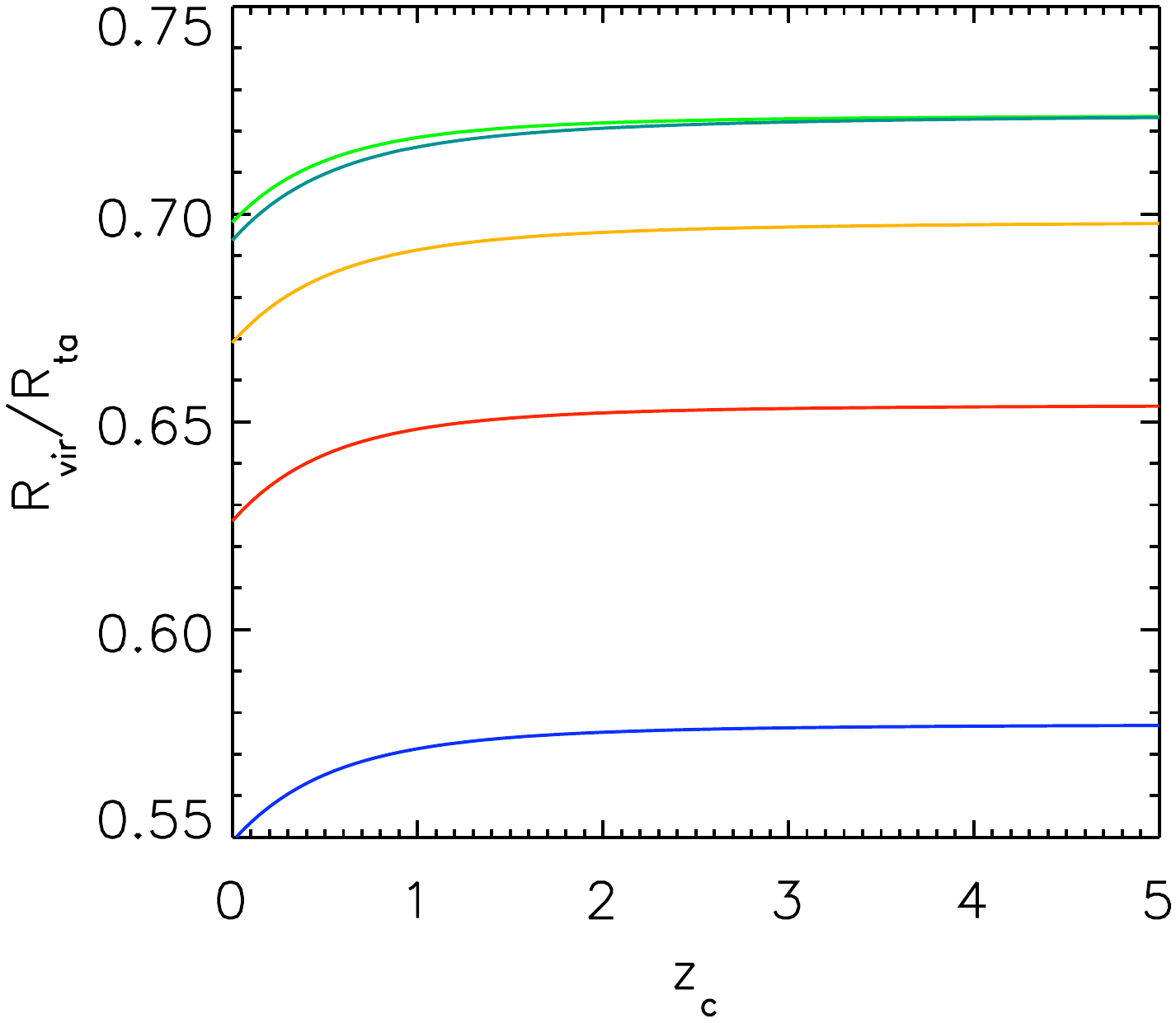}}
\mbox{  \includegraphics[clip, width=3.14in]{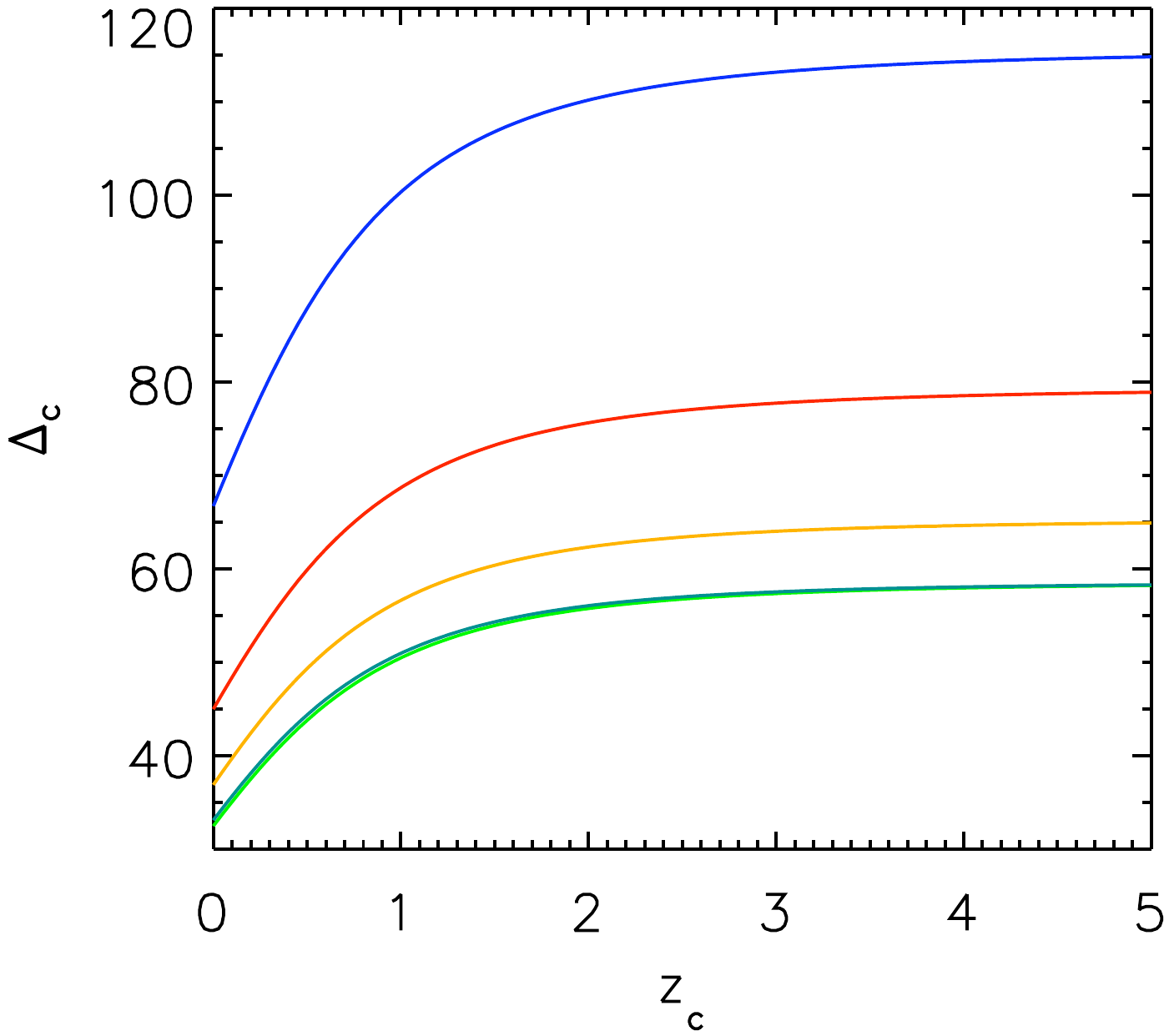}}}
\caption{ \label{fig:results_realistic}$R_{vir}/R_{ta}$ and $\Delta_c$ as functions of the collapse redshift for halos initialized with density profiles calculated from the peak statistics formalism.  The blue, red, orange, teal and light green lines (bottom line to top line in the left panel and top line to bottom line in the right panel) correspond to $(M, M_f)/$M$_\odot h^{-1} = (10^{15}, 10^{10})$, $(10^{13}, 10^8)$, $(10^{11}, 10^7)$, $(10^{9}, 10^5)$ and $(10^{9}, 10^3)$, respectively.} 
\end{figure*}

To calculate $R_{vir}/R_{ta}$ and $\Delta_c$ we may still use Eqns.~\ref{eq:cubic_gen} and~\ref{eq:delta_c_gen} since these equations only assume global conservation of energy (not energy conservation for each particular shell).  We calculate these quantities for halo masses ranging from $M=10^{9}$ to $M=10^{15}$M$_\odot h^{-1}$.  This range in mass corresponds roughly to the halo mass of a small galaxy up to a large galaxy cluster.  To calculate $\mathscr{U}_{vir}^{\Lambda}$ and $\mathscr{U}_{vir}$ we again use an NFW profile at virialization (Eqns.~\ref{eq:u_vir_nfw} and ~\ref{eq:u_vir}) with $c=4$.  We show $R_{vir}/R_{ta}$ and $\Delta_c$ as functions of collapse redshift for different halo masses in Fig.~\ref{fig:results_realistic}.  For $M=10^9$M$_\odot h^{-1}$ we show the results for two different smoothing scales (teal and light green lines).  It is seen that $R_{vir}/R_{ta}$ and $\Delta_c$ have little dependance on smoothing scale.  For halos of larger mass, these quantities have even less dependance on smoothing scale.  We do not plot $\Delta_c$ from the standard calculation since it goes significantly above 120.  It should be kept in mind that, as see in Fig.~\ref{fig:results_gen}, this function starts at about 100 and rises gradually to about 180 at the highest redshifts.  By using these density profiles, $\Delta_c$ is typically smaller by a factor of a few as compared to the standard calculation.

\section{Discussion and Conclusions}
We have calculated the non-linear density of a halo at virialization based on the spherical collapse of a density peak with arbitrary initial and final density profiles.  This is an improvement over the standard result which assumes top-hat profiles in order to simplify the calculation.  For collapsing halos in an E-dS universe, we have used the parametric solution of spherical collapse to solve for the density and velocity profiles at turn-around.  We are thus able to calculate the total potential and kinetic energy at turn-around.  By assuming a density profile at virialization, we are able to employ the virial theorem to calculate the non-linear over-density at the time of collapse.  Using power-law profiles for the initial density and an NFW profile for the virialized density, we find that the over-density at collapse can lower by a factor of 10 relative to the standard value of 178.

We extend our calculation to cosmologies which can include curvature and a cosmological constant.  For moderately peaked initial density profiles, we numerically solve the equations of spherical collapse till turn-around.  For initial profiles which result in shell crossing before turn-around, we implement a one dimensional numerical simulation.  We calculate the over-density for halos in a $\Lambda$CDM cosmology using power-law initial density profiles and profiles calculated from the statistics of a Gaussian random field characterized by a $\Lambda$CDM linear theory matter power spectrum.  For all cases, we find that the non-linear halo density at collapse is significantly smaller by as much as a factor of about 20 (depending on the density profiles used) than that as predicted by the standard top-hat calculation.  We note that for regions in our universe with large-scale over-densities (such as superclusters) or large-scale under-densities (such as voids) the over-density of newly formed halos is the same, regardless of environment.  Even though the dynamics of these regions are effectively governed by a cosmology that includes curvature, \cite{rubin:2012} show that, regardless of the large-scale over/under-density, halos collapsing at the same time must have the same non-linear density.

While our calculation is an improvement over the standard one, it is still based on a highly simplified model of the dynamics of halo collapse.  We have assumed that halos evolve in isolation, so that the gravitational potential of nearby matter can be ignored.  In reality, nearby matter exerts torques on a collapsing halo, inducing angular momentum and breaking the spherical symmetry.  In fact, numerical simulations show that halo collapse is in general ellipsoidal rather than spherical.  The assumption of isolation also ignores the accretion of matter and mergers during collapse.  Additionally, by only considering the final, virialized state of the system, we have swept under the rug all the uncertain physics that occurs during virialization.

%We admit that our calculation is still a great simplification to what is actually happening due to...Granted, we have still simplified the problem.  We do not consider elliptical collapse, virialization beyond the Newtonian limit, or the effects of baryons.  Baryons will affect the virialization of dark matter through their gravitational potential.  Moreover, the physics of the baryons in a collapsing halo is more complicated due to effects such as, shocks, radiative heating and cooling processes and pressure.  isolated halos.  we have considered the halo in isolation and have ignored the gravitational effects of structure around it, as well as halo mergers.

It is interesting to ask how our results may affect analytic halo mass functions when implemented in them.  One such function, the Press-Schechter (PS) \cite{press:1974} mass function, requires a linearized density threshold above which a halo is defined.  This threshold is typically found by using the spherical collapse model to determine the collapse time (defined as twice the turn-around time) of a top-hat density perturbation in the cosmic density field.  Linear theory is then used to calculate the linear over-density at the time of collapse, $\delta_c$, which, for an E-dS universe equals 1.686.  Our results, however, will not affect the PS mass function since our definition of the halo collapse time (twice the turn-around time of the outermost shell) is equivalent to the collapse time of a top-hat perturbation since we do not consider halos whose outermost shell undergoes shell crossing before turn around.  Another  mass function which compares better to numerical simulations is the Sheth-Tormen (ST) mass function \cite{sheth:2001}, the form of which is motivated by ellipsoidal collapse.  Our results have no effect the ST mass function either since its shape is determined by several free parameters which are calibrated with numerical simulations.

\begin{figure}
\centering
\includegraphics[clip, width=6in]{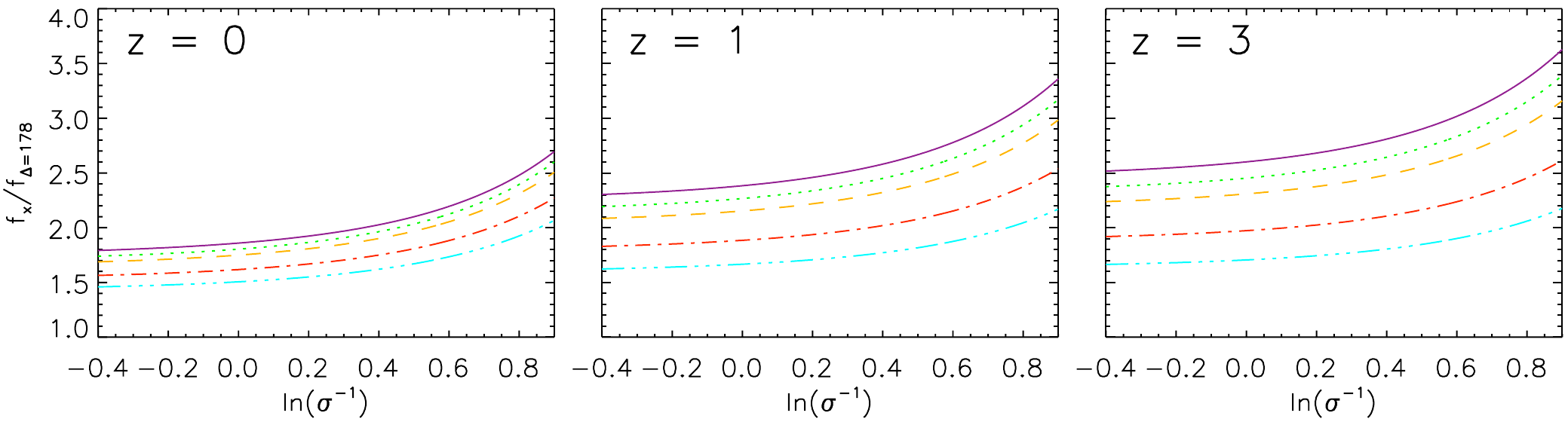}
\caption{ \label{fig:mass_func}The halo mass function using an over-density criterion given by our calculations relative to the halo mass function with the standard 178 over-density criterion.  We use power-law initial density profiles with $\beta = 7$, 3, 1, 0.5, 0.1 (dot-dot-dot dashed turquoise, dot-dashed red, dashed orange, dotted green and solid purple lines respectively).   Here, $f \equiv dN/d \ln \sigma^{-1}$, where $N$ is the halo mass function and $\sigma$ is the cosmic variance.  To make these figures, we use the fitting functions provided by \cite{watson}.} 
\end{figure}

An important implication of our results is how they affect the halo mass function measured from simulations which use halo finders that search for an over-density threshold in the cosmological density field (the spherical over-density method).  For example, \cite{watson} have studied this problem by implementing cosmological simulations and by measuring the halo mass function using different over-density thresholds.  They find that the halo mass function measured with an arbitrary over-density criterion is related to the halo mass function measured with the 178 criterion by a simple scaling relation.  Defining $f$ to be $dN/d \ln \sigma^{-1}$, where $N$ is the halo mass function and $\sigma$ is the cosmic variance of different sized regions, they provide an accurate fitting formula (as a function of over-density criterion, $z$ and $\sigma$) for the scaling relation $f_x/f_{\Delta=178}$.  Using this fitting formula, we show how our results may affect the measured halo mass function in a $\Lambda$CDM cosmology.  Given collapse redshifts of 0, 1 and 3, we calculate the appropriate over-density criterion, $\Delta_c$, assuming power-law initial density profiles with $\beta =$ 0.1, 0.5, 1, 3 and 7 and an NFW profile with $c=5$ at virialization\footnote{We should note that the fitting function for $f_x/f_{\Delta=178}$ at a particular redshift represents the cumulative effect of halo collapse (and halo mergers) at all earlier times.  It is thus not strictly valid to use a single over-density criterion to calculate $f_x/f_{\Delta=178}$ at $z$.  For simplicity, however, we use $\Delta_c(z_c=z)$.}.  Fig.~\ref{fig:mass_func} shows that the mass function measured with our over-density criterion can be several times higher relative to the standard 178 criterion mass function.  The discrepancy is most significant at the highest halo masses.  The overall enhancement of the halo mass function when using our over-density threshold makes qualitative sense since it is easier for density peaks to meet the halo criterion when it is lowered.  This discrepancy is important when analytic models of the halo mass function are assessed by their agreement with the results from numerical simulations.

%They found that, for all redshifts under consideration, threshold values greater than the canonical value of $178$ suppress the halo mass function at all masses and especially the lowest halo masses (see their figure 11).  Threshold values less than the canonical value enhance the halo mass function at all masses and especially the highest halo masses.  The overall suppression/enhancement makes qualitative sense since it is easier for density peaks to meet the halo criterion when it is lowered and harder for density peaks to meet it when it is raised.  

\begin{appendices}
\section{Approximation Formula for $x$ - $y$ Mapping in an E-dS Universe}
\label{app:approx_form}

Numerically calculating the mapping from $x$ to $y$ as given by the formalism in \S~\ref{sec:mapping_eds} becomes impossible for small $x$ at the largest values of $\beta$.  This is because, as seen in Fig.~\ref{fig:delta_v_z}, $\bar \delta_i(x)/\bar \delta_i(R_i)$ comes very close $2^{2/3}$ for small $x$ when $\beta$ is large (for example, see the turquoise line which corresponds to $\beta=7$).  Thus, when the outermost shell turns around, the inner shells have a value of $\Theta_{ta}(x)$ extremely close to $2 \pi$ (i.e., these shells are almost fully collapsed).  Even when using double-precision, the difference between $\Theta_{ta}(x)$ and $2 \pi$ ($\equiv \Delta_{\Theta}$) for these shells becomes impossible to resolve numerically.  Since $y \propto 1- \cos \Theta_{ta}(x) \propto \Delta_{\Theta}^2/2+\mathcal O(\Delta_{\Theta}^4)$ (see Eqn.~\ref{eq:y_function}) when expanded around $2 \pi$, $y$ is also impossible to resolve.  This is a problem in our analysis, because even though $y$ is very small in this regime, these shells still contribute non-negligibly to the integrals used to compute $\mathscr{M}_{ta}$, $\mathscr{U}_{ta}$ and $\mathscr{K}_{ta}$.

In this appendix, we derive a highly accurate approximation formula used to calculate the mapping in this regime.  We start by setting equal the solution for the time of the outermost shell at turn-around to the solution at this time for a shell starting at $x$, found by integrating Eqn.~\ref{eq:eq_of_motion} in an E-dS universe:
\begin{equation}
\label{eq:t_ta_approx}
\int_0^{\mathpzc{X}_{ta}} \frac{dx} {\sqrt{\frac{1}{x}-\frac{5}{3} \frac{\bar \delta_i(R_i)}{a_i}}} = \int_0^{\mathpzc{x}_{\scriptscriptstyle{\mathscr{TA}}}} \frac{dx}{\sqrt{\frac{1}{x}- \frac{5}{3} \frac{\bar \delta_i(r_i)}{a_i}}}+\int_{\frac{y}{x}\mathpzc{X}_{ta}}^{\mathpzc{x}_{\scriptscriptstyle{\mathscr{TA}}}} \frac{dx}{\sqrt{\frac{1}{x}- \frac{5}{3} \frac{\bar \delta_i(r_i)}{a_i}}}.
\end{equation}
On the right hand side of this equation, the first integral represents the amount of time that it takes for a shell starting at $x$ to turn-around, and the second the time between this shell's turn-around and the outermost shell's turn-around.  The lower bound on the second integral $(y/x)\mathpzc{X}_{ta}$ is equal to $\mathpzc{x}_{ta}$ (the normalized position of the shell when the outermost shell turns around), as found with Eqn.~\ref{eq:y_z_mapping_expr}.  

We change the variable of integration on the left hand side of the equation with $x^{\prime} \equiv x (5/3) \bar \delta_i(R_i)/a_i = x \mathpzc{X}_{ta}^{-1}$ and $x^{\prime} \equiv x =x (5/3) \bar \delta_i(r_i)/a_i = x\mathpzc{x}_{\scriptscriptstyle{\mathscr{TA}}}^{-1}$ on the right hand side of the equation:
\begin{align}
\int_0^1 \frac{dx^{\prime}}{\sqrt{\frac{1}{x^{\prime}}-1}} & =  \left [ \frac{\bar \delta_i(r_i)}{\bar \delta_i(R_i)} \right]^{-3/2}  \left [ \int_0^1 \frac{dx^{\prime}}{\sqrt{\frac{1}{x^{\prime}}-1}} +\int_{\frac{y}{x}\frac{\bar \delta_i (r_i)}{\bar \delta_i(R_i)}}^1 \frac{dx^{\prime}}{\sqrt{\frac{1}{x^{\prime}}-1}} \right] \nonumber \\
 & = \left [ \frac{\bar \delta_i(r_i)}{\bar \delta_i(R_i)} \right]^{-3/2} \left [2\int_0^1 \frac{dx^{\prime}}{\sqrt{\frac{1}{x^{\prime}}-1}} - \int_0^{\frac{y}{x}\frac{\bar \delta_i(r_i)}{\bar \delta_i(R_i)}} \left [x^{1/2} + \mathcal O (x^{3/2}) \right]dx \right].
\end{align}
In going from the first line to the second line, we have re-written the second integral on the right hand side as the integral going from 0 to 1 minus the integral going from 0 to $(y/x)(\bar \delta_i(r_i)/\bar \delta_i(R_i))$.  We taylor expand the integrand of the second integral in the second line because its upper bound is much smaller than unity for shells starting at small $x$ ($y\ll x$ for these shells).  The integral $\int_0^1 dx/\sqrt{1/x-1}$ is equal to $\pi/2$, so that integrating all terms and soving for $(y/x)^{3/2}$, we find
\begin{equation}
\label{eq:y_z_stuff}
\left(\frac{y}{x} \right)^{3/2}=\frac{3 \pi}{2}\left \{ \frac{1}{\left [\bar \delta_i(r_i)/\bar \delta_i(R_i)\right ]^{3/2}}-\frac{1}{2} \right \} +\mathcal O\left (\left (\frac{y}{x}\right)^{5/2}\right).
\end{equation}

Since $1/[\bar \delta_i(0)/\bar \delta_i(R_i)]^{3/2}=1/2$, and since $1/[\bar \delta_i(r_i)/\bar \delta_i(R_i)]^{3/2}$ gets very close to this value at small $r_i$ for large $\beta$, the subtraction within the curly brackets becomes impossible to resolve (even with double-precision).  To avoid computing this subtraction, we work instead with the difference between the density at the origin and the density profile,
\begin{equation}
\label{eq:def_of_delta}
\Delta \equiv \frac{\bar \delta_i(0)}{\bar \delta_i(R_i)}-\frac{\bar \delta_i (r_i)}{\bar \delta_i(R_i)},
\end{equation}
and for our adopted initial density profile $\Delta = x^{\beta} (2^{2/3}-1)$ (see Eqn.~\ref{eq:initial_density_profile}).  Substituting Eqn.~\ref{eq:def_of_delta} in Eqn.~\ref{eq:y_z_stuff}, and taylor expanding in $\Delta/2^{2/3}$ we find that
\begin{equation}
\left(\frac{y}{x} \right)^{3/2}=\frac{3 \pi}{2}\left \{ \frac{1}{2} +\frac{3}{4} \frac{\Delta}{2^{2/3}} +\frac{15}{16}\frac{\Delta^2}{2^{4/3}} + \mathcal O (\Delta^3)-\frac{1}{2} \right \} +\mathcal O\left (\left (\frac{y}{x}\right)^{5/2}\right),
\end{equation}
so that the 1/2 cancels out (and thus the problem of resolving the difference from $1/2$ goes away).  We keep terms to second order in $\Delta/2^{2/3}$ since we find that this results in very high accuracy when comparing to our numerical calculation for $y$ as a function of $x$ in the regime in which both the analytic and numerical approaches are valid.  Solving for $y$, our approximation formula is
\begin{equation}
y \cong \left \{ \frac{9 \pi}{8} \frac{\Delta} {2^{2/3}} \left[1+\frac{5}{4} \frac{\Delta}{2^{2/3}} \right] \right \}^{2/3}x.
\end{equation}

Our numerical approach is not an issue for  $\beta \lesssim 1$ because the integrals used to compute $\mathscr{M}_{ta}$, $\mathscr{U}_{ta}$ and $\mathscr{K}_{ta}$ converge in $x$ before $y$ becomes too small to calculate numerically.  For $\beta \approx 1$ we find that the approximation formula is accurate to $\sim 0.01 \%$ for $y \sim 10^{-6}$, and increases in accuracy by several magnitudes for decreasing $y$, and higher values of $\beta$.  Since our numerical mapping fails at $y \sim 10^{-10}$ to $y \sim 10^{-13}$ for $\beta \approx 7$ to $\beta \approx 1$ respectively, we therefore switch to the approximation formula for $y < 10^{-9}$, a regime in which the formula's accuracy is excellent.

\section{Initial density profile of a spherical perturbation}
\label{app:density_profiles}
The formalism to calculate average density profiles around peaks in a Gaussian random field has been derived by \cite{bardeen}.  Subsequent authors have used these equations to initialize realistic density profiles of collapsing dark matter halos (e.g.: \cite{lilje, eisenstein, cupani:2008, cupani:2011}), as well as expanding voids \cite{sheth:2004}.  We now present this formalism, and our prescription to normalize the initial density profile given a halo mass $M$, and collapse redshift, $z_c$.

%For a linear theory matter power spectrum at $z=0$, $P(k)$, smoothed on a comoving scale, $R_f$ with a given window function, the $(l+1)$-th even moment of the smoothed density field is:

The $(l+1)$-th even moment of a density field smoothed on a comoving scale, $R_f$, and described by a linear theory matter power spectrum today, $P(k)$, is given by
\begin{equation}
\sigma_l^2 \equiv \frac{1}{2 \pi^2} \int_0^\infty P(k)| \widetilde W(k R_f)|^2 k^{2(l+1)}dk.
\end{equation}
Consistent with the works mentioned above, we use a Gaussian window function (in real space) so that, that in $k$ space:
\begin{equation}
\label{eq:window_func}
\widetilde W(u) = \exp \left (-\frac{1}{2} u^2\right ).
\end{equation}
The choice of a Gaussian window function was originally made by \cite{bardeen} since a top-hat leads to divergence issues in some of their integrals.  \footnote{We note that the calculations of $\sigma_l^2$ and $\bar \delta_i$ (which also involves a factor of $|\widetilde W(x)|^2$) in both \cite{cupani:2008} and \cite{cupani:2011} contain a factor of $1/2$ mistake.  Their mistake is due to not squaring the fourier transform of the window function (Eqn.~\ref{eq:window_func}), which cancels out the factor of $1/2$ in the exponential.}  According to linear theory, the mass contained within the smoothing scale, $R_f$, is given by
\begin{align}
M_f & = (2 \pi)^{3/2}\bar \rho_m(0) R_f^3 \nonumber \\
& = 4.37 \times 10^{12} \mathrm{M}_{\odot} h^{-1} \Omega_m \left (\frac{R_f}{\mathrm{Mpc}~h^{-1}}\right)^3,
\end{align}
where $\bar \rho_m(0)$ is the co-moving average (matter) density of the universe, and where the pre-factor of $(2 \pi)^{3/2}$ is due to the use of the Gaussian window function.

We use a linear theory matter power spectrum calculated with the CAMB web interface with cosmological parameters consistent with the WMAP7+BAO+$H_0$ parameters of \cite{komatsu}: $(\Omega_m,~\Omega_b,~\Omega_\Lambda,~ \sigma_8,~n,~h)= (0.27,~0.046,~0.73,~0.81,~0.97,~0.7)$.  Note that, for consistency, the values $\Omega_m=0.27$ and $\Omega_\Lambda=0.73$ are the same as those used in \S~\ref{sec:results_gen} to calculate the spherical collapse dynamics to solve for the conditions at turn-around in a $\Lambda$CDM cosmology.  

Lilje and Lahav 1991 show that the spherically (and peak ensemble) averaged density profile associated with a $\nu \sigma_0$ peak in a Gaussian random field, smoothed on a scale $R_f$ and linearly extrapolated to a time, $a_i$, is\footnote{Since we have hitherto been using the letter, $r$, to denote the radial variable in physical coordinates, in the following equations we use the script, $\rr$ to denote the co-moving radial position.}.
\begin{equation}
\label{eq:gauss_field_density}
\delta_i(\rr, a_i, R_f)= \frac{1}{2 \pi^2 \sigma_0(R_f)}\frac{D(a_i)}{D(0)}\int_0^\infty k^2 P(k)e^{-(R_fk)^2}\frac{\sin k \rr}{k \\r}\left [\frac{\nu -\gamma^2 \nu -\gamma \theta}{1-\gamma^2} +\frac{\theta r_\star^2}{3 \gamma(1-\gamma^2)} k^2\right] dk.
\end{equation}
The density profile, when volume averaged with Eqn.~\ref{eq:vol_avg}, is given by:
\begin{equation}
\label{eq:dens_profile_real}
\bar \delta_i(\rr, a_i, R_f)= \frac{3}{2 \pi^2 \sigma_0(R_f) \rr}\frac{D(a_i)}{D(0)}\int_0^\infty k j_1(k \rr)P(k)e^{-(R_fk)^2}\left [\frac{\nu -\gamma^2  \nu -\gamma \theta}{1-\gamma^2} +\frac{\theta r_\star^2}{3 \gamma(1-\gamma^2)} k^2\right] dk,
\end{equation}
where $j_1$ is the 1st order Bessel function.  

The quantities $r_{\star}$ and $\gamma$ are calculated from the moments of the power spectrum, 
\begin{equation}
\label{eq:r_star}
r_\star \equiv \sqrt{3}\frac{\sigma_1(R_f)}{\sigma_2(R_f)}~~\mathrm{and}~~ \gamma \equiv \frac{\sigma_1^2(R_f)}{\sigma_2(R_f) \sigma_0(R_f)},
\end{equation}
$\theta=\theta(\gamma \nu, \gamma)$ and $D$ is the growing mode of the linear theory growth factor.  For a flat cosmology with a cosmological constant, the growth factor is well approximated (to within $\sim 2 \%$ for $\Omega_m>0.1$ \cite{loeb:2012})  by \cite{carroll}:
\begin{equation}
D = \frac{5}{2} \frac{\Omega_m(z)}{(1+z)}  \left \{ \Omega_m^{4/7}(z)- \Omega_{\Lambda}(z)+ \left [1+\frac{\Omega_m(z)}{2} \right ]\left [1+\frac{\Omega_{\Lambda}(z)}{70} \right ] \right \}^{-1},
\end{equation}
with
\begin{equation}
\Omega_m(z) = \frac{\Omega_m (1+z)^3}{\Omega_m (1+z)^3+\Omega_{\Lambda}},
\end{equation}
and
\begin{equation}
\Omega_{\Lambda}(z) = \frac{\Omega_{\Lambda}}{\Omega_m (1+z)^3+\Omega_{\Lambda}}.
\end{equation}
This expression for the growth factor is normalized so that $D=a$ as $\Omega_m(z) \rightarrow 1$ at high redshift.  For the use of the reader, we tabulate values of $r_\star$, $\gamma$ and $\sigma_0$ for our power spectrum for various smoothing scales in Tab.~\ref{tab:mf_stuff}.

%We should point out, however, that the power spectrum we use includes a baryonic component and since $a_i$ is at an early cosmic time, they are coupled to the CMB, and are thus not described by a Gaussian random field.  
%\textbf{We check explicitly that our integrals over the power spectrum converge with the upper and lower integration bound, and the power spectrum resolution.  We check this for the values of $R_f$ that we use, and (in Eqns.~\ref{eq:gauss_field_density} and \ref{eq:dens_profile_real}) for the range of $\nu$, $\gamma$, $\theta$ and $r_{\star}$ that we use.}
%\textbf{We do note one slight inconsistency our power spectrum includes a baryonic compoenent, and the density peaks were only for Gaussian random fields (are the baryons initially a gaussian random field?).  Moreover, in our spherical collapse model, we treat all matter as cold and pressureless, which is not right if there is a small baryonic component.  This treatment will be valid on scales larger than the Jean's scale, but invalid on smaller scales.  In any case, the affect of the baryons will be small, since the baryons are only about 20\% of the total matter budget.}

  \begin{table}[h]
\caption{$r_\star$, $\gamma$ and $\sigma_0$ for various smoothing scales} % title name of the table
\centering % centering table
\begin{tabular}{c lcc} % creating 10 columns
\hline\hline % inserting double-line
$M_f~[\mathrm{M}_\odot h^{-1}] $ & $r_\star~[\mathrm{coMpc} \, h^{-1}] $ & $\gamma$ & $\sigma_0$
\\ [0.5ex]
\hline
  \\[-2.ex]
$10^4$ & 0.00373710 & 0.381189 & 8.99722 \\
$10^5$ & 0.00745276 & 0.400130 & 8.02269 \\
$10^6$ & 0.0155689 & 0.419671 & 6.97921 \\
$10^7$ & 0.0330506 & 0.440261 & 5.94865 \\
$10^8$ & 0.0704251 & 0.463409 & 4.96486 \\
$10^9$ & 0.149908 & 0.489483 & 4.04212 \\
$10^{10}$ & 0.318194 & 0.518820 & 3.19087 \\
$10^{11}$ & 0.672599 & 0.551657 & 2.42195 \\
$10^{12}$ & 1.41376 & 0.587942 & 1.74736 \\
$10^{13}$ & 2.95001 & 0.627038 & 1.17972 \\
$10^{14}$ & 6.10172 & 0.667419 & 0.730102 \\
$10^{15}$ & 12.5073 & 0.706831 & 0.403660 \\ [,5 ex]

 \hline
  \end{tabular}
\label{tab:mf_stuff}
\end{table}
  
Since the function, $\theta(\gamma \nu, \gamma)$ is not straightforward to calculate, most authors have used the fitting function for $\theta$ provided by \cite{bardeen}, which they quote to be accurate to within 1\% in the ranges $0.4 < \gamma < 0.7$ and $1 < \gamma \nu < 3$.  For the calculations presented in this paper, we do not necessarily stay within this range, and therefore calculate $\theta(\gamma \nu, \gamma)$ explicitly.  The function $\theta$ is found by evaluating \cite{bardeen}:
\begin{equation}
\theta(\gamma \nu, \gamma)= \frac{\int_0^\infty  \exp\left[\frac{-(x-\gamma \nu)^2}{2(1-\gamma^2)} \right]xf(x)dx}{\int_0^\infty  \exp\left[\frac{-(x-\gamma \nu)^2}{2(1-\gamma^2)} \right]f(x)dx}-\gamma \nu,
\end{equation}
with
\begin{align}
f(x) & \equiv \frac{x^3-3x}{2}\left \{ \mathrm{erf} \left [ \left (\frac{5}{2}\right )^{1/2}x\right] +  \mathrm{erf} \left [ \left (\frac{5}{2}\right )^{1/2}\frac{x}{2}\right] \right \} \nonumber \\
& +\left ( \frac{2}{5 \pi} \right)^{1/2} \left[ \left( \frac{31 x^2}{4} + \frac{8}{5}\right)e^{-5x^2/8} + \left (\frac{x^2}{2} -\frac{8}{5} \right)e^{-5 x^2/2} \right].
\end{align}
%Indeed, when we compare our exact calculations of $\theta$ with the fitting formula for extreme values of $\gamma$ and $\gamma \nu$ and find that the fitting formula can deviate by ... percent.

%Note that since $\delta_i$ and $\bar \delta_i$ are calculated in linear theory, they will only represent the true density profiles at early times when $\delta_i$ and $\bar \delta_i$ are $\ll 1$.  

For the purposes of this paper, we wish to specify the initial density profile of a halo, normalized by its value at the edge at which the halo is identified, $\bar \delta_i(r_i)/ \bar \delta_i(R_i)$.  To solve for this profile, we first specify a smoothing scale of interest, $M_f$, so that $\sigma_0$, $r_{\star}$ and $\gamma$ may be calculated for a given power spectrum.  We then find the initial seed of a halo, $\bar \delta_i(R_i)/a_i$, given a collapse redshift, $z_c$, utilizing the formalism presented in \S~\ref{sec:gen_cos_formalism} (Eqn.~\ref{eq:fitting_di_ai} for a flat universe).  Finally, given a halo mass, $M$, the initial co-moving radius of a sphere enclosing this mass can be calculated by noting that
\begin{align}
M &= \frac{4 \pi}{3}\bar \rho_m(0) R_{i,co}^3[1+\bar \delta_i(R_{i})] \nonumber \\
& \cong 1.16 \times 10^{12} M_{\odot} h^{-1} \Omega_m \left (\frac{R_{i,co}}{\mathrm{Mpc}~h^{-1}}\right)^3.
\end{align}

We evaluate Eqn.~\ref{eq:dens_profile_real} at $\rr=R_{i, co}$, divide by $D(a_i)$ and take $a_i \rightarrow 0$ ($D(a_i) \rightarrow 1$) so that the left hand side of this equation becomes $\bar \delta_i(R_i)/a_i$.  Inserting the calculated quantities addressed in the previous paragraph ($R_f$, $\sigma_0$, $r_\star$, $\gamma$, $R_{i, co}$, $\bar \delta_i(R_i)/a_i$) leads to a non-linear equation with a single, unknown variable, $\nu$, for which we solve numerically (the parameter $\theta$ is a function of $\nu$ given a value $\gamma$).  We tabulate $\nu$ and $\theta$ calculated under this prescription for several values of $z_c$, $M$ and $M_f$ in Tab.~\ref{tab:nu_table}.  With $\nu$ and $\theta$ known, the profile $\bar \delta_i(r_i)/ \bar \delta_i(R_i)$ can be found using
\begin{equation}
\frac{\bar \delta_i(x)}{\bar \delta_i(R_i)}= \frac{1}{x}\frac{\int_0^\infty k j_1(kxR_{i, co})P(k)e^{-(R_fk)^2}\left [\frac{\nu -\gamma^2  \nu -\gamma \theta}{1-\gamma^2} +\frac{\theta r_\star^2}{3 \gamma(1-\gamma^2)} k^2\right] dk}{\int_0^\infty k j_1(kR_{i, co})P(k)e^{-(R_fk)^2}\left [\frac{\nu -\gamma^2  \nu -\gamma \theta}{1-\gamma^2} +\frac{\theta r_\star^2}{3 \gamma(1-\gamma^2)} k^2\right] dk},
\end{equation}
easily derived from Eqn.~\ref{eq:dens_profile_real}.  A similar formula can be derived from Eqn.~\ref{eq:gauss_field_density} to solve for the (non-volume averaged) normalized initial density profile, $\delta_i(r_i)/ \bar \delta_i(R_i)$.  We show examples of these profiles for several halo masses, smoothing scales, and collapse redshifts in Fig.~\ref{fig:density_profiles_gen}.

%given a halo mass, $M$, redshift of collapse, $z_c$, and smoothing scale $R_f$.  To solve for this function, we first need to calculate several quantities.  Once a smoothing scale is specified, $r_{\star}$ and $\gamma$ can be easily calculated from Eqn.~\ref{eq:r_star} for a given power spectrum.  The initial comoving radius of a sphere enclosing a halo mass, $M$ can be calculated by noting that\footnote{We again add the subscript ``$co$" to the variable representing the initial co-moving radius of the sphere to distinguish it from the initial, physical radius, $R_i$.}
%\begin{align}
%M &= \frac{4 \pi}{3}\bar \rho_m(0) R_{i,co}^3[1+\bar \delta_i(R_{i} |z_c)] \nonumber \\
%& \cong 1.16 \times 10^{12} M_{\odot} h^{-1} \Omega_m \left (\frac{R_{i,co}}{\mathrm{Mpc}~h^{-1}}\right)^3.
%\end{align}
%Finally, given $z_c$, we calculate the quantity $\bar \delta_i(R_i)/a_i$ from Eqns.~\ref{eq:collapse_time} and~\ref{eq:friedmann} (which can equivalently be found from the convenient expression, Eqn.~\ref{eq:fitting_di_ai}).

%\begin{equation}
%\frac{\bar \delta_i(R_{i, co})}{a_i}= \frac{3}{2 \pi^2 \sigma_0 R_{i, co}}\frac{1}{D(0)}\int_0^\infty k j_1(kR_{i, co})P(k)e^{-(R_fk)^2}\left [\frac{\nu -\gamma^2  \nu -\gamma \theta}{1-\gamma^2} +\frac{\theta r_\star^2}{3 \gamma(1-\gamma^2)} k^2\right] dk,
%\end{equation}

  \begin{table}[h]
\caption{$\nu$ and $\theta$ for different halos} % title name of the table
{\centering % centering table
\begin{tabular}{c c lll c lll} % creating 10 columns
\hline\hline % inserting double-line
$z_c$ & $M~[\mathrm{M_\odot} h^{-1}]$ &\multicolumn{3}{c}{$\nu^\mathrm{a}$}&   &\multicolumn{3}{c}{$\theta^\mathrm{a}$}
\\ [0.5ex]
\hline
  \\[-2.ex]
  
& $10^{15}$ & 11.0716 & 14.4091 & 18.1462 & & 0.347416 & 0.296147 & 0.258404\\[-.5 ex]
& $10^{13}$ & 3.47730 & 4.00833 & 4.59198 & & 1.20029 & 1.16248 & 1.11979 \\[-1 ex]
\raisebox{1. ex}{0} & $10^{11}$ & 1.99916 & 2.14021 & 2.28591 & & 1.62999 & 1.63876 & 1.64644\\[-.5 ex]
& $10^{9}$ & 1.46096 & 1.49284 & 1.61482 & & 1.82373 & 1.84741 & 1.75778 \\[1.5 ex]
    
& $10^{15}$ & 14.0725 & 18.3877 & 23.2059 &  & 0.271592 & 0.231193 & 0.201581\\[-.5 ex]
& $10^{13}$ & 4.16247 & 4.86503 & 5.63470 &  & 1.07048 & 1.01764 & 0.963021 \\[-1 ex]
\raisebox{1. ex}{0.5} & $10^{11}$ & 2.28085 & 2.47430 & 2.67408 &  & 1.56644 & 1.56723 & 1.56752\\[-.5 ex]
& $10^{9}$ & 1.61455 & 1.66656 & 1.79871 &  & 1.78976 & 1.81067 & 1.71615 \\[1.5 ex]
     
& $10^{15}$ & 17.6053 & 23.0571 & 29.1345 &  & 0.216289 & 0.183973 & 0.160342\\[-.5 ex]
& $10^{13}$ & 4.97424 & 5.88034 & 6.87037 &  & 0.936227 & 0.872630 & 0.811449 \\[-1 ex]
\raisebox{1. ex}{1} & $10^{11}$ & 2.61115 & 2.86608 & 3.12927 &  & 1.49436 & 1.48646 & 1.47880\\[-.5 ex]
& $10^{9}$ & 1.79424 & 1.86980 & 2.01397 &  & 1.75057 & 1.76832 & 1.66831 \\[1.5 ex]
    
& $10^{15}$ & 25.3402 & 33.2571 & 42.0693 &  & 0.149763 & 0.127298 & 0.110906\\[-.5 ex]
& $10^{13}$ & 6.78250 & 8.14031 & 9.61633 &  & 0.707607 & 0.640391 & 0.582862 \\[-1 ex]
\raisebox{1. ex}{2} & $10^{11}$ & 3.33495 & 3.72473 & 4.12687 & & 1.34587 & 1.32154 & 1.29918\\[-.5 ex]
& $10^{9}$ & 2.18643 & 2.31334 & 2.48431  & & 1.66713 & 1.67834 & 1.56720 \\[1.5 ex]
   
& $10^{15}$ & 33.3940 & 43.8634 & 55.5100  & & 0.113493 & 0.0964424 & 0.0840119 \\[-.5 ex]
& $10^{13}$ & 8.70946 & 10.5399 & 12.5203 &  & 0.548857 & 0.491082 & 0.444443 \\[-1 ex]
\raisebox{1. ex}{3} & $10^{11}$ & 4.09315 & 4.62437 & 5.17210 &  & 1.20463 & 1.16702 & 1.13338\\[-.5 ex]
& $10^{9}$ & 2.59483 & 2.77521 & 2.97478  & & 1.58339 & 1.58827 & 1.46689 \\[1.5 ex]
  
& $10^{15}$ & 41.5547 & 54.6043 & 69.1171  & & 0.0911469 & 0.0774432 & 0.0674569 \\[-.5 ex]
& $10^{13}$ & 10.6947 & 13.0009 & 15.4878  &  & 0.443557 & 0.395722 & 0.357683 \\[-1 ex]
\raisebox{1. ex}{4} & $10^{11}$ & 4.86788 & 5.54379 & 6.24018 & & 1.07563 & 1.02855 & 0.987555\\[-.5 ex]
& $10^{9}$ & 3.00952 & 3.24413 & 3.47361 & & 1.50173 & 1.50073 & 1.37039 \\[1.5 ex]
  
& $10^{15}$ & 49.7604 & 65.4011 & 82.7929  & & 0.0760898 & 0.0646453 & 0.0563071\\[-.5 ex]
& $10^{13}$ & 12.7109 & 15.4920 & 18.4851  & & 0.371276 & 0.330910 & 0.298932 \\[-1 ex]
\raisebox{1. ex}{5} & $10^{11}$ & 5.65413 & 6.47686 & 7.32380  & & 0.960358 & 0.907584 & 0.862900\\[-.5 ex]
& $10^{9}$ & 3.42773 & 3.71698 & 3.97751  & & 1.42290 & 1.41653 & 1.27868 \\[1.5 ex]
   
& $10^{15}$ & 90.9825 & 119.622 & 151.461 &  & 0.0415916 & 0.0353319 & 0.0307728\\[-.5 ex]
& $10^{13}$ & 22.9564 & 28.0998 & 33.6178 &  & 0.203789 & 0.181370 & 0.163696 \\[-1 ex]
\raisebox{1. ex}{10} & $10^{11}$ & 9.70684 & 11.2775 & 12.8858 & & 0.577297 & 0.530356 & 0.494605\\[-.5 ex]
& $10^{9}$ & 5.55144 & 6.11736 & 6.54909 &  & 1.07889 & 1.05418 & 0.901541 \\[.5 ex]

 \hline
  \end{tabular}}
   $\mathrm{^aE}$ach of the three sub-columns corresponds to a smoothing scale of $10^{-4}$, $10^{-5}$, $10^{-6}$ times the halo mass, $M$, from left to right respectively.
\label{tab:nu_table}
\end{table}

We should note that, as seen by the $\nu$ values in Tab.~\ref{tab:nu_table}, halos with certain combinations of $z_c$, $M$ and $M_f$ (according to the prescription above) originate from unrealistically high peaks in the initial cosmic density field.  Specifically, the largest halos collapsing at the earliest times must be spawned from density perturbations with prohibitively large values of $\nu$.  This facet of structure formation, that the largest halos generally form at later cosmic times is well known, and is reflected in such simple models as the Press-Schecter halo mass function.  For perspective when considering the different halos presented in Tab.~\ref{tab:nu_table}, we here calculate averages and standard deviations of $\nu$ and $\theta$ given $z_c$, $M$ and $M_f$ from the formalism in \cite{bardeen}.

%A natural way to average over $\nu$ is by weighting each bin of $\nu+d \nu$ with the number of peaks (that are associated with the type of object for which $\delta_i$ and $\bar \delta_i$ are desired) in that bin.  In this case, $\nu$ and $\theta$ in Eqns. 23 and 26 must be replaced with $\bar \nu$ and $\bar \theta$, where these quantities are defined below.

A natural way to average these quantities is to weight by peak number.  According to \cite{bardeen}, the co-moving number density of peaks in a smoothed Gaussian random field in the range $\nu$ to $\nu+d\nu$ is
\begin{equation}
\label{eq:n_pk}
\mathcal N_{pk} (\nu) d \nu = \frac{1}{(2 \pi)^2 r_\star^3} e^{-\nu^2/2}G(\gamma, \gamma \nu)d\nu,
\end{equation}
with
\begin{equation}
\label{eq:g_func}
G(\gamma, y) = \frac{1}{\sqrt{2 \pi (1-\gamma^2)}} \int_0^\infty \exp\left[\frac{-(x-y)^2}{2(1-\gamma^2)} \right]f(x)dx, 
\end{equation}
so that the averages are:
\begin{equation}
\bar \nu = \frac{\int_{\nu_{th}}^{\infty} \nu \mathcal N_{pk} (\nu) d \nu}{\int_{\nu_{th}}^{\infty} \mathcal N_{pk} (\nu) d \nu}, ~~\mathrm{and}~~\bar \theta = \frac{\int_{\nu_{th}}^{\infty} \theta(\nu) \mathcal N_{pk} (\nu) d \nu}{\int_{\nu_{th}}^{\infty} \mathcal N_{pk} (\nu) d \nu}.
\end{equation}
The corresponding variances are given by:
\begin{equation}
\label{eq:nu_theta_var}
\sigma_\nu^2 = \frac{\int_{\nu_{th}}^{\infty} (\nu-\bar \nu)^2 \mathcal N_{pk} (\nu) d \nu}{\int_{\nu_{th}}^{\infty} \mathcal N_{pk} (\nu) d \nu}, ~~\mathrm{and}~~\sigma_\theta^2 = \frac{\int_{\nu_{th}}^{\infty} [\theta(\nu)-\bar \theta]^2 \mathcal N_{pk} (\nu) d \nu}{\int_{\nu_{th}}^{\infty} \mathcal N_{pk} (\nu) d \nu}.
\end{equation}
As we did for the function $\theta$, we calculate $\mathcal N_{pk}(\nu)$ exactly with Eqns.~\ref{eq:n_pk} and~\ref{eq:g_func}, rather than using the fitting function for $\mathcal N_{pk}(\nu)$ provided by \cite{bardeen}.  In this way, we need not worry about the range of validity of the fitting function.  The paramter, $\nu_{th}$, is a physically motivated peak height threshold to isolate peaks which will eventually turn into the class of objects under consideration (i.e., halos of a certain mass collapsing at a certain redshift $z_c$).  Bardeen \textit{et al} 1986 provide a simple prescription for estimating this value.  They state that, a halo associated with an initial peak height, $\nu<\nu_{th}(z_c)$, will not have collapsed by a redshift $z_c$ if $\nu$ is less than the linear theory over-density at collapse (extrapolated to the present day) in units of $\sigma_0$ today, filtered at the mass scale of interest: 
\begin{equation}
\nu_{th}(z_c) \approx \frac{\bar \delta_{lin}(z=0|z_c)}{\sigma_0(R_f)}.
\end{equation}
For a flat universe with a cosmological constant, the linear theory over-density of a halo at collapse is well approximated by $1.686[\Omega_m(z_c)]^{0.0055}$ (\cite{mo}), so that the linear theory over-density today, given a collapse redshift $z_c$, is:
\begin{equation}
\bar \delta_{lin}(z=0|z_c)= \frac{D(0)}{D(z_c)}1.686 [\Omega_m(z_c)]^{0.0055}.
\end{equation}
Under this prescription, the threshold peak height may therefore be calculated when the halo collapse redshift is specified. 

\begin{figure*}
\centering
\includegraphics[clip, width=5.in]{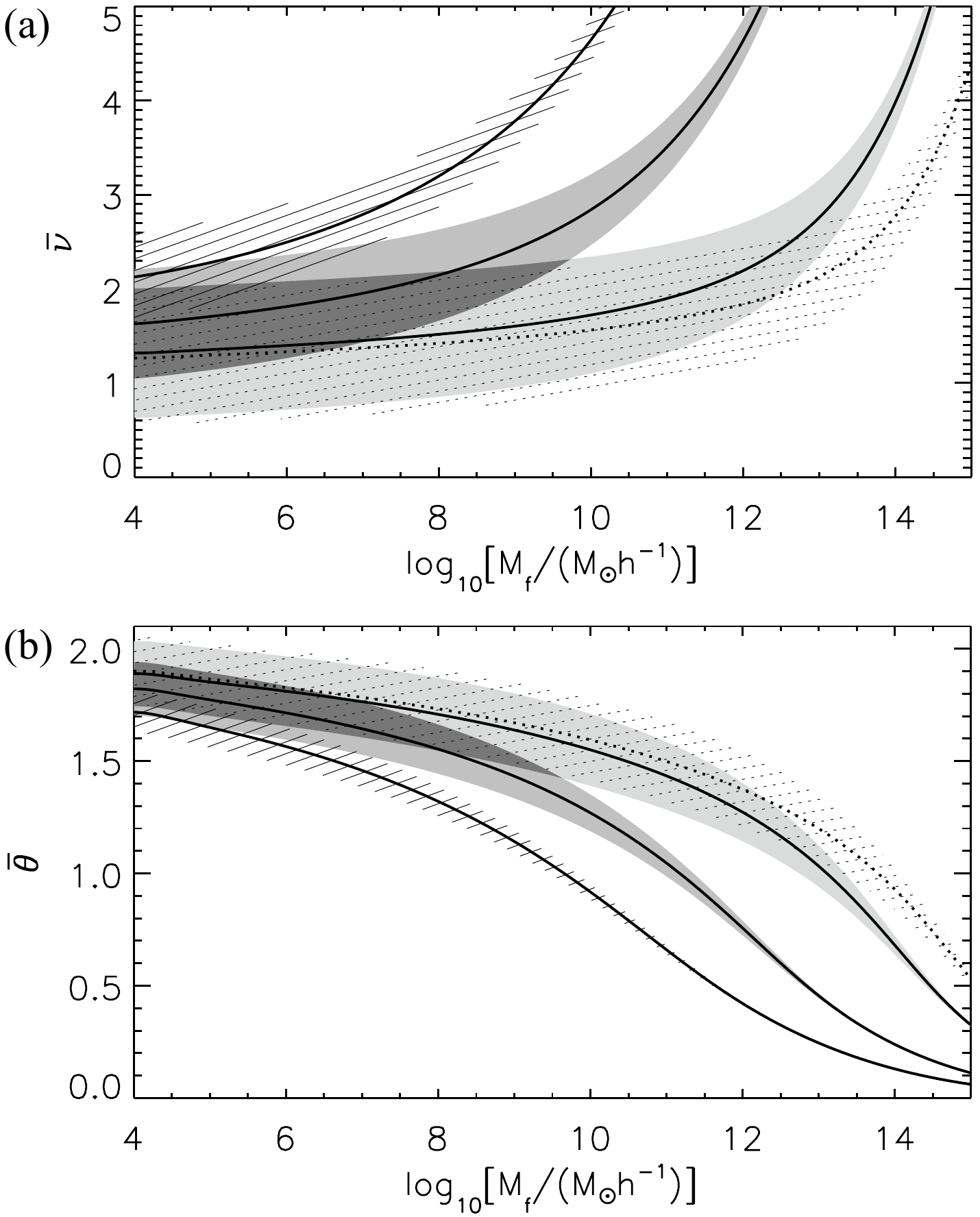}
\caption{ \label{fig:nu_theta}The peak number averaged $\nu$ (a) and $\theta$ (b) values for halos collapsing at different redshifts (different lines) as a function of smoothing scale.  The shaded areas around each line correspond to the 1-$\sigma$ values ($\sigma_\nu$, $\sigma_\theta$).  The lines with the dotted, light grey, dark grey and line-filled 1-$\sigma$ areas correspond to $z_c=0$, 1, 5 and 10, respectively.  For clarity of presentation, the $z_c=0$ line is dotted.} 
\end{figure*}

To illustrate how probable it is to find a halo associated with an initial peak height $\nu$ (and corresponding $\theta$ value) given a collapse redshift $ z_c$ and smoothing scale $M_f$, we plot $\bar \nu$ and $\bar \theta$ in Figs.~\ref{fig:nu_theta}a and b respectively.  We show $\bar \nu$ and $\bar \theta$ as a function of $M_f$ for several values of $z_c$ (lines).  The shaded area around each line corresponds to the 1-$\sigma$ value for $\nu$ and $\theta$, calculated with Eqn.~\ref{eq:nu_theta_var}.  The lines with the dotted, light grey, dark grey and line-filled 1-$\sigma$ areas correspond to $z_c=0$, 1, 5 and 10, respectively.  For clarity of presentation, the $z_c=0$ line is dotted.  These plots show which halos in Tab.~\ref{tab:nu_table} are relatively common (i.e., which halos have $\nu$ and $\theta$ close to the mean) and which are rare (i.e. which halos have $\nu$ and $\theta$ several sigma away from the mean).

\section{Spherical Collapse After Shell Crossing}
\label{sec:simulation}
To follow the dynamics of shells under spherical collapse beyond shell crossing we, employ a one dimensional, Lagrangian shell code with $N_{shells}=10^4$ similar to that used by \cite{thoul}, \cite{lu:2007} and \cite{rubin:2012}.  The code discretizes the density field into a set of concentric, equal mass, shells whose equations of motion must be solved simultaneously since they are coupled via their mutual gravitational attraction.  The equation of motion for an individual shell, labelled as shell ``$j$ '', can be written as two coupled first order differential equations:
\begin{equation}
\label{eq:eom_appendix}
\frac{dv_j}{dt}= -\frac{Gm_j(t)}{r_j^2}+H_o^2\Omega_\Lambda r_j
\end{equation}
and
\begin{equation}
\frac{dr_j}{dt}=v_j,
\end{equation}
where
\begin{equation}
m_j(t) = \sum_{j^{\prime}} \Delta m.
\end{equation}
The symbol, $\Delta m$ represents the mass of an individual shell, and $j^{\prime}$ is the subset of shells that satisfy $r_{j^{\prime}}(t) \le r_{j}(t)$.  Equation~\ref{eq:eom_appendix} could also include an $r^{-3}$ outwardly directed force term due to angular momentum, however, for consistency with the rest of this paper, we choose not to include it.

\subsection{Integration Scheme}
We employ the following definitions to non-dimensionalize our code: $\tilde r \equiv r/R_0$, $\tilde v \equiv v/(R_0 H_o \Omega_m/2)$, $\tilde a \equiv a/(R_0H_o^2\Omega_m/2)$, $\tilde m \equiv m/M_0$ and $\tilde t \equiv t/(H_o^{-1})$.  The variables $R_0$ and $M_0$ refer to the position of the outermost shell, and the mass contained within it at the start time our simulation, $t_0$.  Similar to Eqn.~\ref{eq:mass}, $M_0$ can be written as
\begin{equation}
\label{eq:mo_code}
M_0 = R_0^3 H_o^2  \Omega_m [1+\bar \delta_0(R_0)]/(2 Ga_0^3),
\end{equation}
where the factor, $1+\bar \delta_0(R_0)$ is given by
\begin{equation}
\label{eq:delta_0}
1+\bar \delta_0(R_0) = \mathpzc {X}_0^{-3}a_0^3.
\end{equation}
In the previous equation we have assumed a time when the outermost shell has yet to undergo shell crossing.  The factor, $\mathpzc {X}_0$ ($\equiv R_0 a_i/R_i$), is found by solving Eqn.~\ref{eq:t0_out} when $\tilde t_0$ is specified (as explained in the next section).   

%For the calculation, we use the units given by the Rubin and Loeb 2012 code, which scale dimensional quantities by the conditions at the beginning of the simulation.  We note that Rubin and Loeb 2012 started their simulations at a time $t_i$, when both $a_i$ and $\bar \delta_i$ $\ll 1$.  We start our simulations at a later time, $t_0$, in which shell dynamics are slightly non-linear, so that our definitions to non-dimensionalize velocity, mass, position and time are: $\tilde v \equiv v/(R_0 H_o \Omega_m/2)$, $\tilde m \equiv m/M_0$, $\tilde r \equiv r/R_0$, $t \equiv t/(H_o^{-1})$.  The variables $R_0$ and $M_0$ refer to the position and mass contained within the outermost shell of the collapsing halo at $t_0$, the latter of which can be re-written as $M_0 = R_0^3 H_o^2  \Omega_m [1+\bar \delta_0(R_0)]/(2 Ga_0^3)$.  The factor, $1+\bar \delta_0(R_0)$, can be found from
%\begin{equation}
%\label{eq:delta_0}
%1+\bar \delta_0(R_0) = \mathpzc {X}_0^{-3}a_0^3,
%\end{equation}
%valid for a time when shells have yet to cross the outermost shell ($\mathpzc {X}_0$ can be found by solving Eqn.~\ref{eq:t0_out} when $\tilde t_0$ is specified).   
To integrate each shell's equation of motion, we use the (locally) second order accurate kick-drift-kick leap-frog integration scheme with adaptive time-steps.  Using the units adopted for this calculation, and Eqn.~\ref{eq:mo_code}, the non-dimensionalized update equations are written as:

\begin{equation}
\tilde v_{j}^{n+1/2}=\tilde v_j^n+\tilde a_j^n\frac{\widetilde{\Delta t}}{2},
\end{equation}

\begin{equation}
\tilde r_j^{n+1}=\tilde r_j^n+\tilde v_j^{n+1/2} \widetilde{\Delta t}\frac{\Omega_m}{2},
\end{equation}

\begin{equation}
\tilde a_j^{n+1} = -\frac{\tilde m_j^{n+1}}{(\tilde r^{n+1})^2}\left (\frac{1+\bar \delta_0(R_0)}{a_0^3}\right)+2 \frac{\Omega_\Lambda}{\Omega_m} \tilde r_j^{n+1},
\end{equation}
 and 
\begin{equation}
\tilde v_{j}^{n+1}=\tilde v_j^{n+1/2}+\tilde a_j^{n+1}\frac{\widetilde{\Delta t}}{2},
\end{equation}
where the superscript, ``$n$", indicates the time-step.  

In order to avoid having to resolve the divergence in the force on a shell as it approaches the center, we place a hard inner reflecting sphere at radius $\tilde r_{in}$, a tactic also used by \cite{thoul}.  Clearly, this only an approximation to the full spherical collapse treatment of a collapsing dark matter halo.  However, as long as $\tilde r_{in}$ is sufficiently smaller than all characteristic length scales of the system, the approximation should not significantly affect collapse dynamics.  The relevant length scale of the system is its original size, $R_0$, and we therefore choose $\tilde r_{in} = 0.01$.

The appropriate time scales to consider for choosing the time step, $\widetilde{\Delta t}$, at each iteration are the dynamical time, $\sqrt{\pi^2 r^3/(4 G m)}$, the time it takes for a shell to travel a maximum allowed distance given its velocity, $\ell_{max}/v$,  and the time it takes for a shell to travel a maximum allowed distance given its acceleration, $\sqrt{\ell_{max}/a}$.  The latter two time scales must be considered to ensure that the positions of each shell do not change dramatically across each time-step.  The dynamical time scale of each shell is necessary to consider since the force on each shell blows up as it approaches the center.  By using a time-step much smaller than a shell's dynamical time, we ensure that the shell does not fall too far a distance over which the force changes appreciably.  To time resolve the dynamics of the shells, we therefore choose the time-step at each iteration in the code according to:
\begin{equation}
\widetilde{\Delta t}=\min \{\widetilde{\Delta t}^{dyn}_j,~\widetilde{\Delta t}^v_j,~\widetilde{\Delta t}^a_j, ~\widetilde{\Delta t}_{end} \}.
\end{equation}
Choosing $\ell_{max} =R_0$, and dimensionalizing to the proper units, these time-steps are:
\begin{equation}
\widetilde \Delta t_{dyn} = \min_j \left \{ c_{dyn}\sqrt{\frac{\pi^2 (\tilde r_j^{n})^3 a_0^3}{2 \Omega_m[1+\bar \delta_0(R_0)]\tilde m_j^{n}}},\right \},
\end{equation}
\begin{equation}
\Delta \tilde t_{v} = \min_j \left \{c_v \frac{1}{|v_j^{n}+\epsilon|}\frac{2}{\Omega_m} \right \},
\end{equation}
and
\begin{equation}
\Delta \tilde t_{a}= \min_j \left \{c_a \sqrt{\frac{2}{\Omega_m|a_j^{n+1}+\epsilon|}} \right \}.
\end{equation}
Here $\epsilon$ is a small number in order to keep $\widetilde \Delta t$ from blowing up if the velocities or accelerations are small, and $c_{dyn}$, $c_v$ and $c_a$ are safety constants.  We find that $c_{dyn}=c_{v}=c_a=10^{-4}$ provides adequate time resolution.  We add one last time-step to ensure that the simulation ends exactly when we wish it to end:
\begin{equation}
\widetilde \Delta t_{end}= \tilde t_{end}-\tilde t^{n}.
\end{equation}
Since we are interested in the state of the system at the turn-around time of the outermost shell, we stop the simulation at $\tilde t_{end} = \tilde t_{ta}$, calculated from non-linear theory.

At the end of each simulation, we wish to calculate the non-dimensionalized kinetic energy and binding energies due gravity and dark energy of the system.  Given the definitions utilized to non-dimensionalize our code, it is straightforward to show that 
\begin{equation}
\label{eq:u_ta_sim}
\mathscr{U}_{ta} = \frac{5}{3}\frac{\mathpzc{X}_{ta}}{\mathpzc{X}_0} \widetilde{\Delta m}\sum_j\frac{\tilde m_j^{n=N}}{\tilde r_j^{n=N}},
\end{equation}
\begin{equation}
\mathscr{K}_{ta} = \frac{5}{12} \Omega_m \mathpzc{X}_{ta} \mathpzc{X}_{0}^2 \widetilde{\Delta m}\sum_j \left (\tilde v_j^{n=N} \right)^2,
\end{equation}
and
\begin{equation}
\label{eq:u_ta_lam_sim}
\mathscr{U}_{ta}^{\Lambda} = \frac{5}{3} \frac{\mathpzc{X}_{0}^2}{\mathpzc{X}_{ta}^2} \widetilde{\Delta m}\sum_j \left (\tilde r_j^{n=N} \right)^2.
\end{equation}
Here, $n=N$ refers to the last time step of the calculation.
By taking advantage of Newton's iron shell theorem, we avoid having to gravitationally soften the trajectories of particles that venture too close to each other, as with full, three dimensional simulations.  However, \cite{lu:2006} point out that in our strategy, shells experience an unrealistic discontinuity in force when they cross each other.  This is due to the discretization of the density field and the fact that under the iron shell theorem, shells only feel the gravitational force of other shells at smaller radii.  Indeed, in our simulations we observe a degradation in energy conservation associated with shell crossing events.  To alleviate this effect, we try to ``soften" the crossings using the same tactic as \cite{lu:2006}.  We give each shell a small thickness and assume that the total mass of the shell is spread uniformly across its volume.  Therefore, when two shells undergo crossing, they gradually overlap and the force on either smoothly changes.  We also employ a ``shell crossing time scale" when choosing $\widetilde{\Delta t}$  to properly time resolve the crossing event.  Unfortunately, we find that this strategy does not significantly improve energy conservation, or the convergence of individual shell trajectories.  Specifically, for the steepest initial density profiles we can only obtain reliable convergence with unrealistically high resolution.  In this paper, we therefore only show examples for which we are confident that our final results have converged.  

\begin{figure*}
\centering
\includegraphics[clip, width=6.2in]{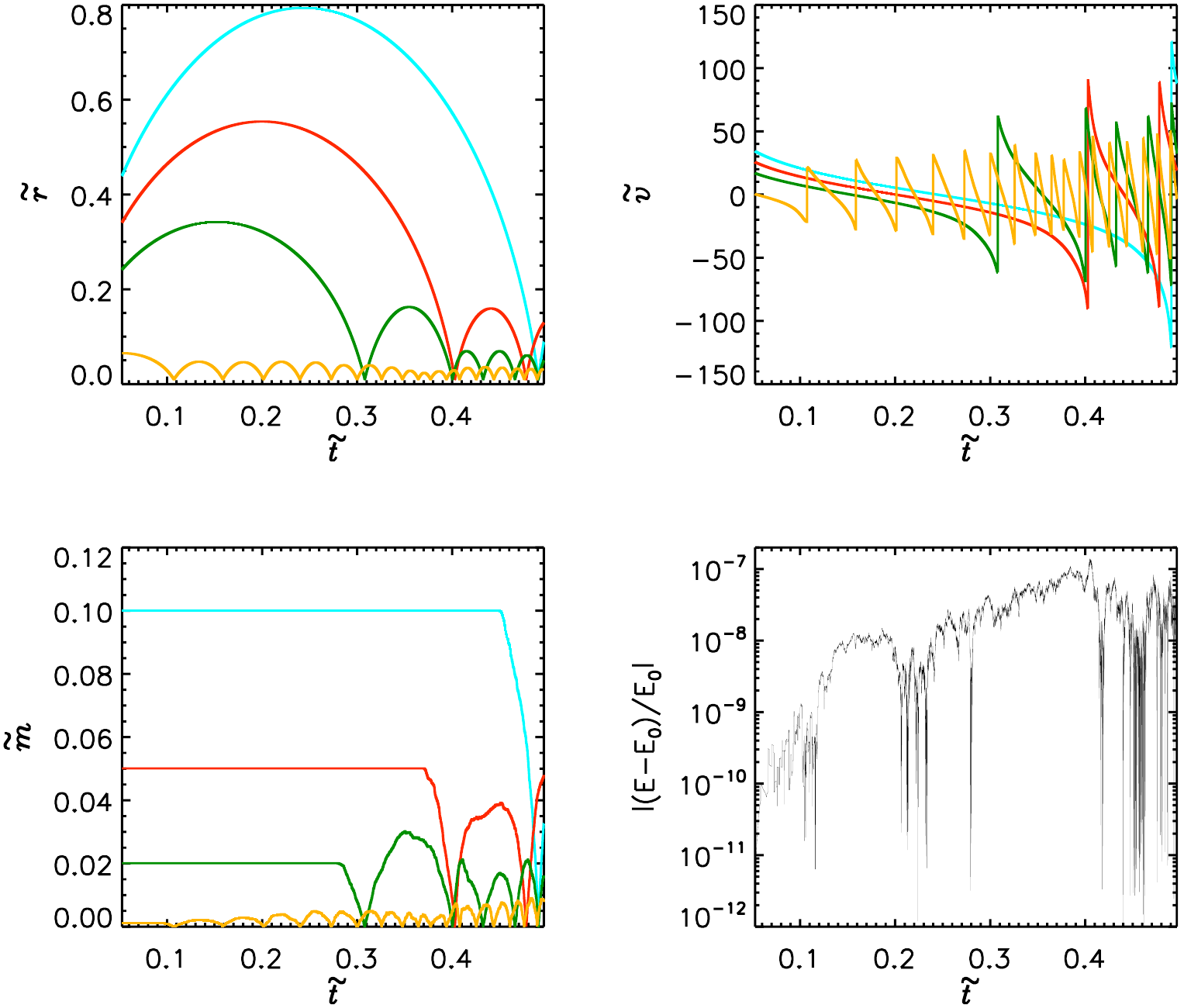}
\caption{ \label{fig:example} Example trajectories, velocity and interior mass profiles for several different shells in a $10^9$M$_\odot h^{-1}$ mass halo collapsing at $z_c=0$.  We also show the fractional difference in energy as a function of time to illustrate our level of energy conservation.} 
\end{figure*}

In Fig.~\ref{fig:example}, we show example trajectories as well as several other quantities for several shells for a $10^9$M$_\odot h^{-1}$ collapsing at $z_c=0$.  The bottom right panel in the figure shows that energy is very well conserved.

%We test for convergence, and find very good convergence with $c_{dyn}=c_{v}=c_a=10^{-4}$.  We also find excellent convergence for $N_s=10^4$ and $\tilde r_{in}=0.01$. 

\subsection{Initial Conditions}

We start our simulations at a time, $\tilde t_0$, corresponding to the time when the innermost shell in the simulation has already turned around and is just bouncing off the inner boundary at $\tilde r_{in}$.  We choose this start time in order to reduce simulation computation time, since it is the latest time at which non-linear, analytic theory is valid.  In order to calculate $\tilde t_0$, we first calculate the initial normalized position of the innermost shell, $x_{j=0}$.  Assuming that no shells cross the first shell between the initial time, $t_i$, and the simulation start time, $t_0$, $m_{0, {j=0}}/M_0 = m_{i, {j=0}}/M_0$ where $m_{0, {j=0}}$ ($m_{i, {j=0}}$) refers to the mass within the innermost shell at time $\tilde t_0$ ($\tilde t_i$).  Since $m_{0, {j=0}}/M_0 = \widetilde{\Delta m}=1/N_{shells}$,
\begin{equation}
\label{eq:N_eq}
\frac{1}{N_{shells}} =\frac{\left (r_{i, {j=0}}\right)^3 a_0^3}{R_0^3 a_i^3}\frac{\left \{1+\bar \delta_i \left (r_{i, {j=0}}\right ) \right \}}{\left \{1+\bar \delta_0(R_0) \right \}}.
\end{equation}
Using Eqn.~\ref{eq:delta_0}, the fact that $\bar \delta_i(r_{i, {j=0}}) \ll 1$, and simplifying, one can show that:
\begin{equation}
\label{eq:initial_z}
z_{j=0} = N_{shells}^{-1/3}.
\end{equation}

By specifying the halo mass and collapse redshift, we find the initial seed, $\bar \delta_i(R_i)/a_i$, and the initial density profile, $\bar \delta_i(x)/\bar \delta_i(R_i)$ (which we evaluate at $x_{j=0}$), given the formalism presented in App.~\ref{app:density_profiles}.  The turn-around radius of the innermost shell, $\mathpzc{x}_{\scriptscriptstyle{\mathscr{TA}}}(x_{j=0})$, is found from Eqn.~\ref{eq:cubic_ta} with 
$\bar \delta_i(r_i)/a_i \rightarrow [\bar \delta_i(x_{j=0})/\bar \delta_i(R_i)][\bar \delta_i(R_i)/a_i]$.  The start time of our simulation can then be found by evaluating the following integral, calculated from the trajectory of the innermost shell:
\begin{equation}
\label{eq:t0_in}
\tilde t_0 = \mathcal I \left[0, \mathpzc{x}_{\scriptscriptstyle{\mathscr{TA}}}(x_{j=0}),  \left (\frac{\bar \delta_i(\mathpzc{x_{j=0}})}{\bar \delta_i(R_i)}\right) \left(\frac{\bar \delta_i(R_i)}{a_i}\right) \right]+\mathcal I \left[\frac{\tilde r_{in}}{x_{j=0}}\mathpzc{X}_0, \mathpzc{x}_{\scriptscriptstyle{\mathscr{TA}}}(x_{j=0}),  \left (\frac{\bar \delta_i(x_{j=0})}{\bar \delta_i(R_i)}\right) \left(\frac{\bar \delta_i(R_i)}{a_i}\right) \right].
\end{equation}
The start time can also be found with
\begin{equation}
\label{eq:t0_out}
\tilde t_0 = \mathcal I \left[0, \mathpzc{X}_0, \left(\frac{\bar \delta_i(R_i)}{a_i}\right) \right],
\end{equation}
calculated from the trajectory of the outermost shell.  Equations~\ref{eq:t0_in} and \ref{eq:t0_out} form a complete set of equations in $\tilde t_0$ and $\mathpzc{X}_0$, which we solve for numerically.  For a flat universe with a cosmological constant, the corresponding scale factor (necessary to calculate $\bar \delta_0(R_0)$) can be found from:
\begin{equation}
a(t) = \left(\frac{\Omega_m}{\Omega_\Lambda} \right)^{1/3} \left [ \sinh \left(\frac{3}{2} H_o t \sqrt{\Omega_\Lambda} \right) \right]^{2/3}.
\end{equation}

We initialize the position and velocity of each shell at $\tilde t_0$ by using non-linear theory.  Similar to Eqn.~\ref{eq:m_int}, one can show that the mass interior to a position $\tilde r_0 \ge \tilde r_{in}$ is given by:
\begin{equation}
\label{eq:m0_initial}
\tilde m_0 (\tilde r_0) = \widetilde{\Delta m}+3\int_{\tilde r_{in}}^{\tilde r_0} x^2(\tilde r_0) \frac{dx}{d\tilde r_0}(\tilde r_0)d\tilde r_0,
\end{equation}
where we calculate the $x$ to $\tilde r_0$ mapping with non-linear theory from a procedure similar to that as presented in \ref{sec:mapping_gen}.  Once this mass profile is calculated, we place shells at positions, $\tilde r_j^{n=0}$, which satisfy the relation $\tilde m_0(\tilde r_j^{n=0})/\widetilde{\Delta m}=j+1$ with $j=0, 1, ..., N_s-1$.  To initialize velocity, it is straightforward to show from Eqn.~\ref{eq:eq_of_motion} with some algebra that the  velocity of each shell at $t_0$ is given by
\begin{equation}
\label{eq:v0_initial}
\tilde v_0 (\tilde r_0) = \pm 2 \left \{ \frac{x^3(\tilde r_0)}{\tilde r_0 \mathpzc{X}_0^3 \Omega_m}+\frac{\Omega_\Lambda}{\Omega_m^2}\tilde r_0^2 -\frac{5}{3} \frac{x^2(\tilde r_0)}{ \mathpzc{X}_0^2 \Omega_m}\left( \frac{\bar \delta_i[x(\tilde r_0)]}{\bar \delta_i(R_i)}\right) \left( \frac{\bar \delta_i(R_i)}{a_i}\right) \right \}^{1/2}.
\end{equation}
The plus sign is chosen for shells that have yet to turn-around and are traveling outward ($\tilde{t}_{0} < \tilde{t}_{\scriptscriptstyle{\mathscr{TA}}}$), and the minus sign is chosen for shells that have already turned around and are traveling inward ($\tilde{t}_{0} > \tilde{t}_{\scriptscriptstyle{\mathscr{TA}}}$).  We set the velocity of the $j=0$ shell to $-\tilde v(\tilde r_{0,j=0} )$ since it is just rebounding off the center boundary.

\end{appendices}

\section*{Acknowledgments}
We thank Eli Visbal, Mark Vogelsberger and Paul Torrey for helpful conversations.  This research was supported in part by NSF grant AST-1312034.

\label{lastpage}


\begin{thebibliography}{99}

\bibitem[\protect\citeauthoryear{Bond et al.}{1991}]{bond} Bond J.~R., Cole S., Efstathiou G., Kaiser N., 1991, ApJ, 379, 440 

\bibitem[\protect\citeauthoryear{Sheth, Mo, \& Tormen}{2001}]{sheth:2001} Sheth R.~K., Mo H.~J., Tormen G., 2001, MNRAS, 323, 1 

\bibitem[\protect\citeauthoryear{Sheth \& Tormen}{2002}]{sheth:2002} Sheth R.~K., Tormen G., 2002, MNRAS, 329, 61 

\bibitem[\protect\citeauthoryear{Sheth \& Tormen}{1999}]{sheth:1999} Sheth R.~K., Tormen G., 1999, MNRAS, 308, 119 

\bibitem[\protect\citeauthoryear{Jenkins et al.}{2001}]{jenkins} Jenkins A., Frenk C.~S., White S.~D.~M., Colberg J.~M., Cole S., Evrard A.~E., Couchman H.~M.~P., Yoshida N., 2001, MNRAS, 321, 372 

\bibitem[\protect\citeauthoryear{Springel et al.}{2005}]{springel} Springel V., et al., 2005, Natur, 435, 629 

\bibitem[\protect\citeauthoryear{Warren et al.}{2006}]{warren} Warren M.~S., Abazajian K., Holz D.~E., Teodoro L., 2006, ApJ, 646, 881 

\bibitem[\protect\citeauthoryear{Reed et al.}{2007}]{reed} Reed D.~S., Bower R., Frenk C.~S., Jenkins A., Theuns T., 2007, MNRAS, 374, 2 

\bibitem[\protect\citeauthoryear{Tinker et al.}{2008}]{tinker} Tinker J., Kravtsov A.~V., Klypin A., Abazajian K., Warren M., Yepes G., 
Gottl{\"o}ber S., Holz D.~E., 2008, ApJ, 688, 709 

\bibitem[\protect\citeauthoryear{Crocce et al.}{2010}]{crocce} Crocce M., Fosalba P., Castander F.~J., Gazta{\~n}aga E., 2010, MNRAS, 403, 1353 

\bibitem[\protect\citeauthoryear{Angulo et al.}{2012}]{angulo} Angulo R.~E., Springel V., White S.~D.~M., Jenkins A., Baugh C.~M., Frenk C.~S., 2012, MNRAS, 426, 2046 

\bibitem[\protect\citeauthoryear{Davis et al.}{1985}]{davis} Davis M., Efstathiou G., Frenk C.~S., White S.~D.~M., 1985, ApJ, 292, 371 

\bibitem[\protect\citeauthoryear{Lacey \& Cole}{1994}]{lacey:1994} Lacey C., Cole S., 1994, MNRAS, 271, 676 

\bibitem[\protect\citeauthoryear{Watson et al.}{2013}]{watson} Watson W.~A., Iliev I.~T., D'Aloisio A., Knebe A., Shapiro P.~R., Yepes G., 2013, MNRAS, 433, 1230 

\bibitem[\protect\citeauthoryear{Loeb}{2012}]{loeb:2012}A. Loeb, \& S. Furlanetto, ``The First Galaxies in the UniverseÓ, Princeton University Press, in press (2012)

\bibitem[\protect\citeauthoryear{Navarro, Frenk, \& White}{1996}]{navarro} Navarro J.~F., Frenk C.~S., White S.~D.~M., 1996, ApJ, 462, 563 

\bibitem[\protect\citeauthoryear{Gunn \& Gott}{1972}]{gunn} Gunn J.~E., Gott J.~R., III, 1972, \emph{ApJ}, {\bf176}, 1 

\bibitem[\protect\citeauthoryear{Lahav et al.}{1991}]{lahav} Lahav O., Lilje P.~B., Primack J.~R., Rees M.~J., 1991, MNRAS, 251, 128 

\bibitem[\protect\citeauthoryear{Bryan \& Norman}{1998}]{bryan} Bryan G.~L., Norman M.~L., 1998, \emph{ApJ}, {\bf495}, 80 

\bibitem[\protect\citeauthoryear{Eke, Cole, \& Frenk}{1996}]{eke} Eke V.~R., Cole S., Frenk C.~S., 1996, MNRAS, 282, 263 

\bibitem[\protect\citeauthoryear{Lacey \& Cole}{1993}]{lacey} Lacey C., Cole S., 1993, MNRAS, 262, 627 

\bibitem[\protect\citeauthoryear{Rubin \& Loeb}{2012}]{rubin:2012} Rubin D., Loeb A., 2013, submitted

\bibitem[\protect\citeauthoryear{Lokas \& Hoffman}{2001}]{lokas} Lokas E.~L., Hoffman Y., 2001, astro, arXiv:astro-ph/0108283 

\bibitem[\protect\citeauthoryear{Loeb}{2006}]{loeb:2006} Loeb A., 2006, astro, arXiv:astro-ph/0603360

\bibitem[\protect\citeauthoryear{Peebles \& Yu}{1970}]{peebles} Peebles P.~J.~E., Yu J.~T., 1970, \emph{ApJ}, {\bf162}, 815 
 
\bibitem[\protect\citeauthoryear{Bertschinger}{1985}]{bertschinger_1} Bertschinger E., 1985, \emph{ApJS}, {\bf58}, 39 

\bibitem[\protect\citeauthoryear{Bertschinger}{1985}]{bertschinger_2} Bertschinger E., 1985, \emph{ApJS}, {\bf58}, 1 

\bibitem[\protect\citeauthoryear{Sheth \& van de Weygaert}{2004}]{sheth:2004} Sheth R.~K., van de Weygaert R., 2004, \emph{MNRAS}, {\bf350}, 517 

\bibitem[\protect\citeauthoryear{Peebles}{1980}]{peebles:1980}Peebles P.~J.~E., ``The large-scale structure of the universe", Princeton University Press (1980)

\bibitem[\protect\citeauthoryear{Zhao et al.}{2009}]{zhao:2009} Zhao D.~H., Jing Y.~P., Mo H.~J., B{\"o}rner G., 2009, ApJ, 707, 354 

\bibitem[\protect\citeauthoryear{Carroll, Press, \& Turner}{1992}]{carroll} Carroll S.~M., Press W.~H., Turner E.~L., 1992, \emph{ARA\&A}, {\bf 30}, 499 

\bibitem[\protect\citeauthoryear{Mo, van den Bosch, \& White}{2010}]{mo} Mo H., van den Bosch F.~C., White S., 2010, gfe..book,  

\bibitem[\protect\citeauthoryear{Komatsu et al.}{2011}]{komatsu} Komatsu E., et al., 2011, ApJS, 192, 18 


\bibitem[\protect\citeauthoryear{Bardeen et al.}{1986}]{bardeen} Bardeen J.~M., Bond J.~R., Kaiser N., Szalay A.~S., 1986, ApJ, 304, 15 

\bibitem[\protect\citeauthoryear{Lilje \& Lahav}{1991}]{lilje} Lilje P.~B., Lahav O., 1991, ApJ, 374, 29 

\bibitem[\protect\citeauthoryear{Eisenstein \& Loeb}{1995}]{eisenstein} Eisenstein D.~J., Loeb A., 1995, ApJ, 439, 520 

\bibitem[\protect\citeauthoryear{Cupani, Mezzetti, \& Mardirossian}{2008}]{cupani:2008} Cupani G., Mezzetti M., Mardirossian F., 2008, MNRAS, 390, 645 

\bibitem[\protect\citeauthoryear{Cupani, Mezzetti, \& Mardirossian}{2011}]{cupani:2011} Cupani G., Mezzetti M., Mardirossian F., 2011, MNRAS, 417, 2554 


\bibitem[\protect\citeauthoryear{Thoul \& Weinberg}{1995}]{thoul} Thoul A.~A., Weinberg D.~H., 1995, ApJ, 442, 480 

\bibitem[\protect\citeauthoryear{Lu \& Mo}{2007}]{lu:2007} Lu Y., Mo H.~J., 2007, MNRAS, 377, 617 

\bibitem[\protect\citeauthoryear{Lu et al.}{2006}]{lu:2006} Lu Y., Mo H.~J., Katz N., Weinberg M.~D., 2006, MNRAS, 368, 1931 


\bibitem[\protect\citeauthoryear{Press \& Schechter}{1974}]{press:1974} Press W.~H., Schechter P., 1974, ApJ, 187, 425 

\end{thebibliography}
\end{document}